\documentclass[a4paper,12pt]{article}
\usepackage{authblk}
\usepackage[dvipsnames]{xcolor}
\usepackage{latexsym,amsmath,amssymb,amsfonts,amsthm,dsfont,enumerate,natbib,bbm,threeparttable,rotating, pdflscape,verbatim,amssymb, setspace, graphicx}
\usepackage{algorithm,algcompatible,algpseudocode,booktabs,xcolor,tikz}
\usetikzlibrary{arrows,shapes,chains}
\usepackage[normalem]{ulem}

\usepackage[colorlinks,linkcolor=red,citecolor=blue,urlcolor=blue,breaklinks]{hyperref}
\usepackage{cleveref}

\usepackage{multirow,multicol}

\usepackage{mathrsfs}

\def\code#1{\texttt{#1}}

\newcommand{\bdelta}{\mbox{\boldmath $\delta$}}

\algnewcommand\algorithmicpara{\textbf{estimate in parallel for}}
\algblockdefx[PARA]{Para}{EndPara}[1]{\algorithmicpara\ #1}{\algorithmicend}
\algnewcommand\algorithmicdivide{\textbf{divide}}
\algnewcommand\algorithmicextract{\textbf{matrix decomposition}}
\algblockdefx[DIVIDE]{Divide}{EndDivide}{\algorithmicdivide}{\algorithmicend}
\algblockdefx[EXTRACT]{Extraction}{EndExtract}{\algorithmicextract}{\algorithmicend}

\algnewcommand{\algorithmicendif}{\textbf{end}}
\algblockdefx[IF]{If}{EndIf}[1]{\algorithmicif\ #1\ \algorithmicthen}{\algorithmicendif}

\newtheorem{definition}{Definition}
\newtheorem{theorem}{Theorem}
\newtheorem{lemma}{Lemma}
\newtheorem{condition}{Condition}
\newtheorem{proposition}{Proposition}

\newcommand{\bM}{\mbox{\bf M}}
\newcommand{\bmm}{\mbox{\bf m}}
\newcommand{\bP}{\mbox{\bf P}}

\newcommand{\ba}{\mbox{\bf a}}
\newcommand{\bb}{\mbox{\bf b}}
\newcommand{\bc}{\mbox{\bf c}}

\newcommand{\bs}{\mbox{\bf s}}
\newcommand{\bu}{\mbox{\bf u}}
\newcommand{\bv}{\mbox{\bf v}}

\newcommand{\bx}{\mbox{\bf x}}
\newcommand{\by}{\mbox{\bf y}}
\newcommand{\bz}{\mbox{\bf z}}
\newcommand{\bA}{\mbox{\bf A}}
\newcommand{\bB}{\mbox{\bf B}}
\newcommand{\bC}{\mbox{\bf C}}
\newcommand{\bD}{\mbox{\bf D}}
\newcommand{\bE}{\mbox{\bf E}}

\newcommand{\bG}{\mbox{\bf G}}

\newcommand{\bI}{\mbox{\bf I}}

\newcommand{\bQ}{\mbox{\bf Q}}

\newcommand{\bU}{\mbox{\bf U}}
\newcommand{\bV}{\mbox{\bf V}}
\newcommand{\bW}{\mbox{\bf W}}
\newcommand{\bX}{\mbox{\bf X}}

\newcommand{\bY}{\mbox{\bf Y}}
\newcommand{\bZ}{\mbox{\bf Z}}

\newcommand{\bzero}{\mbox{\bf 0}}

\newcommand{\balpha}{\mbox{\boldmath $\alpha$}}

\newcommand{\bzeta}{\mbox{\boldmath $\zeta$}}

\newcommand{\bSig}{\mbox{\boldmath $\Sigma$}}

\newcommand{\diag}{\mathrm{diag}}

\newcommand{\bepsilon}{\mbox{\boldmath $\epsilon$}}

\DeclareMathOperator*{\argmin}{argmin}

\RequirePackage{lineno}

\providecommand{\keywords}[1]{\textbf{\textit{Keywords---}} #1}

\begin{document}

\title{Simultaneous Heterogeneity and Reduced-Rank Learning for Multivariate Response Regression \footnote{Jie Wu is Lecturer, School of Big Data and Statistics, Anhui University, Hefei, Anhui 230601, China (E-mail: jiewu@ahu.edu.cn). Bo Zhang is Associate Professor, International Institute of Finance, School of Management, University of Science and Technology of China, Hefei, Anhui 230026, China (E-mail: wbchpmp@ustc.edu.cn). Daoji Li is Associate Professor, College of Business and Economics, California State University, Fullerton, CA, 92831 (E-mail: dali@fullerton.edu). Zemin Zheng is Professor, International Institute of Finance, School of Management, University of Science and Technology of China, Hefei, Anhui, 230026, China (E-mail: zhengzm@ustc.edu.cn).  This work was supported by the National Key R$\&$D Program of China (Grant 2022YFA1008000), the Natural Science Foundation of China (Grants 12401330, 12471268, 72071187, 71731010, and 71921001), the Doctoral Research Start-up Funds Projects of Anhui University (Grant S020318033/005), and the University Natural Science Research Project of Anhui Province (Grant 2023AH050101).}}
\author[1]{Jie Wu}
\author[2]{Bo Zhang}
\author[3]{Daoji Li}
\author[2]{Zemin Zheng}

%
%

\affil[1]{{Anhui University}}

\affil[2]{{University of Science and Technology of China}}

\affil[3]{{California State University, Fullerton}}

\date{
	\medskip
	\medskip
	This manuscript has been published online in \href{https://www.sciencedirect.com/journal/journal-of-multivariate-analysis}{\textit{Journal of Multivariate Analysis}} on December 5, 2025.\\
	\vspace{2mm}
	The final published version is available online at \url{https://doi.org/10.1016/j.jmva.2025.105578}.
}
\maketitle

\begin{abstract}
Heterogeneous data are now ubiquitous in many applications in which correctly identifying the subgroups from a heterogeneous population is critical. Although there is an increasing body of literature on subgroup detection, existing methods mainly focus on the univariate response setting. In this paper, we propose a joint heterogeneity and reduced-rank learning framework to simultaneously identify the subgroup structure and estimate the covariate effects for heterogeneous multivariate response regression. In particular, our approach uses rank-constrained pairwise fusion penalization and conducts the subgroup analysis without requiring prior knowledge regarding the individual subgroup memberships.  We implement the proposed approach by an alternating direction method of multipliers (ADMM) algorithm and show its convergence.  We also establish the asymptotic properties for the resulting estimators under mild and interpretable conditions. A predictive information criterion is proposed to select the rank of the coefficient matrix with theoretical support. The effectiveness of the proposed approach is demonstrated through simulation studies and a real data application.
\end{abstract}

\keywords{Heterogeneity; Latent group; Multi-response; Oracle property; Penalized fusion; Subgroup analysis}


\newpage


\section{Introduction\label{sec:1}}

Heterogeneous data are
ubiquitous in many fields, such as health care, bioinformatics, marketing, and economics. In these fields, correctly identifying subgroups from the heterogeneous data is essential in promoting individual strategies, 
which, in turn, can contribute to a deeper understanding of the research bases, more accurate assessment and better individual prediction. For example, in medical studies, heterogeneity of treatment effects may arise due to underlying differences between groups of patients in terms of the risk, biology, pathology, genetics, severity of disease, and so on. It is necessary to identify subgroups to give precise medical treatments to heterogeneous subgroups of 
patients~\citep{Crump2008, guo2021inference}. 
Similarly, in marketing and economics, homogeneous customers usually have similar behaviors and preferences. Precision marketing that offers personalized customer service can help enterprises increase their profits by identifying different marketing subgroups~\citep{You2015, zhu2023simultaneous}. 

Many statistical methods have been proposed to identify subgroups from heterogeneous data in the past few decades. 
One popular class of subgroup identification methods is based on the finite mixture modeling strategy, which assumes that data come from a mixture of subgroups and each subgroup has its own sets of parameter values. See, for example, 
~\cite{Hastie1996, Wang2016}.  However, these mixture modeling approaches require to know the number of subgroups in the population in advance and also need to specify an underlying distribution for the data, which can be challenging in practice~\citep{Li2010, Kasahara2015}. 
To address these issues, \citet{Shu2017} introduced a pairwise fusion penalized approach for linear regression models to automatically detect heterogeneity and divide the observations into subgroups without the need to know the number of subgroups in advance. The key idea of~\citet{Shu2017} is to define the subgroup structure by group-specific intercepts after adjusting for the effects of covariates and then apply concave penalty functions, such as SCAD~\citep{Fan2001} and MCP~\citep{Zhang2010}, to pairwise differences of the intercepts. 
Such an idea for subgroup identification has been extended to different model settings, including Poisson regression models~\citep{Chen2019}, quantile regression~\citep{Zhangsinica2019,Lu2021}, censored linear regression~\citep{Yan2021}, the Cox model~\citep{Hu2021},  additive partially linear models~\citep{Liu2019, cai2024subgroup}, functional linear regression~\citep{li2021clusterwise, Zhang2022}, functional partial linear regression~\citep{ma2023subgroup}. 
See also~\cite{liu2023fusion} for subgroup analysis of Alzheimer's disease. However, the aforementioned 
methods mainly focus on the univariate response setting. 

In the era of big data, the coexistence of heterogeneity and large number of responses is becoming more prevalent. To characterize the dependence between responses and predictors in multivariate response regression, one commonly used technique is dimension reduction.  In particular, reduced-rank regression~\citep{Anderson1951, Izenman1975} is a popular approach to dimension reduction by constraining the coefficient matrix to be of low rank such that the responses depend on the predictors through a few latent factors. This provides an appealing low-dimensional latent model interpretation for multivariate response regression.  A number of efforts have been made in the literature on reduced-rank regression. See, for example, \citet{Yuan2007}, \citet{Bunea2011}, \citet{Koltchinskii2011}, \citet{Chen2012a}, 
\citet{Chen2012sparse}, \citet{She2017}, \citet{Uematsu2019}, \citet{chen2022fast}, \citet{liu2022multivariate}, and references therein.  See also~\cite{reinsel2022multivariate} for a review of recent developments on multivariate reduced-rank regression.

However, these reduced-rank methods are designed for homogeneous data and are not directly applicable to heterogeneous data with multivariate responses.  In the literature, there has been very limited research on heterogeneity analysis for multivariate response regression. In this paper, we extend the idea of~\citet{Shu2017} to the multivariate response setting and would like to make the first-ever attempt to propose a joint heterogeneity and reduced-rank learning framework to simultaneously identify the subgroup structure and estimate the covariate effects for heterogeneous multivariate response regression. Specifically, we represent the heterogeneity after adjusting for the effects of covariates using group-specific intercepts and assume the coefficient matrix of predictors to be of low rank, yielding a joint modeling of heterogeneous individual effects and common latent factors. A new methodology of reduced-rank learning with subgroup identification is proposed to determine the number of subgroups, identify the group structure, and estimate the coefficient matrix of predictors automatically and simultaneously.  Our approach integrates the reduced-rank regression technique with the pairwise fusion penalized approach, naturally benefiting from the advantages of both.

The main contributions of our work are fourfold.  First, we explicitly introduce a heterogeneous multivariate response regression model to jointly capture data heterogeneity and characterize the dependence between responses and predictors. Unlike most existing methods that treat subgroup identification and low-rank regression separately, our approach performs simultaneous subgroup identification and low-rank regression in a unified framework. To the best of our knowledge, our work is the first attempt to present a joint learning framework for heterogeneity detection and reduced-rank estimation in
the context of multivariate response regression.
Second, we propose a rank-constrained pairwise fusion procedure to automatically identify the subgroup structure without a prior knowledge of the grouping information while simultaneously
uncovering the predictor-response association network.  There is no existing result on subgroup automatic identification for multivariate reduced-rank regression and our approach is new to the literature.
Third, our method extends the model setting considered in~\citet{Shu2017} to the multivariate case and incorporates heterogeneity learning in the context of multivariate response regression. Although such an extension may appear conceptually straightforward, it is nontrivial to develop the corresponding computational algorithm and establish theoretical properties in our more complex model setting, particularly due to the rank constraint.  In particular, the algorithm proposed by~\citet{Shu2017} is not applicable for solving the rank-constrained optimization problem in our paper. We apply the block coordinate descent method from~\cite{Tseng2001}
to address this issue.  We develop an new alternating direction method of multipliers (ADMM) algorithm to implement our proposed approach and show that the algorithm is guaranteed to converge. In addition, we propose a new predictive information criterion to choose the rank of the coefficient matrix of predictors.
Fourth, we establish the theoretical properties of our proposed method under mild and interpretable conditions. The effectiveness of the proposed method is also evidenced by simulation studies and a real data application.

The rest of the paper is organized as follows. Section~\ref{sec2} presents the model setting and describes our new methodology in detail. Section~\ref{sec3} establishes the theoretical properties of the proposed approach. Section~\ref{sec4} verifies the theoretical results empirically and demonstrate the advantages of our approach via simulation studies. Section~\ref{real data} provides a real data example. Section~\ref{sec.dicus} includes discussions and extensions of our work. The proofs of main results are relegated to the Appendix. Additional 
technical details are provided in the supplementary material.


\section{Simultaneous heterogeneity and reduced-rank learning}\label{sec2}


\subsection{Model setting}

Suppose that we observe $n$ independent pairs $(\by_i, \bx_i)$, $i \in \{1, \dots, n\}$, where $\by_i\in\mathbb{R}^q$ is the multivariate response vector for the $i$th subject and $\bx_i\in\mathbb{R}^p$ is the corresponding vector of covariates. Assume that there are $K\geq 1$ potential subgroups $\psi_1, \dots, \psi_K$ 
where $K$, the number of subgroups, is \textit{unknown} but bounded. We consider the following heterogeneous multivariate response regression model
\begin{align}\label{HTE model}
	\by_i = \ba_i  + \bB^{\top} \bx_i + \bepsilon_i,
\end{align}
where $\ba_i\in \mathbb{R}^{q}$ is the unknown subgroup-specific intercept vector, 
$\ba_i=\bc_k$ for all $i\in \psi_k$ with $k\in\{1,  \ldots, K\}$, $\bB \in \mathbb{R}^{p \times q}$ is the unknown coefficient matrix which is assumed to be of low-rank, i.e., $\mbox{rank}(\bB)=r^{*}$ with $r^{*}\leq \min\{p, q\}$, and $\bepsilon_i \in \mathbb{R}^q$ is the random error vector with zero mean.  
It can be seen from model \eqref{HTE model} that the 
intercept vector $\ba_i$ 
can vary across different subgroups while the coefficient matrix $\bB$
is the same across all subjects.  
This means that after accounting for the effects of the predictors $\bx_i$, the heterogeneity of the population 
can be modeled through subgroup-specific intercept vectors $\ba_i$'s.  

In matrix form, model~\eqref{HTE model} can be rewritten as
\begin{align}\label{model2}
	\bY =  \bA + \bX \bB + \bE,
\end{align}
where $\bY = (\by_1, \dots, \by_n)^{\top} \in \mathbb{R}^{n \times q}$ denotes the response matrix, 
$\bA = (\ba_1, \dots, \ba_n)^{\top} \in \mathbb{R}^{n \times q}$ is the stack matrix of intercept vectors for all observations, $\bX = (\bx_1, \dots, \bx_n)^{\top} \in \mathbb{R}^{n \times p}$ is the predictor matrix, and $\bE = (\bepsilon_1, \dots, \bepsilon_n)^{\top} \in \mathbb{R}^{n \times q}$ is the random error matrix. 
Since there are only $K$ subgroups and 
$\ba_i=\bc_k$ for all $i\in \psi_k$ with $k\in\{1,  \ldots, K\}$, we know that $\ba_i$ can 
only take values from $ \{\bc_1, \dots, \bc_K\}$, where $\bc_k$ is the common vector for the $\ba_i$'s from subgroup $\psi_k$.  
Let $\bC = (\bc_1, \dots, \bc_K)^{\top}\in\mathbb{R}^{K\times q}$. Then we have
\begin{equation} \label{C decomposition}
	\bA =
	\left(
	\begin{array}{l}
		\ba_1^{\top} \\
		\ba_2^{\top} \\
		\,\vdots \\
		\ba_n^{\top}
	\end{array}
	\right)
	=
	\left(
	\begin{array}{cccc}
		w_{11} &w_{12} &\cdots&w_{1K} \\
		w_{21} &w_{22} &\cdots&w_{2K} \\
		\vdots&\vdots&\ddots&\vdots\\
		w_{n1} &w_{n2} &\cdots&w_{nK} 
	\end{array}
	\right)
	\left(
	\begin{array}{l}
		\bc_1^{\top} \\
		\bc_2^{\top} \\
		\,\vdots  \\
		\bc_K^{\top}
	\end{array}
	\right)
	=
	\bW \bC,
\end{equation}
where $\bW = (w_{ik})$ is an $n \times K$ matrix with $w_{ik}=1$ if $i\in\psi_k$ and $w_{ik}=0$ otherwise.   
Substituting the decomposition in~\eqref{C decomposition} into model (\ref{model2}) yields
\begin{align}\label{model3}
	\bY =  \bW \bC + \bX \bB + \bE.
\end{align}
Note that the matrices $\bW$, $\bB$ and $\bC$, as well as the number of subgroups $K$, are unknown. The subgroup indicator matrix $\bW$ is also  unobservable. Our goal is to simultaneously identify the subgroups $\psi_1, \dots, \psi_K$, determine the number of subgroups $K$ in the data, and accurately estimate the  
matrices $\bB$ and $\bC$.

\subsection{Rank-constrained penalized estimation via pairwise fusion}\label{subsec: method}

In this subsection, we introduce our rank-constrained pairwise fusion penalized procedure for simultaneous subgroup identification and coefficient matrix estimation. 
To concurrently induce grouping and dimension reduction, we use a concave pairwise fusion penalty and impose a rank constraint in our procedure.   
To be more specific, for a given $\lambda>0$ and a positive integer $r$, define 
\begin{align}\label{initial goal}
	(\widehat{\bA},\,\widehat{\bB})
	= & \arg \min_{\mathbf{A}, \,\mathbf{B}} \left\{2^{-1} \sum_{i = 1}^{n} \|\by_i - \bB^{\top} \bx_i - \ba_i  \|_2^2 + \sum_{1 \leq i < j \leq n} p_{\gamma} ( \|\ba_i - \ba_j \|_2, \lambda)\right\}\nonumber\\
	\quad &\mbox{subject to}\,\, \mbox{rank}(\bB) \leq r,
\end{align}
where $\widehat{\bA}=(\widehat{\ba}_1, \dots, \widehat{\ba}_n)^{\top}$ and $p_{\gamma} (t, \lambda)$ is a penalty function with tuning parameter $\lambda$ that controls the penalty level on $\|\ba_i - \ba_j\|_2$.  Let $\{\,\widehat{\bc}_1, \dots, \widehat{\bc}_{\widehat{K}}\}$ be 
the distinct values of $\widehat{\ba}_1, \dots, \widehat{\ba}_n$. Then the estimator of $\bC$ 
is given by $\widehat{\bC} = (\widehat{\bc}_1, \dots, \widehat{\bc}_{\widehat{K}})^{\top}$ and the identified subgroup structure is $\widehat{\psi}_k=\{i:\widehat{\ba}_i=\widehat{\bc}_k,1\le i\le n\}$ for $1\leq k\leq \widehat{K}$. 
Since the $L_1$ penalty $p_{\gamma} (t, \lambda)=\lambda |t|$~\citep{tibshirani1996regression} applies the same thresholding to all pairwise differences $\|\ba_i - \ba_j\|_2$, it leads to biased estimates and may 
fail to correctly identify subgroups, which is conﬁrmed in our simulation studies in Section~\ref{sec4}. Therefore, we use 
the concave penalties, such as SCAD~\citep{Fan2001} with 
\begin{align*}
	p_{\gamma}(t, \lambda)=\lambda\int_0^{|t|} \min\{1, \, (\gamma-t/\lambda)_{+}/(\gamma-1)\} \, dx,\,\,\gamma>2
\end{align*}
and MCP~\citep{Zhang2010} with 
\begin{align*}
	p_{\gamma}(t, \lambda)=\int_{0}^{|t|} (\gamma\lambda-x)_+ / \gamma \,dx,\,\,\gamma>1.
\end{align*}

Note that the estimators $\widehat{\bA}$,  $\widehat{\bB}$, $\widehat{\bC}$, and $\widehat{K}$ depend on the tuning parameter $\lambda$ and the rank $r$.  Both $\lambda$ and $r$ can be selected using 
an appropriate criterion.  In our paper, we introduce a predictive
information criterion (PIC), given in \eqref{PIC} later in Section~\ref{sec3}, to select the tuning parameter $\lambda$ and the rank $r$. 
In particular, we partition the support of $\lambda$ into a grid of $\lambda_{\min}=\lambda_0<\lambda_1<\cdots<\lambda_J=\lambda_{\max}$ and set a maximum value for the rank (say, $r_{\max}$).
Then for each $(r, \lambda)$ with $r = \{1, \dots, r_{\max}\}$ and $\lambda\in\{\lambda_0, \dots, \lambda_J\}$, we compute a solution path for $\widehat{\bA}(r, \lambda)$ and $\widehat{\bB}(r, \lambda)$, and
obtain the estimated matrix $\widehat{\bC}(r, \lambda)$ and the number of subgroups $\widehat{K}(r, \lambda)$. Then the rank and the tuning parameter are selected by minimizing a data-driven PIC, that is, $(\hat{r}, \hat{\lambda})=\argmin \limits_{1\leq r\leq r_{\max}, \, \lambda\in\{\lambda_0, \lambda_1, \dots, \lambda_J\}} \mbox{PIC}(r, \lambda)$. 
Subsequently, our final estimators for $\bA$ and $\bB$ are given by
$\widehat{\bA}=\widehat{\bA}(\hat{r}, \hat{\lambda})$ and $\widehat{\bB}=\widehat{\bB}(\hat{r}, \hat{\lambda})$.
Then the final estimators $\widehat{\bC}$, $\widehat{K}$, and $\widehat{\psi}_k$ for $1\leq k\leq \widehat{K}$ can be obtained accordingly.

\subsection{Algorithm}
In this subsection, we present the algorithm to solve the optimization problem~\eqref{initial goal} and obtain the estimators $\widehat{\mathbf{A}}$ and $\widehat{\mathbf{B}}$. By introducing a new set of parameters $\bdelta_{ij} = \ba_i - \ba_j$, the optimization problem~\eqref{initial goal} can be reformulated as 
the rank-constrained minimization problem
\begin{align*}
	L_0 (\bA, \bB, \bdelta) = & 2^{-1} \sum_{i = 1}^{n} \|\by_i - \bx_i^{\top} \bB - \ba_i  \|_2^2 + \sum_{1 \leq i < j \leq n} p_{\gamma} ( \|\bdelta_{ij} \|_2, \lambda), \\ 
	& \mbox{subject to}\,\, \ \ba_i - \ba_j - \bdelta_{ij} = \mathbf{0}\,\,\mbox{and}\,\, \mbox{rank}(\bB) \leq r,
\end{align*}
where $\bdelta = \{\bdelta_{ij}, i \leq j\}^{\top}\in\mathbb{R}^{[n(n-1)/2 ]\times q}$ with $\bdelta_{ij}$ a $q$-dimensional vector.
Thus, the augmented Lagrangian function is given by
\begin{align*}
	L(\bA, \bB, \bdelta, \bV)= & L_0 (\bA, \bB, \bdelta) + \sum_{1 \leq i < j \leq n} \langle  \bv_{ij}, (\ba_i - \ba_j - \bdelta_{ij}) \rangle +  \frac{\vartheta}{2}\sum_{1 \leq i < j \leq n} \| \ba_i - \ba_j - \bdelta_{ij} \|_2^2\\
	& \mbox{subject to}\,\, \ \ba_i - \ba_j - \bdelta_{ij} = \mathbf{0}\,\,\mbox{and}\,\, \mbox{rank}(\bB) \leq r,
\end{align*}
where $\bV = \{\bv_{ij}, i < j \}^{\top}$ with each dual variable $\bv_{ij}\in\mathbb{R}^q$ being a Lagrange multiplier, $\vartheta$ is a penalty parameter, and $\langle \ba, \bb \rangle = \ba^{\top} \bb$ denotes the inner product of two vectors $\ba$ and $\bb$. 
We then compute the estimators of 
$(\bA, \bB, \bdelta, \bV)$
through the following ADMM algorithm. 

Given the parameter values $\bA^{(m)}, \bB^{(m)}, \bdelta^{(m)}$, $\bV^{(m)}$ at the $m$th iteration, our algorithm proceeds as follows:
\begin{align}\label{step1}
	\mbox{Step 1:}\quad & (\mathbf{A}^{(m+1)},\mathbf{B}^{(m+1)} )= \arg \min_{\mathbf{A}, \,\mathbf{B}} L( \mathbf{A}, \mathbf{B}, \bdelta^{(m)}, \mathbf{V}^{(m)}) \,\,\mbox{subject to} \ \mbox{rank}(\mathbf{B}) \leq r, \\ \label{step2}
	\mbox{Step 2:}\quad& \bdelta^{(m + 1)} = \arg \min_{\bdelta} L (\bA^{(m +1)}, \bB^{(m +1)}, \bdelta, \bV^{(m)}), \\ \label{step3}
	\mbox{Step 3:}\quad&\bv_{ij}^{(m + 1)} = \bv_{ij}^{(m)} + \vartheta (\ba_i^{(m + 1)} - \ba_j^{(m +1)} - \bdelta_{ij}^{(m + 1)}).
\end{align}
In Step 1,
the objective function $L( \mathbf{A}, \mathbf{B}, \bdelta^{(m)}, \mathbf{V}^{(m)})$ in the optimization problem \eqref{step1} can be simplified as
\begin{align*}
	f(\bA, \bB) & =  2^{-1} \sum_{i = 1}^{n} \|\by_i - \bx_i^{\top} \bB -  \ba_i \|_2^2  + 2^{-1}\vartheta \sum_{1 \leq i < j \leq n} \|\ba_i - \ba_j - \bdelta_{ij}^{(m)} + \vartheta^{-1} \mathbf{v}_{ij}^{(m)} \|_2^2 + C_n, 
\end{align*}
where $C_n$ is a constant independent of $(\bA,\bB)$. 
In matrix notation, we can rewrite $f(\bA, \bB)$ as 
\begin{align*} 
	f(\bA, \bB) & =  2^{-1} \|\bY - \bX \bB - \bA \|_F^2 
	+ 2^{-1}\vartheta \|\mathbf{\Delta} \bA - \bdelta^{(m)} + \vartheta^{-1} \mathbf{V}^{(m)} \|_F^2 + C_n,   
\end{align*}
where $\|\cdot\|_F$ denotes the Frobenius norm of a given matrix,
$\mathbf{\Delta} = \{(\mathbf{e}_i - \mathbf{e}_j), i< j \}^{\top}\in\mathbb{R}^{[n(n-1)/2 ]\times n}$, and $\mathbf{e}_i$ 
is an $n \times 1$ unit vector 
with $1$ for the $i$th element and $0$ for all other elements.
Therefore
the rank-constrained minimization problem \eqref{step1} is equivalent to the optimization problem  
\begin{align}\label{Delta} 
	\arg \min_{\mathbf{A}, \,\mathbf{B}}
	f(\bA, \bB)\quad  \mbox{subject to}\,\, \mbox{rank}(\bB) \leq r, 
\end{align}

Due to the rank constraint on $\bB$, the ADMM algorithm proposed by~\citet{Shu2017} is not directly applicable for solving 
the optimization problem \eqref{Delta}. Therefore,
we apply the block coordinate descent method from~\cite{Tseng2001} here. The main idea is to alternately update the matrices $\bA$ and $\bB$, keeping one matrix fixed while updating the other. Similar techniques have been used in~\cite{Bunea2012} and~\cite{She2017}.  
Specifically, $\bA^{(m+1)}$ and $\bB^{(m+1)}$ are obtained by 
\begin{align}\label{C update}
	&\mathbf{A}^{(m + 1)} = ( \bI_n + \vartheta \mathbf{\Delta}^{\top}\mathbf{\Delta})^{-1} [ \mathbf{Y} - \mathbf{X} \mathbf{B}^{(m)} + \vartheta \mathbf{\Delta}^{\top} (\bdelta^{(m)} - {\vartheta}^{-1} \bV^{(m)})],    \\ \label{B update}
	&\mathbf{B}^{(m + 1)} = (\mathbf{X}^{\top} \mathbf{X})^{-1} \mathbf{X}^{\top}(\mathbf{Y} - \mathbf{A}^{(m+1)}) \bQ_{V(\mathbf{X},\, \mathbf{Y} - \mathbf{A}^{(m+1)},\, r)}.
\end{align}
Here $\bI_n$ is an $n\times n$ identity matrix
and $V(\mathbf{X},\, \mathbf{Y} - \mathbf{A}^{(m+1)},\, r)$ is formed by the leading $r$ eigenvectors of $(\bY - \bA^{(m+1)})^{\top} \bQ_{\mathbf{X}} $ $(\bY - \bA^{(m+1)})$ with $\bQ_{\mathbf{X}}= \mathbf{X}(\mathbf{X}^{\top} \mathbf{X})^{-1} \mathbf{X}^{\top}$. 
Updating $\bA$ in \eqref{C update} requires simple matrix operations while updating $\bB$ in \eqref{B update} performs reduced-rank regression on the adjusted response matrix $\bY - \bA^{(m+1)}$.

In Step 2, 
after discarding the terms independent of $\bdelta$, the objective function in \eqref{step2} is
\begin{align}\label{solution-delta}
	\sum_{1 \leq i < j \leq n} \langle \bv^{(m)}_{ij}, (\ba^{(m+1)}_i - \ba^{(m+1)}_j - \bdelta_{ij}) \rangle + 2^{-1}\vartheta\|\ba^{(m+1)}_i - \ba^{(m+1)}_j - \bdelta_{ij} \|_2^2 + p_{\gamma} (\| \bdelta_{ij}\|_2, \lambda).
\end{align}
It is easy to update $\bdelta_{ij}$ for the $L_1$, MCP and SCAD penalties. 
Denote by $\bzeta_{ij}^{(m)} = (\ba_i^{(m+1)} - \ba_j^{(m+1)}) + \vartheta^{-1} \bv_{ij}^{(m)}$. For the $L_1$ penalty, the estimate of $\bdelta_{ij}$ is updated as
\begin{align*}
	\bdelta^{(m+1)}_{ij} = S (\bzeta^{(m)}_{ij}, \lambda /\vartheta),
\end{align*}
where $S(\bz, t) = (1 - t/\| \bz\|_2)_+ \bz$ is a groupwise soft thresholding operator. For the MCP with $\gamma > 1/\vartheta$, $\bdelta^{(m+1)}_{ij}$ is 
\begin{equation*}
	\bdelta^{(m+1)}_{ij} =
	\left\{
	\begin{array}{ll}
		S (\bzeta^{(m)}_{ij}, \lambda /\vartheta)/[1 - 1/(\gamma \vartheta)],  &  \ \|\bzeta_{ij}^{(m)}\|_2 \leq \gamma \lambda, \\
		\bzeta_{ij}^{(m)}, &  \ \|\bzeta_{ij}^{(m)}\|_2 > \gamma \lambda.
	\end{array}
	\right.
\end{equation*}
For the SCAD penalty with $\gamma > 1/\vartheta + 1$, $\bdelta^{(m+1)}_{ij}$ is given by 
\begin{equation*}
	\bdelta^{(m+1)}_{ij} =
	\left\{
	\begin{array}{ll}
		S (\bzeta^{(m)}_{ij}, \lambda /\vartheta),
		& \ \|\bzeta_{ij}^{(m)}\|_2 \leq \lambda + \lambda/\vartheta, \\
		S (\bzeta^{(m)}_{ij}, \gamma \lambda / ((\gamma - 1)\vartheta))/[1 - 1/((\gamma -1 )\vartheta)], &  \ \lambda + \lambda/\vartheta \leq \|\bzeta_{ij}^{(m)}\|_2 \leq \gamma \lambda, \\
		\bzeta_{ij}^{(m)}, &  \ \|\bzeta_{ij}^{(m)}\|_2 > \gamma \lambda .
	\end{array}
	\right.
\end{equation*}

In Step 3, we update the dual variable $\bv_{ij}$ using (\ref{step3}). Note that $\mathbf{r}^{(m+1)} = \mathbf{\Delta} \bA^{(m+1)} - \bdelta^{(m+1)}$ and $\mathbf{s}^{(m+1)} = \vartheta \mathbf{\Delta}^{\top} (\bdelta^{(m)} - \bdelta^{(m+1)})$ are 
the primal and dual residuals, respectively, in the ADMM algorithm.
The iterations continue until the stopping criterion is reached. We stop the algorithm when the primal residual $\mathbf{r}^{(m+1)} = \mathbf{\Delta} \bA^{(m+1)} - \bdelta^{(m+1)}$ is sufficiently close to zero such that $\| \mathbf{r}^{(m+1)} \|_F < \varepsilon$ for some small $\varepsilon>0$. 

We minimize the following ridge fusion criterion
\begin{align*}
	L_R (\bA, \bB) = 2^{-1} \| \bY-  \bA  - \bX \bB \|_F^2 + 2^{-1}\lambda^* \sum_{1 \leq i \leq j \leq n} \| \ba_i - \ba_j\|_2^2,
\end{align*}
to obtain the initial values of the parameters in our algorithm, 
where $\lambda^*$ is a small positive number. We use $\lambda^* = 0.001$ in our simulation studies.
Using the definition of $\mathbf{\Delta}$ in Section~\ref{subsec: method} and matrix notation, we can write 
$L_R (\bA, \bB)$ as $L_R (\bA, \bB) = 2^{-1} \| \bY- \bA  - \bX \bB \|_F^2 + 2^{-1}\lambda^*\| \mathbf{\Delta}  \bA \|_F^2$,
which yields initial values $\bA^{(0)} = [ \bI_n - \bQ_{\mathbf{X}} + \lambda^* \mathbf{\Delta}^{\top} \mathbf{\Delta}]^{-1} (\bI_n - \bQ_{\mathbf{X}}) \bY$ and 
$\bB^{(0)} = (\bX^{\top} \bX)^{-1} \bX^{\top} (\bY - \bA^{(0)})$. Then we take 
$\bdelta^{(0)} = \mathbf{\Delta}\bA^{(0)}$, and $\bV^{(0)}= \mathbf{0}$.

\subsection{Convergence of the algorithm}
\smallskip
\begin{proposition}\label{pro1}
	For both MCP and SCAD penalties, 
	we have $\lim_{m \rightarrow \infty} \| \mathbf{r}^{(m)}\|^2_F = 0$ and $\lim_{m \rightarrow \infty} \|\mathbf{s}^{(m)}\|^2_F = 0$.
\end{proposition}
The proof of Proposition~\ref{pro1} is given in the second part of 
the supplementary material. Proposition \ref{pro1} shows that the primal feasibility and dual feasibility are achieved by the algorithm such that the optimization problem can converge to a minimum point. The implementation of our rank-constrained penalized estimation procedure 
is summarized in Algorithm~\ref{alg}.

%
%
\begin{algorithm}[H]
	\caption{Simultaneous heterogeneity and reduced-rank learning}
	\label{alg}
	\begin{algorithmic}[1]
		\Require Data $\mathbf{Y} \in \mathbb{R}^{n \times q}$ and $\mathbf{X} \in \mathbb{R}^{n \times p}$, maximum rank $r_{\max} = \min(p, q)$, and termination accuracy $\varepsilon$
		\State \textbf{Initialize:} $\mathbf{A}^{(0)}, \mathbf{B}^{(0)}, \delta^{(0)}, \mathbf{V}^{(0)}$
		\For{$r = 1, 2, \dots, r_{\max}$}
		\For{$\lambda = \lambda_0, \dots, \lambda_J$}
		\For{$m = 0, 1, \dots$}
		\State Update $\mathbf{A}^{(m)}$ using  \eqref{C update}.
		\State Update $\mathbf{B}^{(m)}$ using \eqref{B update}.
		\State Update $\delta^{(m)}$ using \eqref{solution-delta}.
		\State Update $\mathbf{V}^{(m)}$ using \eqref{step3}.
		\State Compute $\mathbf{r}^{(m)} = \Delta \mathbf{A}^{(m)} - \delta^{(m)}$
		\If{$\|\mathbf{r}^{(m+1)}\|_F < \varepsilon$}
		\State Stop and obtain $(\widehat{\mathbf{A}}(r, \lambda), \widehat{\mathbf{B}}(r, \lambda), \widehat{\mathbf{C}}(r, \lambda), \widehat{\mathbf{K}}(r, \lambda))$ from the last iteration.
		\EndIf
		\EndFor
		\EndFor
		\EndFor
		\State Select the optimal rank $\hat{r}$ and tuning parameter $\hat{\lambda}$ by the predictive information criterion \eqref{PIC}.
		\State \textbf{Output:} $\mathbf{A} = \widehat{\mathbf{A}}(\hat{r}, \hat{\lambda}), \mathbf{B} = \widehat{\mathbf{B}}(\hat{r}, \hat{\lambda}), \mathbf{C} = \widehat{\mathbf{C}}(\hat{r}, \hat{\lambda}), \mathbf{K} = \widehat{\mathbf{K}}(\hat{r}, \hat{\lambda})$
	\end{algorithmic}
\end{algorithm}

\smallskip

In our paper, the sequence $\{\lambda_0, \ldots, \lambda_J\}$ for the tuning parameter $\lambda$ is generated by a data-driven approach. To be more specific, for each rank $r\in\{1,\ldots,r_{\max}\}$, we first determine $\lambda_J$  and $\lambda_0$ based on our observed data ($\bX, \bY$), and then obtain the sequence $\{\lambda_0, \ldots, \lambda_J\}$ using the formula $\exp(seq( \mathrm{ln} (\lambda_0),\mathrm{ln}(\lambda_J), n_{\lambda}))$ where $seq(a,b,n_{\lambda})$ represents a sequence with $n_{\lambda}$ elements, evenly spaced from $a$ to $b$. In our numerical studies, for each rank $r\in\{1, \ldots, r_{\max}\}$, we first conduct a classical reduced rank regression on the observed data $(\mathbf{X}, \mathbf{Y})$ to obtain an estimator of the intercept matrix $\bA$, which is given by $\widehat{\bA}^{R} = \mathbf{Y} - \bX \widehat{\bB}^{R} = (\hat{\ba}^R_1, \dots, \hat{\ba}^R_n)^\top$, where $\widehat{\bB}^{R}$ denotes the reduced-rank estimator. Then we can set $\lambda_J$ to be the maximum pairwise row norm, that is, $\lambda_J=\max_{i,j} \| \hat{\ba}^{R}_i - \hat{\ba}^{R}_j \|_2$, and $\lambda_0$ to be $\lambda_0 = 10^{-3} \lambda_{J}$. We take $n_{\lambda} = 20$ in our numerical studies.


\section{Theoretical properties}\label{sec3}
In this section, we study the theoretical properties of the proposed method. To facilitate the technical presentation, we first introduce some definitions and notations.
\begin{definition}
	(Matrix subspace) Define
	$\mathcal{M}_{\psi} = \{ \bA = (\ba_1, \dots, \ba_n)^{\top} \in \mathbb{R}^{n \times q}: \ba_i = \ba_j, \, \ba_i, \ba_j \in \mathbb{R}^{ q}, \ i,j \in \psi_k, 1 \leq k \leq K \}$. Then, for any $\bA \in \mathcal{M}_{\psi}$, it can be written as $\bA = \bW \bC$, where $\bW$ is an indicator matrix defined in \eqref{C decomposition} and $\bC$ is the common matrix.
\end{definition}
\begin{definition} 
	(Subgroup size) Let $| \psi_k|$ denote the number of elements in the subgroup $\psi_k$. Let 
	$|\psi_{\min}| = \min_{1 \leq k \leq K}|\psi_k|$ and $| \psi_{\max}| = \max_{1 \leq k \leq K} | \psi_k|$ represent the minimum and maximum subgroup sizes, respectively.
\end{definition}

By the definition of  
$\bW$ in \eqref{C decomposition}, we have $\bW^{\top}\bW = \diag(| \psi_1|$ , $\dots, |\psi_K|)$. Let $\|\cdot\|_{\infty}$ denote the maximum absolute entry norm for a given vector or matrix. For any vector $\bu=(u_j)$, we write $\| \bu\|_{2} = (\sum_{j}u_j^2)^{1/2}$.
For any matrix $\bD$, let $\|\bD\|_2$ be the largest singular value of  $\bD$. Let $\lambda_{\min} (\cdot)$ and $\lambda_{\max} (\cdot)$ be the smallest and largest eigenvalues of a given matrix, respectively. 
Let $r(\cdot)$ denote the rank of a given matrix.
We denote by $\rho(t) = \lambda^{-1} p_{\gamma} (t,\lambda)$ the scaled penalty function and $\rho'(t)$ the derivative function of $\rho(t)$.  
A random variable $\eta$ is sub-Gaussian if there exist constants $C, c>0$ such that $\mathrm{Pr}(|\eta|>t)\leq C\exp(-ct^2)$ for any $t>0$.
A random vector $\bepsilon\in\mathbb{R}^q$ is said to be a sub-Gaussian random vector with $\psi_2$-norm (also called scale) bounded by
$\sigma$ if all one-dimensional
marginals $\langle\bb, \bepsilon\rangle$ are sub-Gaussian and satisfy $\|\langle\bb, \bepsilon\rangle\|_{\psi_2}\leq \sigma\|\bb\|_2$ for any $\bb\in \mathbb{R}^q$, where $\|\cdot\|_{\psi_2}=\inf\{t>0: \mathbb{E}\exp[(\cdot)/t)]^2\leq 2\}$.
Throughout this paper, $\tilde{c}$, $\tilde{c}_1$, $\tilde{c}_2$, $c$, $c_1$, $c_2$, $C$, $C_1$ and $C_2$ are generic positive constants, which may vary from line to line.

\begin{condition} \label{con1} 
	The function $p_{\gamma} (t, \lambda)$ is symmetric with respect to $t$. It is non-decreasing and concave in $t \in [0,\infty)$. For some constant $a_1 >0$, $\rho(t)$ is constant when $t \geq a_1 \lambda$ and $\rho (0) = 0$. Besides, $\rho' (t)$ exists and is continuous except for a finite number of values of $t$ and $\rho' (0+) = 1$.
\end{condition}

\begin{condition}\label{con2}
	Each row of the noise matrix $\bE$ is a sub-Gaussian random vector with mean zero and $\psi_2$-norm bounded by $\sigma$. 
\end{condition}

\begin{condition}\label{con3}
	There exists a positive constant $C_1$ such that $\lambda_{\min} ((\bW, \bX)^{\top} (\bW, \bX)) \geq C_1 | \psi_{\min}|$.
\end{condition}

\begin{condition}\label{con4}
	There exist a positive constant $C_2$ such that $\sup_i \| \bx_i \|_2$ $ \leq C_2 \sqrt{p}$. 
\end{condition}

Condition~\ref{con1} ensures that the penalties used in Algorithm~\ref{alg} enjoy the soft-thresholding properties and is commonly used in the subgroup analysis literature~\citep{Shu2017, Zhangsinica2019, Ma2020}. It is clear that both concave penalties, MCP and SCAD, satisfy Condition~\ref{con1}. The sub-Gaussian assumption in Condition~\ref{con2} is used to control the tail behavior of the random error vector. The same tail assumption has been used in multivariate response regression; see, for example, \cite{She17}, \cite{She2017}, and \cite{dong2021parallel}. Condition~\ref{con3} imposes a constraint on the smallest eigenvalue of the new matrix $(\bW, \bX)$ and shares a similar spirit as the restricted eigenvalue condition in~\cite{Bickel2009}. This condition is initially introduced by~\cite{Shu2017} and also used in~\cite{Zhangsinica2019}, \cite{Ma2020}, \cite{Yan2021}, and references therein. Condition~\ref{con4} puts a mild assumption on the lengths of the covariate vectors. The upper bound on $\| \bx_i\|_2$ is reasonable and consistent with the dimension of the covariate vector.

Next, we introduce the oracle estimators of the matrices $\bA$, $\bB$ and $\bC$, which assume that the true subgroups and the true rank $r^{*}$ of the coefficient matrix $\bB$ are known.  These oracle estimators are not available in practice, but can be used to derive the theoretical properties of our proposed method later.

\begin{definition}(Oracle estimators)
	When the true subgroups $\psi_1, \dots, \psi_K$ and the true rank of the matrix $\bB$ are known in advance, the oracle estimators for $\bB$ and $\bA$ are defined as
	\begin{align}\label{oracle}
		(\widehat{\mathbf{B}}^{or}, \widehat{\mathbf{A}}^{or}) = \arg \min_{\mathbf{A} \in \mathcal{M}_{\psi},\, \mathbf{B} \in \mathbb{R}^{p \times q}} 2^{-1}\| \bY - \bX \bB - \bA \|^2_F \,\,  \mbox{subject to}\,\, \mbox{rank}(\bB) \leq  r^*,
	\end{align}
	and correspondingly, the oracle estimators for the common coefficient matrix $\bC$ and the coefficient matrix $\bB$ are given by
	\begin{align}\label{oracle2}
		(\widehat{\mathbf{B}}^{or}, \widehat{\mathbf{C}}^{or}) = \arg \min_{\mathbf{C} \in \mathbb{R}^{K \times q}, \,\mathbf{B} \in \mathbb{R}^{p \times q}} 2^{-1} \| \bY - \bX \bB - \bW \bC\|^2_F\,\,  \mbox{subject to}\,\, \mbox{rank}(\bB) \leq  r^*.
	\end{align}
\end{definition}

Let $K$ be the true number of subgroups and $\bC^*=(\bc_1^*, \ldots, \bc_K^*)^{\top}$, where $\bc_k^*$ is the true common intercept vector for subgroup $\psi_k$, with $k = \{1, \ldots, K\}$. Then $\ba_i^*=\bc_k^*$ for all $i\in\psi_k$ with $k = \{1, \ldots, K\}$ and thus $\bA^*=(\ba_1^*, \ldots, \ba_n^*)^{\top}=\bW\bC^*$ is the true value of $\bA$. Let $\bB^*$ denote the true value of the coefficient matrix $\bB$ with $r(\bB^*) = r^*$. Now we are ready to present our main results.

\begin{theorem}\label{theo1}
	
	Under Conditions \ref{con2}-\ref{con4}, the following oracle inequality 
	\begin{align*}
		\| \bX(\widehat{\bB}^{or} - \bB^* ) + \bW (\widehat{\bC}^{or}- \bC^* ) \|_F^2 \leq c_1 \sigma^2 [  (p+q-r^*)(r^* + K) + qK + \mathrm{ln} (n)]
	\end{align*}
	holds with probability at least $1 - \tilde{c}_1 \exp\{-c \mathrm{ln} (n)\}$, where $c_1$, $\tilde{c}_1$ and $c$ are positive constants. Suppose $|\psi_{\min}| \gg \sigma^2 [  (p+q-r^*)(r^* + K) + qK + \mathrm{ln} (n)]$. Then under Conditions \ref{con2}-\ref{con4}, with probability at least $1- \tilde{c}_1\exp\{-c \mathrm{ln} (n)\}$, we have
	\begin{align}\label{estimation bound}
		\| ((\widehat{\bB}^{or} - \bB^* )^{\top}, (\widehat{\bC}^{or}- \bC^* )^{\top})^{\top} \|_F^2 \leq \phi_n^2\quad\mbox{and}\quad \|  \widehat{\bA}^{or}- \bA^* \|_F^2 \leq |\psi_{\max}| \phi_n^2,
	\end{align}
	where $\phi_n=\sqrt{c_1 C_1^{-1} \sigma^2 [  (p+q-r^*)(r^* + K) + qK + \mathrm{ln} (n)]/|\psi_{\min}|}$.
\end{theorem}

Theorem~\ref{theo1} establishes the prediction and estimation error bounds for the oracle estimators defined in \eqref{oracle}-\eqref{oracle2}. Note that both estimation error bounds for $(\widehat{\bB}^{or \top}, \widehat{\bC}^{or \top})^{\top}$ and $\widehat{\bA}^{or}$ depend on the quantity $\phi_n^2$. The numerator of $\phi_n^2$ involves $(p+q-r^*)(r^* + K) + qK$ and $\mathrm{ln} (n)$.  The term $(p+q-r^*)(r^* + K) + qK$ represents the number of free parameters in the heterogeneous reduced-rank regression model and will be dominated by the term $\mathrm{ln} (n)$ when the sample size $n$ is sufficiently large. Recall that the term $|\psi_{\min}|$, which is the denominator of $\phi_n^2$, is defined as the minimum size of the subgroups. Thus, we have $|\psi_{\min} | \leq n/K$. Without loss of generality, we assume $| \psi_{\min}| = \kappa n /K$ for some $\kappa \in (0,1]$. Then with the assumption $|\psi_{\min}| \gg \sigma^2 [  (p+q-r^*)(r^* + K) + qK + \mathrm{ln} (n)]$, the quantity $\phi_n^2$ is of the order of
$O(K[  (p+q-r^*)(r^* + K) + qK + \mathrm{ln} (n)] /n)$, indicating that $\phi_n^2 \rightarrow 0$ as $n \rightarrow \infty$. Therefore,
the oracle estimators can achieve consistency with probability approaching one.

To gain more insights into Theorem~\ref{theo1}, we compare the estimation error bound for $(\widehat{\bB}^{or \top}, \widehat{\bC}^{or \top})^{\top}$ in~\eqref{estimation bound}  with that for the univariate response case in~\cite{Shu2017}. To do so, we can take $q = 1$ and $r^* = 1$. Then the upper bound $\phi_n^2$ for 
$\| ((\widehat{\bB}^{or} - \bB^* )^{\top}, (\widehat{\bC}^{or}- \bC^* )^{\top})^{\top} \|_F^2$ in \eqref{estimation bound} simplifies to the order of $|\psi_{\min}|^{-1} [p + K + pK + \mathrm{ln} (n)]$.  Since the inequality  $\|\cdot\|_{\infty}\leq \|\cdot\|_F$ holds true for any given matrix, our upper bound for $\| ((\widehat{\bB}^{or} - \bB^* )^{\top}, (\widehat{\bC}^{or}- \bC^* )^{\top})^{\top} \|_{\infty}$ is the order of $\sqrt{ [p + K + pK + \mathrm{ln} (n)]/|\psi_{\min}|}$. The estimation error bound obtained for the heterogeneous univariate response regression in~\cite{Shu2017} is of the order of 
$|\psi_{\min}|^{-1} (p + K) \sqrt{n \mathrm{ln} (n)}$. Since $|\psi_{\min} | \leq n/K$, we can assume $| \psi_{\min}| = \kappa n /K$ for some $\kappa \in (0,1]$.  
Consequently, it follows that when $K$ and $p$ are fixed, the two error bounds are of the same order as $\sqrt{n^{-1}\mathrm{ln} (n) }$, which is asymptotically
vanishing.

For $K \geq 2$, let
\begin{align}\label{def.bn}
	b_n = \min_{i \in \psi_k,\, j \in \psi_{k'},\, k \neq k^{'}} \|\ba^* _i - \ba^* _j \|_2= \min_{ k \neq k'}\| \bc_k^* - \bc_{k'}^*\|_2
\end{align}
be the minimal difference of the common  
intercept vectors between two subgroups. Then we have the following result.

\begin{theorem}\label{theo-pro}
	Suppose that Conditions~\ref{con1}-\ref{con4} hold. If $b_n > a_1 \lambda$ and $\lambda \gg \phi_n$,
	where $a_1>0$ is given in Condition~\ref{con1}, then as $n \rightarrow \infty$, a local minimizer $(\widehat{\bA}, \widehat{\bB})$ of the objective function (\ref{initial goal}) with its  constrained rank equals to the true rank of the matrix $\mathbf{B}$, satisfying
	\begin{align*}
		\mathrm{Pr}\left( (\widehat{\bA}, \widehat{\bB}) = (\widehat{\bA}^{or}, \widehat{\bB}^{or}) \right) \rightarrow 1.
	\end{align*}
\end{theorem}

Theorem~\ref{theo-pro} reveals that the rank-constrained oracle estimator defined in~\eqref{oracle} is indeed a local minimizer of the objective function~\eqref{initial goal} 
when the constrained rank in \eqref{initial goal} equals the true rank of $\mathbf{B}$. 
Then, combined with the results established in Theorem~\ref{theo1}, which show that the oracle estimators 
asymptotically converge to the true parameter matrices, we can conclude that the proposed estimator $(\widehat{\bA}, \widehat{\bB})$, obtained by minimizing \eqref{initial goal} with the true rank constraint, can converge to the true parameter matrices with probability approaching one.
Similar results have been established in 
\cite{Shu2017} for 
univariate response regression. However, compared with \cite{Shu2017}, the presence of the rank-constraint in our 
multivariate response regression requires more delicate technical analyses.

The conditions $b_n > a_1 \lambda$ and $\lambda \gg \phi_n$ are required in Theorem~\ref{theo-pro}. Recall that the tuning parameter $\lambda$ in~\eqref{initial goal} controls the degree of shrinkage on the pairwise fusion term $\|\ba_i -\ba_j\|_2$. Thus, it is reasonable to assume that the minimum signal difference $b_n$ between two subgroups is greater than $a \lambda$ to ensure the identifiability of the heterogeneous intercept vectors.
The condition $\lambda \gg \phi_n$ is assumed mainly for theoretical analysis. It can be easily satisfied as long as
$\lambda$ is set relatively larger than the magnitude of $\phi_n$, which is of the order $\sqrt{n^{-1}\mathrm{ln} (n)}$ when $K$ and $p$ are fixed, as discussed following Theorem~\ref{theo1}.

As mentioned in Section~\ref{subsec: method} and Algorithm~\ref{alg}, we need to choose the rank $r$ and the tuning parameter $\lambda$. For this purpose, we propose the following predictive information criterion (PIC) 
\begin{align}\label{PIC}
	\mbox{PIC}(r, \lambda)=\mathrm{ln} \left(\| \bY - \bX \widehat{\bB} - \widehat{\bW} \widehat{\bC}\|_F^2\right) +  \left\{A_1 \left[(p + q - r)(r+\widehat{K}) + \widehat{K}q\right] + A_2 \mathrm{ln} (n) \right\}/(nq),
\end{align}
where $\widehat{\bB}=\widehat{\bB}(r, \lambda)$, $\widehat{\bW}=\widehat{\bW}(r, \lambda)$, $\widehat{\bC}=\widehat{\bC}(r, \lambda)$, $\widehat{K}=\widehat{K}(r, \lambda)$,
$\| \bY - \bX \widehat{\bB} - \widehat{\bW} \widehat{\bC}\|_F^2$ is the residual sum of squared errors, 
and $A_1$ and $A_2$ are  positive constants. 
Following~\cite{She17}, we take $A_1=7$ and $A_2=2$.

\begin{theorem}\label{theo3}
	Let $P(\mathbf{B}, \mathbf{C}) = A_1[(p + q - r)(r+k) + kq] + A_2\mathrm{ln} (n)$. 
	Suppose that the true model is parsimonious in the sense that $P(\mathbf{B}^*, \mathbf{C}^*) \leq nq/A_3$ for some constant $A_3 > 0$. Let $\delta (\mathbf{B}, \mathbf{C}) = AP(\mathbf{B}, \mathbf{C}) / (nq)$ where $A$ is a positive constant satisfying $A < A_3$ such that $\delta (\mathbf{B}^*, \mathbf{C}^*) < 1$. Then for sufficiently large values of $A_3$ and $A$, any $(\widehat{\mathbf{B}}, \widehat{\mathbf{C}})$ that minimizes $ \mathrm{ln}\left(\| \bY - \bX \bB -\mathbf{W}\bC\|_F^2\right) + \delta(\mathbf{B}, \mathbf{C})$ subject to $\delta(\mathbf{B}, \mathbf{C}) < 1$ satisfies
	\begin{align*}
		\| \mathbf{X} (\widehat{\mathbf{B}}-\mathbf{B}^*) +  ( \widehat{\mathbf{W}}\widehat{\mathbf{C}} - \mathbf{W}\mathbf{C}^* )\|_F^2 \lesssim \sigma^2\left[(p + q - r^*)(r^*+K) + Kq + \mathrm{ln} (n) \right]
	\end{align*}
	with probability at least $1 - \tilde{c}_1 n^{-c_1} - \tilde{c}_2 \exp(- c_2 nq)$ for some constants $\tilde{c}_1, c_1, \tilde{c}_2, c_2> 0$, where $\widehat{\mathbf{W}}$ is the estimated indicator matrix determined by $\widehat{\bA}=\widehat{\bW}\widehat{\bC}$ and $\lesssim$ means that the inequality holds up to a multiplicative constant.
\end{theorem}

Theorem~\ref{theo3} shows that 
the underlying matrices $\bB^*$ and $\bC^*$ can be accurately recovered by minimizing the predictive information criterion~\eqref{PIC} and the prediction error $\| \mathbf{X} (\widehat{\mathbf{B}}-\mathbf{B}^*) +  ( \widehat{\mathbf{W}}\widehat{\mathbf{C}} - \mathbf{W}\mathbf{C}^* )\|_F^2$ can be controlled within the rate $ \sigma^2 \left[(p + q - r^*)(r^*+K) + Kq + \mathrm{ln} (n) \right]$.


\section{Simulation studies}\label{sec4} 

In this section, we examine the finite-sample performance of our proposed method through two simulation examples. Note that our method can be implemented with different penalties.  
To simplify the presentation, our methods implemented with the MCP, SCAD, and $L_1$ penalties are denoted as SR-MCP, SR-SCAD, and SR-Lasso, respectively. In each example, we compare our methods with several approaches that focus on either reduced-rank regression or subgroup analysis alone. For reduced-rank regression alone, we consider the classical reduced rank regression (RRR), which is implemented using the \code{R} package \code{rrpack}~\citep{rrpack}. Since existing subgroup analysis methods are mainly designed for the univariate response setting, to conduct subgroup learning for multivariate response regression, we remove the rank constraint from the objective \eqref{initial goal} and use a similar ADMM algorithm as in \cite{Shu2017}. The resulting subgroup-learning-only procedures implemented with the MCP, SCAD, and $L_1$ penalties are denoted as S-MCP, S-SCAD, and S-Lasso, respectively. We also include two oracle procedures, Oracle.s and Oracle.sr, as benchmarks in our comparisons. The Oracle.s procedure uses the true subgroup information only, while the Oracle.sr procedure utilizes both the true subgroup information and true rank.

The rank $r$ and the tuning parameter $\lambda$ in our methods (
SR-MCP, SR-SCAD, and SR-Lasso) are jointly selected using the predictive information criterion (PIC) defined in \eqref{PIC}. For  subgroup analysis methods S-MCP, S-SCAD, and S-Lasso, there is only one tuning parameter $\lambda$.  Following the idea in~\cite{Shu2017}, we use the modified Bayesian information criterion (BIC), defined as $\mbox{BIC}(\lambda) = \mathrm{ln}  \left[  \sum_{i=1}^n (\by_i - \ba_i - \bx_i^{\top} \bB)^2/(nq)\right] + C_n (\widehat{K} (\lambda) + pq) n^{-1}\mathrm{ln} (n)$, to select the tuning parameter $\lambda$ in these methods. The cross-validation is used to choose the rank in the methods RRR and Oracle.s. 
As pointed out by one referee, since 
the Oracle.sr procedure uses both the true subgroup indicator matrix $\bW$ and true rank $r^*$, the estimators $\widehat{\mathbf{B}}$ and $\widehat{\mathbf{C}}$ for Oracle.sr can be obtained directly by minimizing $\|\bY- \bX\bB - \bW \bC\|_F^2$ subject to $r(\bB) \leq r^*$.

We evaluate the performance of each method by different measures. The first two measures are the relative estimation errors, $\rm{Err}(\widehat{\bB}) =\| \bB^* - \widehat{\bB}\|_F^2/\| \bB^* \|^2_F$ and $\rm{Err} (\widehat{\bA}) = \| \bA^* - \widehat{\bA}\|_F^2 / \| \bA^* \|_F^2$, which quantify the estimation performance of matrices $\bB$ and $\bA$, respectively. The third measure is the prediction accuracy, defined as Pre$(\widehat{\bB}, \widehat{\bC}) = \|\bY_{\rm{test}} - \bX_{\rm{test}} \widehat{\bB} - \bW_{\rm{test}} \widehat{\bC} \|_F^2/(n_{\rm{test}}q)$, computed from an independent test sample of size $n_{\rm{test}}=90$ and 
using the true subgroup indicator matrix $\bW_{\rm{test}}$ for the test sample. For rank estimation accuracy, we report the estimated rank (Rank). 
To examine the subgroup identification performance, 
we report the estimated number of subgroups ($\rm{\hat{K}}$) and the squared error for each of the $K$ different subgroups, defined as $\mathrm{Err}(\widehat{\ba}_k) = \| \ba_k - \widehat{\ba}_k \|_2^2 / q$, for $1 \leq k \leq K$. 
We also report the percentage of correct rank identification (Rank$\%$) and the percentage of correct subgroup identification (K$\%$), 
both computed from 100 replications.

\subsection{Simulation Example 1}\label{subsection: Example1}

In this example, we consider a heterogeneous multivariate regression model with three subgroups (i.e., $K=3$). 
Each row of the covariate matrix $\bX$ is independently generated from a Gaussian distribution $N(\bzero_p, \bSig_{X})$, where $\bSig_{X}$ is a $p\times p$ matrix with 1 on the diagonal and 0.5 off-diagonal. Each row of the error matrix $\bE$ (i.e., $\bepsilon_i$) is independently drawn from $N(\bzero_q,\sigma^2 \bSig_E)$,
where $\bSig_{E}$ is a $q\times q$ matrix with 1 on the diagonal and 0.5 off-diagonal and the noise level $\sigma^2$ is set such that the signal-to-noise ratio (SNR) equals a given value.
Here SNR is defined as the $r^{*}$th singular value of $\bX \bB^{*}$ divided by $\|\bE\|_{F}$.
The true coefficient matrix $\bB^*$ is constructed as $\bB^* = \bB_1 \bB^{\top}_2$, where all entries of $\bB_1\in \mathbb{R}^{p \times r^*}$ and $\bB_2 \in \mathbb{R}^{q \times r^*}$ are independently generated from $N(0, 1)$.
Recall that the true intercept matrix 
$\bA^*=(\ba_1^*, \ldots, \ba_n^*)^{\top}$. Each row of $\bA^*$ is independently and identically generated from three different vectors, $\bc_1^*$, $\bc_2^*$, and $\bc_3^*$ with equal probability, where $\bc_1^*=(\mu, \dots, \mu)^\top  \in \mathbb{R}^q$, $\bc_2^* = - \bc_1^*$, and $\bc_3^* = \bzero_q$. In other words, there are three different intercept vectors $\bc_1^*$, $\bc_2^*$, and $\bc_3^*$ in this simulation example, and $\mathrm{Pr}(\ba^*_i = \bc_1^*) = \mathrm{Pr}(\ba^*_i =\bc_2^*) = \mathrm{Pr}(\ba^*_i =\bc_3^*) = 1/3$ for $i = \{1, \dots, n\}$. We set 
$(n,p,q,r^*) = (100,12,8,3)$.
As mentioned earlier, after obtaining $\widehat{A}$, $\widehat{B}$, and $\widehat{C}$ based on the simulated data, we generate an independent test sample of size $n_{\rm{test}}=90$ to compute Pre$(\widehat{\bB}, \widehat{\bC})$.   
We consider two different settings: 
(i) $\mu$ is randomly generated from the standard Gaussian distribution $N(0, 1)$ and SNR varies across $\{1, 1.25, 1.5\}$;
(ii) $\mu$ takes different values $0.5, 1$, and $2$ with SNR fixed at $1.25$.
The first setting is designed to evaluate the performance of our methods under different 
signal-to-noise ratios
while the second setting is suggested by one referee to illustrate the impact of the minimal difference of signals between subgroups $b_n$ (which is defined in \eqref{def.bn}). 
It is clear that $b_n=\sqrt{q} \mu$ in this simulation example, meaning that 
$b_n$ increases as $\mu$ becomes larger.
Thus, we can expect that, when $\mu$ takes larger values, it is more easier to identify subgroups in setting (ii).

\begin{table}
	\caption{Means and standard errors (in parentheses) of performance measures for various methods under different signal-to-noise (SNR) ratios over 100 replications in Simulation Example 1 with setting (i). RRR: reduced-rank regression; S-MCP, S-SCAD, and S-Lasso: subgroup-learning-only procedures with MCP, SCAD, and $L_1$ penalties, respectively; SR-MCP, SR-SCAD, and SR-Lasso: our methods with the MCP, SCAD, and $L_1$ penalties, respectively; Oracle.s: oracle procedure using the true subgroup information only; Oracle.sr: oracle procedure using both the true subgroup information and true rank. The details of these methods and the simulation setting are provided in Section \ref{subsection: Example1}.  
	}
	
	\centering
	\vskip.3cm
	\scalebox{0.62}[0.62]{ 	
		\begin{tabular}{lcccccccccc}
			\hline
			Method          & $\rm{Err}(\widehat{\bB})$ & $\rm{Err}(\widehat{\bA})$   & $\rm{Pre}(\widehat{\bB},\widehat{\bC} )$ &  Rank &  Rank(\%)      &   $\rm{Err}(\widehat{\balpha}_1)$ & $\rm{Err}(\widehat{\mathbf{\balpha}}_2)$ & $\rm{Err}(\widehat{\mathbf{\balpha}}_3)$  &  $\rm{\hat{K}}$ &  K (\%)   \\
			\hline
			 \multicolumn{11}{c}{$\mbox{SNR} = 1$} \\
			\addlinespace[0.4em]
			 RRR    &  0.076 (0.009) &  0.408 (0.026)  &   0.513 (0.010)  &   4.25 (0.435)  & 0   &------  & ------    &------      & ------        &  ------   \\
			S-MCP  & 0.230 (0.041) & 0.321 (0.033)  & 0.190 (0.015)  &  8 (0)  & 0  &  0.010 (0.002)  &  0.101 (0.009)  &  0.071 (0.017)  &   2.26 (0.527)   &  48  \\
			S-SCAD & 0.236 (0.033)  &  0.386 (0.146) & 0.191 (0.012)  & 8 (0)     & 0  &  0.009 (0.007) &  0.047 (0.064)  & 0.060 (0.063)  &  2.18 (0.523)    & 48  \\
			S-Lasso &  0.324 (0.001)  & 1.000 (0.000)    & 0.562 (0.031)   & 8 (0)    & 0  & ------  & ------   & ------  & 1 (0)       & 0  \\	
			SR-MCP  & 0.068 (0.030) &  0.203 (0.113)  &  0.142 (0.005) &  3.72 (1.230) & 68 &  0.220 (0.373) &  0.076 (0.009)  &  0.108 (0.198)  &  2.83 (0.562)   &  62  \\  
			SR-SCAD    &  0.035 (0.009)  &  0.106 (0.035)  &  0.152 (0.006)  & 4.32 (1.558)  &   72  &   0.118 (0.062) &  0.011 (0.001)  & 0.012 (0.003)  &  3.04 (0.198)  &   56 \\
			SR-Lasso     & 0.053 (0.005)  &  1.000 (0.000)  & 0.514 (0.016)  & 3  (0)      & 100       & ------   &------  &------  & 1  (0)      & 0   \\
			Oracle.s    & 0.006 (0.002)  &  0.003 (0.001)  &  0.013 (0.000)  &  3.04 (0.197) &  96  &  0.004 (0.003)  & 0.004 (0.002)  & 0.003 (0.002)  & 3 (0) & 100  \\
			Oracle.sr    & 0.006 (0.002)  & 0.006 (0.003)  & 0.012 (0.000)  &  3 (0) & 100  &  0.003 (0.003)  &  0.003 (0.003)   & 0.003 (0.002)  &  3 (0) & 100 \\ 
			
		   \multicolumn{11}{c}{$\mbox{SNR} = 1.25$} \\
			\addlinespace[0.4em]
			 RRR  & 0.050 (0.004)  &  0.272 (0.015)    &  0.505 (0.012)  &  4.19 (0.394) & 0     &------  & ------    &------      & ------        &  ------   \\
			S-MCP   & 0.094 (0.013) & 0.093 (0.009) &  0.039 (0.004) &  8 (0)  & 0  &   0.008 (0.002) &  0.001 (0.001)  &  0.006 (0.002) &  3 (0)  & 100 \\
			S-SCAD     & 0.146 (0.063)  & 0.055 (0.015)   & 0.018 (0.005)  & 8 (0)     & 0  & 0.002 (0.002)  &  0.001 (0.001)  &  0.006 (0.002) &  3 (0)  &  100 \\
			S-Lasso  &  0.143 (0.009)  &  1.000 (0.000)  & 0.573 (0.039)   & 8 (0)    & 0  & ------  & ------   & ------  & 1 (0)       & 0 \\	
			SR-MCP    & 0.027 (0.013) &  0.033 (0.082)  &  0.018 (0.004) & 3.62 (1.008)  & 76 &  0.008 (0.006)  &  0.005 (0.008)  &  0.016 (0.009)     &  3.36 (0.631)   &  76 \\ 
			SR-SCAD    &0.005 (0.005) &  0.034 (0.067)   &  0.010 (0.001)  &  3.02 (1.412)   &  78 &  0.003 (0.003)  & 0.002 (0.003)    & 0.003 (0.002)  & 3.12 (0.558)    & 85  \\
			SR-Lasso     &  0.015 (0.002)  &  1.000 (0.000)  & 0.508 (0.007)  & 3  (0)      & 100       & ------   &------  &------  & 1  (0)      & 0   \\
			Oracle.s    &  0.004 (0.002) &  0.005 (0.002)  & 0.010 (0.000)   & 3.03 (0.197)   & 97   &  0.003 (0.002)  &  0.002 (0.001)  & 0.002 (0.002)  & 3 (0) & 100 \\
			Oracle.sr    &  0.004 (0.002) & 0.005 (0.002)  & 0.010 (0.000)  &  3 (0) & 100  &  0.003 (0.002)  &   0.002 (0.002)  &  0.002 (0.002)  &  3 (0) & 100 \\
			
			 \multicolumn{11}{c}{$\mbox{SNR} = 1.5$} \\
			\addlinespace[0.4em]
			RRR  &  0.042 (0.004) & 0.248 (0.024)    &   0.505 (0.010)    &   4.29 (0.456) & 0     &------  & ------    &------      & ------        &  ------  \\
			S-MCP   & 0.043 (0.025) & 0.035 (0.017)  & 0.015 (0.003)  &  8 (0)  & 0  & 0.002 (0.001)   &  0.001 (0.001)  &  0.006 (0.004) &  3.02 (0.141)  & 99   \\
			S-SCAD  &  0.039 (0.018) & 0.033 (0.013)   &  0.014 (0.002) & 8 (0)     & 0  & 0.001 (0.001)  &  0.001 (0.001)  & 0.003 (0.003)  &  3 (0)  & 100  \\
			S-Lasso  &  0.628 (0.014)  &  1.000 (0.000)  & 0.545 (0.039)   & 8 (0)    & 0  & ------  & ------   & ------  & 1 (0)       & 0 \\	
			SR-MCP  & 0.005 (0.001) & 0.003 (0.001)   & 0.008 (0.000)  &  3.52 (1.182) & 78 & 0.002 (0.001)  &  0.001 (0.001)  &   0.002 (0.001) &  3 (0)   &  100 \\  
			SR-SCAD    &  0.004 (0.002)  &  0.006 (0.007)   &  0.009 (0.000)  &  3.38 (0.967)   &  82  &  0.002 (0.002)  &  0.002 (0.001)   &  0.002 (0.002) &  3.06 (0.24)   &  94 \\
			SR-Lasso     &  0.007 (0.001)  &  1.000 (0.000)   & 0.502 (0.009)  & 3  (0)      & 100       & ------   &------  &------  & 1  (0)      & 0   \\
			Oracle.s  &  0.003 (0.001) &  0.001 (0.001)   &   0.003 (0.000)  &  3 (0)  &  100  &  0.001 (0.001)  &  0.001 (0.001)  &  0.001 (0.001) & 3 (0) & 100  \\
			Oracle.sr  &  0.002 (0.000) &  0.003 (0.002)  &  0.008 (0.000) &  3 (0) & 100  &   0.001 (0.001) &   0.001 (0.001)   &   0.001 (0.001)  &  3 (0) & 100 \\
			\addlinespace[0.4em]			
			\hline
		\end{tabular}
	}\label{t1}
\end{table}

\begin{table}
	\caption{Means and standard errors (in parentheses) of performance measures for various methods 
		over 100 replications in Simulation Example 1 with setting (ii). RRR: reduced-rank regression; S-MCP, S-SCAD, and S-Lasso: subgroup-learning-only procedures with MCP, SCAD, and $L_1$ penalties, respectively; SR-MCP, SR-SCAD, and SR-Lasso: our methods with the MCP, SCAD, and $L_1$ penalties, respectively; Oracle.s: oracle procedure using the true subgroup information only; Oracle.sr: oracle procedure using both the true subgroup information and true rank. The details of these methods and the simulation setting are provided in Section \ref{subsection: Example1}. 
		Note that the minimal difference of signals between subgroups $b_n$, defined in \eqref{def.bn}, equals to $\sqrt{q} \mu$ in this simulation example.
	}
	\centering
	\vskip .3cm
	\scalebox{0.62}[0.62]{ 	
		\begin{tabular}{lcccccccccc}
			\hline
			Method          & $\rm{Err}(\widehat{\bB})$ & $\rm{Err}(\widehat{\bA})$  & $\rm{Pre}(\widehat{\bB},\widehat{\bC} )$ &  Rank &  Rank(\%)      &   $\rm{Err}(\widehat{\balpha}_1)$ & $\rm{Err}(\widehat{\mathbf{\balpha}}_2)$ & $\rm{Err}(\widehat{\mathbf{\balpha}}_3)$  &  $\rm{\hat{K}}$ &  K (\%)   \\
			\hline
			\multicolumn{11}{c}{$\mu = 0.5$}                      \\
			\addlinespace[0.4em]
		 RRR         &  0.027 (0.003) &  0.699 (0.022)  &   0.154 (0.003) &  4.08 (0.506)   & 0    &------  & ------    &------      & ------        &  ------   \\
		 S-MCP   & 0.188 (0.036) & 0.718 (0.317) & 0.092 (0.007)  &  8 (0)  & 0  &  0.020 (0.022)  &   0.032 (0.025) & 0.115 (0.107)  &  2.24 (0.894)  & 56  \\
		 S-SCAD & 0.095 (0.027)   & 0.285 (0.094)  &  0.039 (0.007)  & 8 (0)     & 0  &  0.007 (0.002)  & 0.002 (0.001) &  0.005 (0.002)  &  2.28 (0.454) &   58 \\
		S-Lasso  &  0.049 (0.002)  &  0.999 (0.001)  &  0.167 (0.009)   & 8 (0)    & 0  & ------  & ------   & ------  & 1 (0)       & 0 \\	
		SR-MCP    & 0.012 (0.005) &  0.343 (0.075) &  0.010 (0.001)  & 3.92 (1.322)  & 64  &  0.040 (0.104) &  0.031 (0.053)  &  0.045 (0.122) &   2.88 (0.746)  &  66 \\  
		SR-SCAD    &  0.006 (0.003)  & 0.402 (0.076)   &  0.019 (0.001)  &   4.24 (1.623)  &   58 &  0.083 (0.203)  &  0.033 (0.060)   & 0.052 (0.116)  & 2.72 (1.07)    &  60  \\
		SR-Lasso     & 0.016 (0.001)   &  1.000 (0.000) &   0.153 (0.003)  & 3  (0)      & 100       & ------   &------  &------  & 1  (0)      & 0 \\
		Oracle.s  & 0.005 (0.002)  &  0.013 (0.007)  &  0.010 (0.000)  &  3.02 (0.141)  & 98   &  0.002 (0.001)  &  0.002 (0.002)  & 0.003 (0.002)  & 3 (0) & 100 \\
		Oracle.sr    & 0.005 (0.002)  &  0.015 (0.007) &  0.010 (0.000) &  3 (0) & 100  &  0.002 (0.001)  &    0.002 (0.002)  &   0.003 (0.002)  &  3 (0) & 100\\
			\multicolumn{11}{c}{$\mu = 1$}                      \\
			\addlinespace[0.4em]
			
			RRR         & 0.040 (0.004)  &  0.214 (0.007)  &    0.606 (0.011) & 4.12 (0.64) &  0     &------  & ------    &------      & ------        &  ------  \\
			S-MCP   & 0.781 (0.097) & 0.382 (0.100) &  0.201 (0.020) &  8 (0)  & 0  &  0.007 (0.003)  &  0.055 (0.036)  & 0.226 (0.202)  &  3.28 (0.621)  & 75  \\
			S-SCAD     &  0.793 (0.065)  &  0.153 (0.082) & 0.058 (0.014)  &  8 (0) & 0 & 0.025 (0.014)   &   0.004 (0.005) & 0.006 (0.004)    &    2.82 (0.438)   & 78  \\
			S-Lasso  &  0.750 (0.027)  &  0.999 (0.003)  &  0.717 (0.059) & 8 (0)    & 0  & ------  & ------   & ------  & 1 (0)       & 0 \\	
			SR-MCP    & 0.005 (0.003) & 0.094 (0.031)  & 0.010 (0.002)  &  4.58 (1.605)  &  56 & 0.012 (0.010)  &  0.007 (0.012) &  0.003 (0.003)  & 3.2   (0.571)     &  83 \\  
			SR-SCAD    &  0.032 (0.021)  &  0.204 (0.066)   &   0.017 (0.003) &  4.1 (1.282) &  58  &  0.009 (0.011)  &  0.012 (0.017)   &  0.009 (0.011) &  3.18 (0.596)   &  78    \\
			SR-Lasso     &  0.025 (0.005)  &  1.000 (0.000)   & 0.627 (0.015)  & 3  (0)      & 100       & ------   &------  &------  & 1  (0)      & 0  \\
			Oracle.s    & 0.005 (0.002)  & 0.004 (0.002)   &  0.010 (0.000)  & 3.03 (0.171)   &  97  &   0.002 (0.002) &  0.002 (0.002)  & 0.003 (0.002)  & 3 (0) & 100 \\
			Oracle.sr    &  0.004 (0.001) & 0.004 (0.002)   &  0.010 (0.000) &  3 (0) & 100  &  0.002 (0.002)  &   0.002 (0.002)  & 0.003 (0.002)   &  3 (0) & 100 \\
			
			\multicolumn{11}{c}{$\mu = 2$}  \\
			\addlinespace[0.4em]
			
			RRR         &  0.093 (0.008) &   0.093 (0.002)   &  2.384 (0.038)  &   4.13 (0.65)  &  0    & ------  & ------    &------      & ------        &  ------ \\
			S-MCP   & 1.235 (0.092) & 0.126 (0.006) & 0.209 (0.020) &  8 (0)  & 0  &  0.049 (0.009)  & 0.010 (0.005)   &  0.052 (0.006) &  3 (0)  & 100  \\
			S-SCAD     &  1.235 (0.099) &  0.126 (0.006) &  0.209 (0.021)  & 8 (0)     & 0  &  0.050 (0.009) & 0.010 (0.004)   &   0.052 (0.007) & 3 (0)   & 100  \\
			S-Lasso  &  1.837 (0.026)  & 1.000 (0.000)   & 2.758 (0.208)   & 8 (0)    & 0  & ------  & ------   & ------  & 1 (0)       & 0  \\	
			SR-MCP     & 0.038 (0.019) & 0.053 (0.024)  &  0.025 (0.003)    &  4.02 (1.342) & 62 & 0.015 (0.014) & 0.003 (0.004)  & 0.055 (0.034)  & 3.3 (0.463)   & 85  \\ 
			SR-SCAD    &  0.030 (0.018)  &  0.073 (0.028)  &  0.024 (0.004)  &      4.18 (1.371)  &  68  &  0.023 (0.032)   & 0.017 (0.022)  &   0.014 (0.015)  &  3.08 (0.444) & 96  \\
			SR-Lasso     &   0.347 (0.002) & 1.000 (0.000)  & 2.444 (0.054)  & 3  (0)      & 100       & ------   &------  &------  & 1  (0)      & 0   \\
			Oracle.s    &  0.005 (0.002) &  0.001 (0.001)   &   0.010 (0.000)  &  3.03 (0.171)  &  97  &  0.002 (0.002)  &  0.002 (0.002)   &   0.003 (0.002) & 3 (0) & 100  \\
			Oracle.sr    &  0.004 (0.001) &  0.001 (0.001) & 0.010 (0.000)  &  3 (0) & 100  &  0.002 (0.002)  &   0.002 (0.002)  & 0.003 (0.002)   &  3 (0) & 100 \\
			\addlinespace[0.4em]			
			\hline
		\end{tabular}
	}
	\label{t2}
\end{table}

Table~\ref{t1} lists the results of different methods under setting (i) in Simulation Example 1. It can be seen that our methods with the MCP and SCAD penalties (SR-MCP and SR-SCAD) perform better than other methods and the performance of our methods improves when the SNR becomes larger.
First, we can observe that both S-Lasso and SR-Lasso, which use the $L_1$ penalty for subgroup learning, tend to yield no subgroups. This is not surprising because that the $L_1$ penalty applies the same thresholding to all pairs $\|\ba_i-\ba_j\|_2$ and leads to biased estimates for the intercept matrix $\bA$. A similar finding is also reported by \cite{Shu2017} for the univariate response setting.  However, for S-MCP, S-SCAD, SR-MCP, and SR-SCAD, which use the concave penalties, can correctly recover subgroups. In fact, we observe that the average values of $\hat{K}$ across the 100 replications for these four methods are close to the true number of subgroups under different signal-to-noise ratios. 
Moreover, we observe that the percentage of correct subgroup identification (K$\%$) for our methods SR-MCP and SR-SCAD become larger as the SNR increases.
Since the RRR method does not incorporate subgroup analysis, the corresponding performance measures for subgroup analysis are not included. 
Second, compared with the subgroup-learning-only procedures (S-MCP and S-SCAD) or the reduced-rank-only procedure RRR, our methods SR-MCP and SR-SCAD uniformly yield smaller values of the estimation error Err$(\widehat{\bB})$ and the prediction error Pre$(\widehat{\bB}, \widehat{\bC})$. This indicates that our methods SR-MCP and SR-SCAD outperform the subgroup-learning-only procedures and the reduced-rank-only procedure in estimating the matrix $\bB$ and predicting the response vector.
Third, the performance of our methods SR-MCP and SR-SCAD, in terms of the four performance measures—Err$(\widehat{\bB})$, Err$(\widehat{\bA})$, 
Pre$(\widehat{\bB}, \widehat{\bC})$, and K$\%$—approaches that of the two benchmarks, Oracle.s and Oracle.sr, as the SNR increases.

The results of different methods under setting (ii) in Simulation Example 1 are summarized in Table~\ref{t2}. We can observe similar conclusions as those for setting (i) in Simulation Example 1. In addition, as expected, we observe that for each of these four methods (S-MCP, S-SCAD, SR-MCP, and SR-SCAD), as $\mu$ increases, there is a gradual decrease in the estimation error of $\bA$ (i.e., Err$(\widehat{\bA})$) and an increasing trend in the percentage of correct subgroup identification (i.e., K$\%$).

\subsection{Simulation Example 2}\label{subsection: Example2}

In our second simulation example, we consider a homogeneous multivariate response regression model 
$\by_i = \bc_1  + \bB^{\top} \bx_i + \bepsilon_i$
where $\bc_1=(\mu, \dots, \mu)^\top  \in \mathbb{R}^q$ and $\mu$ is randomly generated from the standard Gaussian distribution $N(0, 1)$ and SNR varies across $\{1, 1.25, 1.5\}$. In other words, the number of subgroups $K=1$ and each row of the true intercept matrix $\bA^*$ is equal to $(\mu, \dots, \mu)^\top  \in \mathbb{R}^q$ for $i = \{1, \dots, n\}$ in this example. The covariate matrix $\bX$, the error matrix $\bE$, and the true coefficient matrix $\bB^*$ are simulated in the same way as in Simulation Example 1. The corresponding simulation results over 100 replications are reported in Table~\ref{tab: Example2} in the supplementary material. Again, the performance measures for subgroup analysis using the RRR method are not included, as the method does not incorporate subgroup analysis.
We observe that the mean of the estimated number of subgroups ($\hat{K}$) is 1 for all other methods. In addition, the percentage of correct subgroup identification (K$\%$) is $100\%$ for all other methods. Moreover, the performance of our methods SR-MCP and SR-SCAD is very close to that of the benchmark methods, Oracle.s and Oracle.sr, and exceeds that of RRR, S-MCP, and S-SCAD in rank selection, estimating the matrix $\bB$ and predicting the response vector. This is consistent with those reported in Example 1.

\begin{table}[H] 
	\caption{Means and standard errors (in parentheses) of performance measures for various methods under different signal-to-noise (SNR) ratios over 100 replications in Simulation Example 2. RRR: reduced-rank regression; S-MCP, S-SCAD, and S-Lasso: subgroup-learning-only procedures with MCP, SCAD, and $L_1$ penalties, respectively; SR-MCP, SR-SCAD, and SR-Lasso: our methods with the MCP, SCAD, and $L_1$ penalties, respectively; Oracle.s: oracle procedure using the true subgroup information only; Oracle.sr: oracle procedure using both the true subgroup information and true rank. The details of these methods and the simulation setting are provided in Section \ref{subsection: Example2}} 
	\centering
	\vskip.3cm
	\scalebox{0.8}[0.8]{ 	
		\begin{tabular}{lcccccccc}
			\hline
			Method          & $\rm{Err}(\widehat{\bB})$ & $\rm{Err}(\widehat{\bA})$   & $\rm{Pre}(\widehat{\bB},\widehat{\bC} )$ &  Rank &  Rank(\%)      &   $\rm{Err}(\widehat{\balpha}_1)$  &  $\rm{\hat{K}}$ &  K (\%)   \\
			\hline
			 \multicolumn{9}{c}{$\mbox{SNR} = 1$} \\
			\addlinespace[0.4em]
			RRR   &  0.030 (0.004) &  0.148 (0.011) & 0.019 (0.001)  &  4.44 (0.656)  & 0     &------  & ------        &  ------   \\
			S-MCP   & 0.012 (0.003) & 0.000 (0.000)  & 0.026 (0.001)   & 8 (0) & 0 & 0.000 (0.000)  &  1 (0)  &  100  \\
			S-SCAD   & 0.015 (0.002) &  0.000 (0.000)  &  0.019 (0.001)  & 8 (0)  &  0 & 0.000 (0.000)  &  1 (0)  & 100 \\
			S-Lasso  & 0.008 (0.009) & 0.000 (0.000) & 0.010 (0.000) & 8 (0) & 0  &  0.000 (0.000)   & 1 (0) & 100 \\	
			SR-MCP   & 0.006 (0.002)  & 0.000 (0.000)  &  0.009 (0.001)  &  3.02 (0.141) & 98 &  0.000 (0.000) & 1 (0)   & 100 \\  
			SR-SCAD  & 0.007 (0.002) &  0.000 (0.000) &  0.008 (0.000)  & 3.06 (0.24)  & 94 &  0.000 (0.000) & 1 (0)   & 100 \\
			SR-Lasso    & 0.012 (0.000) &  0.000 (0.000) & 0.006 (0.000) & 3 (0) &  100 &  0.000 (0.000) & 1 (0) & 100  \\
			Oracle.s  & 0.007 (0.004) & 0.000 (0.000) & 0.006 (0.001)  &  3.08 (0.274)  &  92 & 0.000 (0.000)  & 1 (0) & 100 \\
			Oracle.sr    &  0.006 (0.001) & 0.000 (0.000)  &   0.006 (0.000) &  3 (0) & 100  & 0.000 (0.000)  & 1 (0) & 100 \\ 
			\multicolumn{9}{c}{$\mbox{SNR} = 1.25$} \\
			\addlinespace[0.4em]
			 RRR  & 0.027 (0.003)  & 0.105 (0.006)  &  0.013 (0.001) &  4.28 (0.494)  & 0     &------  & ------     &  ------   \\
			S-MCP   & 0.009 (0.001) &  0.000 (0.000) &  0.016 (0.000)  &  8 (0) & 0  & 0.000 (0.000)  &  1 (0)  & 100  \\
			S-SCAD   & 0.009 (0.001)  & 0.000 (0.000)  &  0.016 (0.000)  &  8 (0)  & 0 &  0.000 (0.000) &  1 (0)  & 100 \\
			S-Lasso  & 0.010 (0.001) & 0.000 (0.000) & 0.013 (0.000) & 8 (0) & 0  & 0.000 (0.000)  & 1 (0) &  100 \\	
			SR-MCP   & 0.006 (0.002) & 0.000 (0.000)  & 0.008 (0.001)   & 3 (0)  & 100 & 0.000 (0.000)  &  1 (0)  & 100 \\  
			SR-SCAD  & 0.006 (0.006) & 0.000 (0.000)  &  0.007 (0.000)  &   3 (0) & 100 &  0.000 (0.000) & 1 (0)   & 100 \\
			SR-Lasso    & 0.004 (0.001) & 0.000 (0.000) & 0.008 (0.000) & 3 (0) &  100 & 0.000 (0.000)  & 1 (0) &  100 \\
			Oracle.s  & 0.005 (0.002)  & 0.000 (0.000) & 0.010 (0.000) & 3.08 (0.274)  &  92   & 0.000 (0.000)  & 1 (0) & 100 \\
			Oracle.sr    & 0.006 (0.002)  &  0.000 (0.000) &  0.007 (0.000)  &  3 (0) & 100  &  0.000 (0.000) &   1 (0) & 100 \\
			\multicolumn{9}{c}{$\mbox{SNR} = 1.5$} \\
			\addlinespace[0.4em]
			RRR   &  0.025 (0.003) & 0.083 (0.004)  & 0.010 (0.000)  &  4.22 (0.44)  & 0     &------  & ------       &  ------   \\
			S-MCP   & 0.007 (0.008) & 0.000 (0.000)  & 0.108 (0.000) & 8 (0) & 0 &  0.000 (0.000) &  1 (0)  & 100  \\
			S-SCAD   & 0.007 (0.001) & 0.000 (0.000)  & 0.011 (0.000)   &  8 (0)   & 0 & 0.000 (0.000)  &  1 (0)  & 100 \\
			S-Lasso  & 0.004 (0.000) & 0.000 (0.000) & 0.008 (0.000) & 8 (0) & 0  & 0.000 (0.000)  & 1 (0) & 100 \\	
			SR-MCP   & 0.003 (0.000) & 0.000 (0.000)  &  0.009 (0.001)  &   3 (0)  & 100 & 0.000 (0.000)  &   1 (0) & 100 \\  
			SR-SCAD  & 0.003 (0.001) & 0.000 (0.000)  &  0.007 (0.000) &   3 (0) & 100 &   0.000 (0.000)  &  1 (0)  & 100 \\
			SR-Lasso    & 0.004 (0.001) &  0.000 (0.000) & 0.008 (0.000) & 3 (0) &  100 & 0.000 (0.000)  & 1 (0) &  100 \\
			Oracle.s  & 0.003 (0.002) & 0.000 (0.000) &  0.007 (0.000)  & 3.06 (0.24)  &  94   & 0.000 (0.000)  & 1 (0) & 100 \\
			Oracle.sr    &  0.003 (0.001) & 0.000 (0.000)  &  0.007 (0.000)  &  3 (0) & 100  & 0.000 (0.000)  &      1 (0) & 100  \\
			\addlinespace[0.4em]
			\hline
		\end{tabular}
	}\label{tab: Example2}
\end{table}


\section{Data analysis}\label{real data}

In this section, we further demonstrate our methods using Arabidopsis thalian dataset, which is a gene expression dataset obtained from Affymetrix GeneChip microarray experiments on the plant Arabidopsis thaliana~\citep{Wille2004}. It is well known that isoprenoids play a key role in plant biochemical functions such as photosynthesis, growth regulation, and defense against pathogens. To examine the regulatory control mechanisms in the isoprenoid gene network in Arabidopsis thaliana, \cite{Wille2004} conducted a genetic association study under various experimental conditions. In particular, they conducted $n=118$ GeneChip microarray experiments to monitor gene expression levels, which are represented by continuous fluorescence signal intensity values captured by a confocal scanner.  To analyze the connections between downstream pathways and isoprenoid biosynthesis pathways, we consider a multivariate response regression model, where the expression levels of $p=18$ genes from one isoprenoid biosynthesis pathway, known as the non-mevalonate pathway (or MEP), serve as predictors $\bX$, and the expression levels of $q=28$ genes from two downstream pathways (carotenoid and glutamate-glutamine) serve as the responses $\bY$.
We applied a log transformation to the responses to reduce skewness and standardized the predictors to make them comparable.

\begin{figure}[h]
	\centering
	\includegraphics[width=7cm,height=6.2cm]{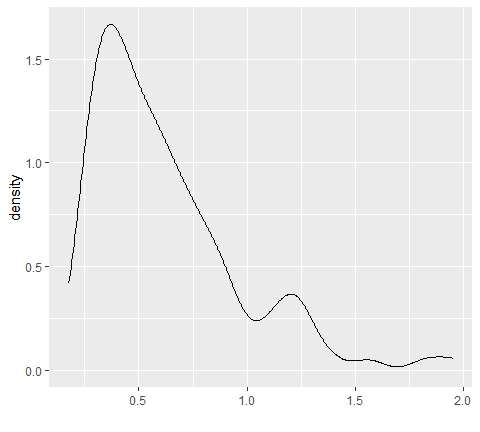}
	\caption{Density plot of $\|\hat{\bepsilon}_i\|_2/\sqrt{q}$, where $\hat{\bepsilon}_i^{\top}$ is the $i$th row of the residual matrix after controlling for the effects of the predictors in the Arabidopsis thalian data.}
	\label{fig:densityOLS}
\end{figure}

To examine whether heterogeneity is present in this dataset, we first fit a homogeneous multivariate regression model $\bY = \bX \bB + \bE$. Then the error matrix $\bE$ can be estimated by $ \widehat{\bE}^\textrm{ols} = \bY - \bX \widehat{\bB}^{\textrm{ols}} $, where $\widehat{\bB}^{\textrm{ols}}$ is the estimated coefficient matrix of $\bB$ obtained using the ordinary least squares (OLS) method. Fig~\ref{fig:densityOLS} plots the kernel density estimates of $\|\hat{\bepsilon}_i\|_2/\sqrt{q}$, where $\hat{\bepsilon}_i^{\top}$ is the $i$th row of $\widehat{\bE}^\textrm{ols}$. Obviously, there are two modes, indicating that unobserved heterogeneous effects may still exist even after adjusting for the effects of the covariates. 
Therefore, it is not suitable to fit a homogeneous  multivariate regression model. 
This motivates us to consider 
the heterogeneous  
model~\eqref{model2} and identify subgroups using our proposed procedure.

Note that Oracle.sr and Oracle.s are no longer applicable, as the true subgroup information and true rank are unknown in this real dataset.   
Therefore, Table~\ref{tab: RealData} does not include the results of these two methods for this dataset.
The in-sample 
mean squared error $\mbox{MSE}=\| \bY - \bX \widehat{\bB} - \widehat{\bA}\|_F^2/(nq)$
is used to evaluate the goodness of fit of different methods.
We also report the
estimated rank (Rank) and the estimated number
of subgroups ($\hat{K}$). 
Our findings from Table~\ref{tab: RealData} can be summarized as follows. First, our proposed methods, SR-MCP and SR-SCAD, perform similarly. Both identified two subgroups and five latent factors. 
Second, compared to RRR, which cannot capture the latent heterogeneous structure, our methods SR-MCP and SR-SCAD significantly reduces the prediction error.
Third, although S-MCP and S-SCAD also identified two subgroups, the selected rank of 18 makes the estimated model difficult to interpret in practice.

\begin{table}[h]	
	\caption{Estimated rank (Rank), estimated number of subgroups ($\hat{K}$), in-sample mean squared error (MSE) of different methods for the Arabidopsis thalian dataset. S-MCP, S-SCAD, and S-Lasso: subgroup-learning-only procedures with MCP, SCAD, and $L_1$ penalties, respectively; SR-MCP, SR-SCAD, and SR-Lasso: our methods with the MCP, SCAD, and $L_1$ penalties, respectively; RRR: reduced-rank regression.
	}
	\centering
	\vskip.3cm	
	\begin{tabular}{cccccccccc}
		\hline
		Method      & S-MCP    &S-SCAD & S-Lasso     &  SR-MCP      & SR-SCAD  & SR-Lasso   &RRR        \\
		\hline
		Rank  & 18     &18     &  18 & 5   & 5      & 5      &9            \\
		$\hat{K}$   & 2  & 2    &  1 & 2    & 2  & 1  & ---       \\
		$\mbox{MSE}$ & 0.134      & 0.126        & 0.135      &   0.136     & 0.148  & 0.164   &  0.509  \\
		\hline
	\end{tabular}
	\label{tab: RealData}
\end{table}

Next, we further analyze the results of this real dataset from two perspectives: subgroup analysis and low-rank association network analysis. First, as shown in Table~\ref{tab: RealData}, our methods, SR-MCP and SR-SCAD, identified two subgroups. To uncover this heterogeneity, we compare the estimated common intercept vectors between the two identified subgroups. Clearly, the greater the absolute difference between the intercept vectors, the greater the heterogeneity between the subgroups. The downstream genes in the response matrix $\bY$ associated with the five largest absolute differences between corresponding elements of the intercept vectors for the two identified subgroups are AT5G35630, AT5G04140, AT1G66200, AT5G18170, and AT5G07440. All these five genes come from the Glutamate-Glutamine pathway.  Previous studies have confirmed heterogeneity in the expression of genes involved in Glutamate–Glutamine metabolism~\citep{liao2022}, and our results are, to some extent, consistent with these findings.

Second, to gain further insights into the regulatory mechanisms within the isoprenoid gene association network, we examine the loading coefficients of the estimated factor and select those larger than $ \pm \sigma^{X \widehat{B} + \widehat{A}}/q^{-1/2}$, where the estimated noise variance $\sigma^{X \widehat{B} + \widehat{A}}$ is given by $\|\bY - \bX \widehat{\bB} - \widehat{\bA}\|_F/\sqrt{nq}$.  Let $\widehat{\bZ} \widehat{\bD} \widehat{\bV}^\top$ be the singular value decomposition of $\bX \widehat{\bB}$, where $\widehat{\bZ} \in \mathbb{R}^{p \times 5}$ denotes the estimated factor matrix containing five orthogonal factors, and $\widehat{\bD} \widehat{\bV}^\top \in \mathbb{R}^{5 \times q}$ represents the associated factor loading coefficients.
It is evident that a higher coefficient reflects stronger regulatory influence of the predictor on downstream genes. We list the genes selected by our method SR-MCP in Table \ref{table.factor}. For comparison, we also report the results of RRR for the first five latent factors. Due to the large number of genes selected by RRR, we report only those that overlap with the genes selected by SR-MCP to save space. It can be seen that for our method SR-SCAD, three genes, namely AT3G10230, AT5G17230, and AT1G77670, are regulated by all latent factors. Both AT3G10230 and AT5G17230 are downstream genes in the carotenoid pathway. AT3G10230 encodes a protein with lycopene $\beta$-cyclase activity, while AT5G17230 encodes phytoene synthase, the rate-limiting enzyme in the carotenoid biosynthetic pathway.  It has been experimentally confirmed that the carotenoid pathway is strongly linked to the MEP pathway \citep{Wille2004}. factors. 
Moreover, compared to RRR, we observe a stronger regulatory mechanism between the predictors and the downstream genes AT2G22250 and AT1G80360. Our methods identify that both genes are regulated by three latent factors, whereas the RRR method finds that AT2G22250 is regulated by one latent factor and AT1G80360 by two. AT2G22250 encodes a prokaryotic-type plastidic aspartate aminotransferase with glutamate activity, while AT1G80360 encodes a methionine-specific aminotransferase that enables VAS1 to coordinate both auxin and ethylene biosynthesis. This may reflect a meaningful biological relationship with the isoprenoid biosynthesis pathway, warranting further investigation.

\begin{table}[h]
	\caption{The genes selected by SR-MCP in each latent factor and the genes selected by RRR in the top five factors, along with their associated pathways, for the Arabidopsis thaliana dataset. Note that for RRR, we report only the genes that overlap with those selected by SR-MCP to save space.
	} \centering
	\vskip.3cm
	\scalebox{0.56}[0.8]{ 
		\begin{tabular}{lllcccclccccc}
			\hline  
			& Pathway &  \multicolumn{5}{c}{SR-MCP} & \multicolumn{5}{c}{RRR}\\
			
			\cmidrule(lr){2-2} \cmidrule(lr){3-7}	\cmidrule(lr){8-12}
			\addlinespace[0.3em]
			1th factor & carotenoid  & AT3G10230 &  AT3G04870  & AT4G14210  &  AT5G17230 &  &   AT3G10230& AT3G04870  & AT4G14210  &  AT5G17230         \\
			& GlutamateGlutamine 	   &  AT4G31990 & AT1G77670  &   AT1G80360  & AT1G62800  &         &    AT4G31990 & AT1G77670  &   AT1G80360  & AT1G62800     \\
			\addlinespace[0.3em]				
			2th factor & carotenoid & AT3G10230 & AT4G14210   &  AT5G17230  &    &   & AT5G17230   & & & &    \\
			& GlutamateGlutamine  &  AT1G77670  &    AT1G80360 & AT2G22250   &  &   &   AT1G80360 & AT2G22250  & &      \\
			\addlinespace[0.3em]
			3th factor & carotenoid& AT5G57030   &  AT3G10230 & AT5G17230 &     &      &  & & &      \\
			& GlutamateGlutamine  & AT4G31990  & AT1G77670  & AT1G62800 &  AT2G22250 &      & AT4G31990 & AT1G62800 &     &       \\
			\addlinespace[0.3em]  
			4th factor & carotenoid & AT5G57030   & AT3G10230 &  AT3G04870 & AT4G14210 & AT5G17230  &     AT5G57030 & AT3G10230 & AT3G04870 &        \\
			& GlutamateGlutamine   & AT4G31990  &  AT1G77670 & AT1G62800 & AT2G22250 &     &  AT1G77670  & & &       \\
			\addlinespace[0.3em]   
			5th factor & carotenoid & AT5G57030   & AT3G10230 &  AT3G04870    & AT5G17230  &   & AT5G57030  & & &      \\
			&  GlutamateGlutamine & AT1G77670 &  AT1G80360 & &  &         & AT1G77670 & & &        \\
			\addlinespace[0.3em]
			\hline
		\end{tabular}
	}
	\label{table.factor}
\end{table}


\section{Discussion}\label{sec.dicus}

In this paper, we propose a novel heterogeneity and reduced-rank learning framework that simultaneously identifies subgroup structures and estimates covariate effects in heterogeneous multivariate response regression. 
The algorithm proposed by~\cite{Shu2017} is not applicable for solving the rank-constrained optimization problem in our paper. We apply the block coordinate descent method from~\cite{Tseng2001} to address this issue and implement the proposed approach using a new ADMM algorithm. It can be shown that the computational complexity of Step 1 in our algorithm is of the order $O(n^3q + n(p + q)^2)$, indicating that the computational cost becomes substantial when the sample size $n$ is very large. Thus, developing a more scalable algorithm to implement the proposed approach would be a valuable direction for future work.

In addition, to simplify our presentation and introduce our main idea, we focus on studying model~\eqref{HTE model} in which population heterogeneity is captured through subgroup-specific intercept vectors, implying that heterogeneity comes from unobserved latent factors after accounting for the effects of the predictors. Our joint heterogeneity and reduced-rank learning framework can be readily extended to the following heterogeneous multivariate response model 
\begin{align}\label{HTE-mod3} 
	\by_i = \bA_i^{\top} \bz_i + \bB^{\top} \bx_i + \bepsilon_i,
\end{align}
where $\bz_i$ is a $d$-dimensional vector of observed variables and the population heterogeneity is characterized by coefficient matrices $\bA_i$'s.  In other words, in this model some covariates can also contribute to the population heterogeneity. 
This model also includes model~\eqref{HTE model} as a special case when $\bz_i=1$. Our ADMM algorithm can be easily modified to obtain the estimators of $\bA$ and $\bB$ in this model; see the supplementary material for details.

Another possible extension is to consider a more general model setting in which the covariance structure of the random error vector $\bepsilon_i$ varies across different subgroups. This is an interesting but challenging problem, as it requires significant additional effort to account for different covariance structures across subgroups and to establish the theoretical properties of the estimators. 
Finally, in many applications, the responses are categorical or count data. Therefore, it is of practical importance to extend our joint heterogeneity and reduced-rank learning framework to multi-response generalized linear models. These possible extensions are beyond the scope of the current paper and will be interesting topics for future research.


\section*{Appendix A: Proofs of Theorems~\ref{theo1}-\ref{theo3}}

\setcounter{equation}{0}
\renewcommand{\theequation}{A.\arabic{equation}} %

Throughout the rest of the paper, with slight abuse of notation, we use $\langle\cdot, \cdot\rangle$ to denote either the inner product of two vectors or the Frobenius inner product of two matrices.
Let $RS(\cdot)$ and $CS(\cdot)$ denote the row space and column space of a given matrix, respectively. 
Denote by $\mathbf{P}_{\mathbf{X}}$ the orthogonal projection matrix onto $CS(\bX)$, that is, $\mathbf{P}_{\mathbf{X}}= \mathbf{X}(\mathbf{X}^\top \mathbf{X})^{-}\mathbf{X}^{\top}$, where $``-"$ stands for
the Moore–Penrose inverse. Given $\mathbf{P}_{\mathbf{X}}$, let $\mathbf{P}^{\perp}_{\mathbf{X}}$ be the projection onto its orthogonal complement.  For two real numbers $x$ and $y$, denote by $x \vee  y =
\max\{x, y\}$ and $x \land  y =\min\{x, y\}$. 
Recall that throughout this paper, $\tilde{c}$, $\tilde{c}_1$, $\tilde{c}_2$, $c$, $c_1$, $c_2$, $C$, $C_1$ and $C_2$ are generic positive constants, which may vary from line to line.

\vspace{4mm}


\noindent{{\bf Proof of Theorem~\ref{theo1}}}. 
Since $(\widehat{\bB}^{or}, \widehat{\bC}^{or})$ is a global minimizer of \eqref{oracle2}, for any $(\bB, \bC)$ with $\mbox{rank}(\bB) \leq r^*$, we have
\begin{align}\label{eq17-new}
	\| \bY - \bX \widehat{\bB}^{or} - \bW \widehat{\bC}^{or} \|_F^2 \leq \|  \bY - \bX \bB - \bW \bC \|_F^2.
\end{align}
Since $\bY = \bX \bB^* + \bW \bC^* + \bE$, the left-hand side of~\eqref{eq17-new} can be decomposed into 
$\|\bE \|_F^2- 2 \langle \bE, \bX(\widehat{\bB}^{or} - \bB^*) + \bW (\widehat{\bC}^{or} - \bC^*) \rangle + \| \bX (\widehat{\bB}^{or} - \bB^*) + \bW (\widehat{\bC}^{or} - \bC^*) \|_F^2 $. 
Similarly, the right-hand side of~\eqref{eq17-new} can be rewritten as
$\|\bE \|_F^2- 2 \langle \bE, \bX(\bB - \bB^*) + \bW (\bC - \bC^*) \rangle + \| \bX (\bB - \bB^*) + \bW (\bC - \bC^*) \|_F^2 $.
After some algebra, we have
\begin{align}\label{theo-1} 
	& \| \bX (\widehat{\bB}^{or} - \bB^*  ) +  \bW (\widehat{\bC}^{or} - \bC^* ) \|_F^2 
	\leq \| \bX (\bB - \bB^*  ) +  \bW (\bC - \bC^* ) \|_F^2
	+ 2\langle \bE, \bX \mathbf{\Delta}^B +  \bW \mathbf{\Delta}^C \rangle,
\end{align}
where $\mathbf{\Delta}^B = \widehat{\bB}^{or} - \bB$ and $\mathbf{\Delta}^C = \widehat{\bC}^{or} - \bC$. Since 
$\mbox{rank}(\widehat{\bB}^{or}) \leq r^*$ and $\mbox{rank}(\bB) \leq r^*$, we have $r(\mathbf{\Delta}^B) \leq 2r^*$. Define the event $$\mathscr{A}=\left\{2 \langle \bE, \bX \mathbf{\Delta}^B + \bW \mathbf{\Delta}^C \rangle - a^{-1}\| \bX \mathbf{\Delta}^B + \bW \mathbf{\Delta}^C \|_F^2 - 4a L_0 \sigma^2 [(p+q-r^*)(r^* + K) + qK]\geq 4a\sigma^2 t\right\}$$ where $L_0>0$ and $a>0$ are two constants, and $t>0$. Since
$4[(p+q-r^*)(r^* + K) + qK]
>(p+q-2r^*)(2r^* + K) + qK$, we have 
$$\mathscr{A}\subseteq
\left\{2 \langle \bE, \bX \mathbf{\Delta}^B + \bW \mathbf{\Delta}^C \rangle - a^{-1}\| \bX \mathbf{\Delta}^B + \bW \mathbf{\Delta}^C \|_F^2 - a L_0 \sigma^2 [(p+q-2r^*)(2r^* + K) + qK]\geq 4a\sigma^2 t\right\}$$
and 
\begin{align*}
	& \mathrm{Pr}(\mathscr{A}) \\
	\leq & \mathrm{Pr}\left(2 \langle \bE, \bX \mathbf{\Delta}^B + \bW \mathbf{\Delta}^C \rangle - a^{-1}\| \bX \mathbf{\Delta}^B + \bW \mathbf{\Delta}^C \|_F^2 - a L_0 \sigma^2 [(p+q-2r^*)(2r^* + K) + qK]\geq 4a\sigma^2 t\right)\\
	\leq & \tilde{c}\exp(-c t),
\end{align*}
where the last inequality is obtained by applying Lemma~\ref{lem1} to $\mathbf{\Delta}^B$ and $\mathbf{\Delta}^C$. 
This leads to 
\begin{align*}
	\mathrm{Pr}(\mathscr{A}^c)=1-\mathrm{Pr}(\mathscr{A})\geq 1-\tilde{c}\exp(-c t)
\end{align*}
for any $t>0$, where $\mathscr{A}^c$ is the complement of event $\mathscr{A}$. In other words, the following inequality 
\begin{align}\label{eq20}
	&2 \langle \bE, \bX \mathbf{\Delta}^B + \bW \mathbf{\Delta}^C \rangle - a^{-1}\| \bX \mathbf{\Delta}^B + \bW \mathbf{\Delta}^C \|_F^2 - 4a L_0 \sigma^2 [(p+q-r^*)(r^* + K) + qK] \nonumber\\
	<& 4a\sigma^2 t 
\end{align}
holds with probability at least $1-\tilde{c}\exp(-c t)$ for any $t>0$.

On the other hand, 
using the Cauchy-Schwarz inequality and applying the inequality $2xy\leq x^2+y^2$ for two real numbers $x$ and $y$, we have
\begin{align*}
	& 2\langle \bX(\widehat{\bB}^{or} -   \bB^* ) + \bW (\widehat{\bC}^{or} - \bC^* ), \,\bX(\bB^*-\bB ) + \bW (\bC^*-\bC) \rangle \\
	\leq & 2\| \bX(\widehat{\bB}^{or} - \bB^* ) + \bW (\widehat{\bC}^{or} - \bC^* )\|_F\cdot \| \bX(\bB^*-\bB ) + \bW (\bC^*-\bC) \|_F \\
	\leq & \| \bX(\widehat{\bB}^{or} - \bB^* ) + \bW (\widehat{\bC}^{or} - \bC^* )\|_F^2 +  \| \bX(\bB^*-\bB ) + \bW (\bC^*-\bC) \|_F^2 \\
	= & \| \bX(\widehat{\bB}^{or} - \bB^* ) + \bW (\widehat{\bC}^{or} - \bC^* )\|_F^2 +  \| \bX(\bB -\bB^* ) + \bW (\bC-\bC^*) \|_F^2 
\end{align*}
and thus
\begin{align*}
	&\| \bX \mathbf{\Delta}^B + \bW \mathbf{\Delta}^C \|_F^2
	\leq 2\| \bX(\widehat{\bB}^{or} - \bB^* ) + \bW (\widehat{\bC}^{or} - \bC^* )\|_F^2 +  2\| \bX(\bB^*-\bB ) + \bW (\bC^*-\bC) \|_F^2.
\end{align*}
This, together with \eqref{theo-1} and \eqref{eq20}, leads to 
\begin{align*}
	& (1 -  2a^{-1})\| \bX (\widehat{\bB}^{or} - 
	\bB^*  ) +  \bW (\widehat{\bC}^{or} - \bC^* ) \|_F^2\\ 
	\leq & (1+2a^{-1})\| \bX (\bB - \bB^*  ) +  \bW (\bC - \bC^* ) \|_F^2  +4a L_0 \sigma^2 \left[(p+q-r^*)(r^* + K) + qK\right]+ 4a\sigma^2 t
\end{align*}
holds with probability at least $1-\tilde{c}\exp(-c t)$ for any $t>0$. Setting $\bB = \bB^* $, $\bC = \bC^* $, $a = 4$ and $t= \mathrm{ln} (n)$ yields
that
\begin{align}\label{eq: theo1-pt1-bound}
	\|\bX(\widehat{\bB}^{or} - \bB^* ) + \bW (\widehat{\bC}^{or}- \bC^* )  \|_F^2 \leq c_1 \sigma^2 \left[ (p+q-r^*)(r^* + K) + qK + \mathrm{ln} (n) \right], 
\end{align}
holds with probability at least $1- \tilde{c}_1\exp( - c \ln (n)) $,
where $c_1=\max\{32L_0, 32\}$ is a positive constant. This completes the proof of the first part of Theorem~\ref{theo1}.

We now proceed to prove the second part of Theorem~\ref{theo1}. 
In view of \eqref{eq: theo1-pt1-bound}, we have $\mathrm{Pr}(\Omega_1)\geq 1- \tilde{c}_1\exp\{ - c \mathrm{ln} (n)\}$, where 
$$\Omega_1=\left\{\|\bX(\widehat{\bB}^{or} - \bB^* ) + \bW (\widehat{\bC}^{or}- \bC^* )  \|_F^2 \leq c_1 \sigma^2 \left[ (p+q-r^*)(r^* + K) + qK + \mathrm{ln} (n) \right]\right\}.$$ 
Thus, to prove the second part of Theorem~\ref{theo1}, it suffices to show that, conditional on the event $\Omega_1$, the following two inequalities 
\begin{align*}
	\| ((\widehat{\bB}^{or} - \bB^* )^{\top}, (\widehat{\bC}^{or}- \bC^* )^{\top})^{\top} \|_F^2 \leq \phi_n^2\quad\mbox{and}\quad \|  \widehat{\bA}^{or}- \bA^* \|_F^2 \leq |\psi_{\max}| \phi_n^2
\end{align*}
hold, where $\phi_n=\sqrt{c_1 C_1^{-1} \sigma^2 [  (p+q-r^*)(r^* + K) + qK + \mathrm{ln} (n)]/|\psi_{\min}|}$. 

In what follows, our analysis will be conditional on the event $\Omega_1$. 
Let $\mathbf{F} = (\bX, \bW )$, $ \breve{\bX} = \bI_q \otimes \mathbf{F} $, $\mathbf{\Delta}^{BC} =  \left[(\widehat{\bB}^{or} - \bB^*)^{\top}, (\widehat{\bC}^{or} - \bC^*)^{\top} \right]^{\top}$ and $\bdelta^{BC} = \rm{vec}(\mathbf{\Delta}^{BC})$, where 
$\bI_q$ is the $q\times q$ identity matrix and 
$``\rm{vec}"$ is the standard vectorization operator. 
The singular values of $ \breve{\bX}$ are the singular values of $\mathbf{F}$, each repeated $q$ times. Then by  Condition~\ref{con3}, we have
\begin{align*}
	\|\bX(\widehat{\bB}^{or} - \bB^* ) + \bW (\widehat{\bC}^{or}- \bC^* )  \|_F^2 =\| \mathbf{F} \mathbf{\Delta}^{BC} \|_F^2  = \|\breve{\bX} \bdelta^{BC}\|_2^2 \geq C_1 | \psi_{\min}|\|  \bdelta^{BC} \|_2^2 = C_1 | \psi_{\min}| \|  \mathbf{\Delta}^{BC} \|_F^2.
\end{align*}
Thus, conditional on the event
$\Omega_1$, we have 
$$C_1 | \psi_{\min}|\cdot \|  \mathbf{\Delta}^{BC} \|_F^2\leq c_1 \sigma^2 \left[ (p+q-r^*)(r^* + K) + qK + \mathrm{ln} (n) \right]$$
which yields 
\begin{align}\label{BC}
	\| ((\widehat{\bB}^{or} - \bB^* )^{\top}, (\widehat{\bC}^{or}- \bC^* )^{\top})^{\top} \|_F^2 = \|  \mathbf{\Delta}^{BC} \|_F^2
	\leq \phi_n^2.
\end{align}
Moreover,  
we have
\begin{align*}
	\| \widehat{\bA}^{or} - \bA^*  \|_F^2  
	= \sum_{k=1}^K \left(|\psi_k|\cdot\|\widehat{\bc}^{or}_k - \bc^* _k\|_2^2\right) 
	\leq |\psi_{\max}| \sum_{k=1}^K \|\widehat{\bc}^{or}_k - \bc^* _k \|_2^2 
	= |\psi_{\max}|\cdot \| \widehat{\bC}^{or} - \bC^* \|_F^2.  
\end{align*}
This, together with \eqref{BC}, leads to 
\begin{align*}
	\| \widehat{\bA}^{or} - \bA^*  \|_F^2 \leq |\psi_{\max}| \phi_n^2
	\quad\mbox{and}\quad
	\sup_i \|\widehat{\ba}^{or}_i - \ba^* _i\|_2 = \sup_k \|\widehat{\bc}^{or}_k - \bc^* _k \|_2 \leq \|\widehat{\bC}^{or} - \bC^* \|_F \leq \phi_n.
\end{align*}
This completes the proof of the second part of Theorem~\ref{theo1}.

\hfill $\square$

\bigskip


\noindent{{\bf Proof of Theorem~\ref{theo-pro}}}. 
First, we define
\begin{align*}
	& L_n (\bA, \bB) = 2^{-1}\|\bY - \bA - \bX \bB \|_F^2, \ P_n (\bA) = \lambda \sum_{1 \leq i < j \leq n} \rho (\| \ba_i - \ba_j \|_2),\\
	& L_n^{\psi} (\bB, \bC) = 2^{-1} \|\bY - \bW \bC - \bX\bB \|_F^2, \ P_n^{\psi} (\bC) = \lambda \sum_{1 \leq k < k^{'} \leq K} |\psi_k|\cdot |\psi_{k^{'}}| \cdot\rho (\| \bc_k - \bc_{k^{'}} \|_2).
\end{align*}
Then the corresponding objective functions can be written as
\begin{align*}
	&\bQ_n (\bA, \bB) = L_n (\bA,\bB) + P_n (\bA) \,\,\mbox{subject to}\,\, \mbox{rank}(\bB) \leq r, \\
	&\bQ_n^{\psi} (\bC, \bB) = L_n^{\psi} (\bC,\bB) + P_n^{\psi} (\bC)\,\,\mbox{subject to}\,\,   \mbox{rank}(\bB) \leq r.
\end{align*}

Let $T: \mathcal{M}_{\psi} \rightarrow \mathbb{R}^{K \times q}$ be the mapping such that $T(\bA)$ is a $K \times q$ matrix whose $k$th row  equals to the common vector of $\ba_i$ for $i \in \psi_k$. Let $T^* : \mathbb{R}^{n  \times q} \rightarrow \mathbb{R}^{K \times q}$ be the mapping such that $T^*(\bA) = \{|\psi_k|^{-1} \sum_{i \in \psi_k} \ba_i, k = 1, \dots, K\} ^{\top}$. Obviously, when $\bA \in \mathcal{M}_{\psi} $, we have $T(\bA) = T^*(\bA)$ and $P_n (\bA) = P_n^{\psi} (T(\bA))$. For any $\bC \in \mathbb{R}^{K \times q}$, we also have $P_n (T^{-1}(\bC)) = P_n^{\psi} (\bC)$. Therefore, we have
\begin{align}\label{Q relationship}
	Q_n (\bA, \bB) = Q_n^{\psi} (T(\bA), \bB)\quad\mbox{and}\quad Q_n^{\psi}(\bC, \bB) = Q_n (T^{-1}(\bC), \bB).
\end{align}
Consider the neighborhood of the true coefficient matrices $(\bA^* , \bB^* )$
\begin{align*}
	\Theta = \left\{(\bA^{\top}, \bB^{\top})^{\top}: \bA \in \mathbb{R}^{n \times q}, \,\bB \in \mathbb{R}^{p \times q}, r(\bB) \leq r, \,\|\bB - \bB^* \|_F \leq \phi_{n}\,\,\mbox{and}\,\, \sup_{i} \|\ba_i - \ba^*_i \|_2 \leq \phi_n \right\}.
\end{align*}
It follows from Theorem \ref{theo1} that there exists an event $E_1$ with $\mathrm{Pr}(E_1)\leq \tilde{c}_1\exp\{-c\mathrm{ln} (n)\}$ such that, conditional on $E_1$, we have
\begin{align*}
	\|\widehat{\bB}^{or} - \bB^* \|_F \leq \phi_{n}\,\,\,\mbox{and}\,\,\, \sup_{i} \|\widehat{\ba}_i^{or} - \ba^*_i \|_2 \leq \phi_n.
\end{align*}
This indicates that $(\widehat{\bA}^{or}, \widehat{\bB}^{or}) \in \Theta$ on the event $E_1$.
For any $\bA \in \mathbb{R}^{n \times q}$, let $\bA^0 = T^{-1} (T^* (\bA))$. Next, we will show that $(\widehat{\bA}^{or}, \widehat{\bB}^{or})$ is a strictly local minimizer of the objective function \eqref{initial goal} with probability approaching one through the following two steps.
\begin{description}
	\item[Step 1.] On the event $E_1$, for any $((\bA^0)^{\top}, \bB^{\top})^{\top} \in \Theta$ and $ ((\bA^0)^{\top}, (\bB)^{\top})^{\top} \neq  ((\widehat{\bA}^{or})^{\top},$ $  (\widehat{\bB}^{or})^{\top})^{\top}$, we have $Q_n (\bA^0, \bB) > Q_n (\widehat{\bA}^{or}, \widehat{\bB}^{or})$.
	\item[Step 2.] There exists an event $E_2$ such that $\mathrm{Pr}(E_2^C) \leq 2 qn^{-2}$, where $E_2^C$ is the complement of event $E_2$. On $E_1 \cap E_2$, there is a neighborhood of $((\widehat{\bA}^{or})^{\top}, (\widehat{\bB}^{or})^{\top})^{\top}$, denoted by $\Theta_n$, such that $Q_n(\bA, \bB ) \geq Q_n (\bA^0, \bB)$ for any $(\bA^{\top}, \bB^{\top})^{\top} \in \Theta \cap \Theta_n$ for sufficiently large $n$.
\end{description}
To save space, the proofs of the results in these two steps are provided in the supplementary material. Hence, by the results in these two steps, we have $Q_n(\bA, \bB ) > Q_n(\widehat{\bA}^{or}, \widehat{\bB}^{or})$ for any $(\bA^{\top}, \bB^{\top})^{\top}$ $\in \Theta \cap \Theta_n$ and $ (\bA^{\top}, \bB^{\top})^{\top} \neq  ((\widehat{\bC}^{or})^{\top}, (\widehat{\bB}^{or})^{\top})^{\top}$ in $E_{1} \cap E_2$, so that $((\widehat{\bC}^{or})^{\top},(\widehat{\bB}^{or})^{\top})^{\top}$ is a local minimizer of \eqref{initial goal} over the event $E_1 \cap E_2$ with $\mathrm{Pr}(E_1 \cap E_2) \geq 1 - \tilde{c}_1\exp\{-c \mathrm{ln} (n)\} - 2 q n^{-2}$ for sufficiently large $n$. This completes the proof of Theorem~\ref{theo-pro}.

\hfill $\square$

\bigskip


\noindent{{\bf Proof of Theorem~\ref{theo3}}}. Before presenting the detailed proofs, we recall some notation to facilitate calculations.
Specifically, $P(\mathbf{B}, \mathbf{C}) = A_1[(p + q - r)(r+k) + kq] + A_2\mathrm{ln} (n)$ and $\delta (\mathbf{B}, \mathbf{C}) = AP(\mathbf{B}, \mathbf{C}) /(nq)$ where $A_1$ and $A_2$ are positive constants, $A$ is a positive constant satisfying $A < A_3$ such that $\delta (\mathbf{B}^*, \mathbf{C}^*) < 1$. Note that $P(\mathbf{B}, \mathbf{C}) = A_1[(p + q - r)(r+k) + kq] + A_2\mathrm{ln} (n)$ has the same order as $(p + q - r)(r+k) + kq + \mathrm{ln} (n)$.

Let $h(\mathbf{B}, \mathbf{C}; A) = 1/ [nq - AP(\mathbf{B}, \mathbf{C})]$. 
For any $(\widehat{\mathbf{B}}, \widehat{\mathbf{C}})$ that minimizes $ \mathrm{ln} \left( \| \bY - \bX \bB -\mathbf{W}\bC\|_F^2\right) + \delta(\mathbf{B}, \mathbf{C})$ subject to $\delta(\mathbf{B}, \mathbf{C}) < 1$, 
by the fact that 
$1/ (1- \delta/2) \leq \exp( \delta)$ for any $0 \leq \delta < 1$, we have  
\begin{align*}
	(nq) \| \mathbf{Y} - & \mathbf{X} \widehat{\mathbf{B}} - \widehat{\mathbf{W}}  \widehat{\mathbf{C}}\|_F^2\cdot  h(\widehat{\mathbf{B}}, \widehat{\mathbf{C}}; A/2)
	\leq \| \mathbf{Y} - \mathbf{X} \widehat{\mathbf{B}} - \widehat{\mathbf{W}} \widehat{\mathbf{C}}\|_F^2 \exp\{\delta (\widehat{\mathbf{B}}, \widehat{\mathbf{C}})\} \\
	\leq & \| \mathbf{Y} - \mathbf{X} \mathbf{B}^* -  \mathbf{W} \mathbf{C}^*\|_F^2 \exp \{ \delta (\mathbf{B}^*, \mathbf{C}^*) \}
	\leq \| \mathbf{Y} - \mathbf{X} \mathbf{B}^* - \mathbf{W} \mathbf{C}^*\|_F^2\cdot h(\mathbf{B}^*, \mathbf{C}^*; A) \cdot (nq),
\end{align*}
where the last inequality follows from the fact that  $\exp(\delta) \leq 1/(1-\delta)$ for any $0 \leq \delta < 1$. 
Since $h(\widehat{\mathbf{B}}, \widehat{\mathbf{C}}; A/2) > 0$, we have
\begin{align*}
	\| \mathbf{Y} - \mathbf{X} \widehat{\mathbf{B}} -  \widehat{\mathbf{W}}  \widehat{\mathbf{C}}\|_F^2 \leq \| \mathbf{Y} - \mathbf{X} \mathbf{B}^* -  \mathbf{W} \mathbf{C}^*\|_F^2 h(\mathbf{B}^*, \mathbf{C}^*; A) / h(\widehat{\mathbf{B}}, \widehat{\mathbf{C}}; A/2).
\end{align*}
Recall that $\bY = \bX \bB^* + \bW \bC^* + \bE$. After some algebra, we obtain
\begin{align}\label{eq: PE}
	& \| \mathbf{X} (\widehat{\mathbf{B}}-  \mathbf{B}^*) +  \widehat{\mathbf{W}} \widehat{\mathbf{C}} -  \mathbf{W} \mathbf{C}^* \|_F^2 \nonumber \\
	\leq & \| \mathbf{E}\|_F^2 \left[h(\mathbf{B}^*, \mathbf{C}^*; A) / h(\widehat{\mathbf{B}}, \widehat{\mathbf{C}}; A/2) - 1 \right] + 2 \langle  \mathbf{E},\, \mathbf{X} (\widehat{\mathbf{B}}-\mathbf{B}^*) +  \widehat{\mathbf{W}}  \widehat{\mathbf{C}} -  \mathbf{W} \mathbf{C}^* \rangle \nonumber\\
	\leq & \frac{ A \| \mathbf{E}\|_F^2}{ nq \sigma^2  - A \sigma^2 P(\mathbf{B}^*, \mathbf{C}^*)} \sigma^2 P(\mathbf{B}^*, \mathbf{C}^*) - \frac{0.5A \| \mathbf{E}\|_F^2}{ nq \sigma^2} \sigma^2 P(\widehat{\mathbf{B}}, \widehat{\mathbf{C}})   \nonumber\\
	&\quad\quad + 2 \langle  \mathbf{E}, \mathbf{X} (\widehat{\mathbf{B}}-\mathbf{B}^*) +  \widehat{\mathbf{W}} \widehat{\mathbf{C}} -   \mathbf{W}  \mathbf{C}^* \rangle.
\end{align}

Next, we first show that the term $2\langle  \mathbf{E}, \mathbf{X} (\widehat{\mathbf{B}}-\mathbf{B}^*) + \widehat{\mathbf{W}} \widehat{\mathbf{C}} -  \mathbf{W} \mathbf{C}^* \rangle$ can be bounded by $P(\mathbf{B}^*, \mathbf{C}^*)$ and $P(\widehat{\mathbf{B}}, \widehat{\mathbf{C}})$ up to a multiplicative constant with high probability using 
arguments similar to those in the proof of Lemma \ref{lem 2}. With a slight abuse of notation, let $\mathbf{\Delta}^{\widehat{B}} =  \widehat{\mathbf{B}} -\mathbf{B}^*$, $\mathbf{\Delta}^{WC} =  \widehat{\mathbf{W}} \widehat{\mathbf{C}} - \mathbf{W} \mathbf{C}^*$, $\mbox{rank}(\mathbf{B}^*) = r^*$, and $\mbox{rank}(\widehat{\mathbf{B}}) = \widehat{r}$. Let $RS( \mathbf{W} \mathbf{C}^*)$ be the row space of $ \mathbf{W} \mathbf{C}^*$, $\mathbf{P}_{RS(\mathbf{C}^*)}$ be the orthogonal projection onto the row space of $ \mathbf{W} \mathbf{C}^*$, and $\mathbf{P}_{RS(\mathbf{C}^*)}^{\bot}$ be the orthogonal complement of $\mathbf{P}_{RS(\mathbf{C}^*)}$. Similarly, let $\mathbf{P}_{RS(\mathbf{B}^*, \mathbf{C}^*)}$ be the orthogonal projection onto the row space of $ \mathbf{X} \mathbf{B}^* +  \mathbf{W} \mathbf{C}^*$ and $\mathbf{P}_{RS(\mathbf{B}^*, \mathbf{C}^*)}^{\bot}$ be the orthogonal complement of $\mathbf{P}_{RS(\mathbf{B}^*, \mathbf{C}^*)}$. Then
\begin{align*}
	& \mathbf{X} (\widehat{\mathbf{B}}- \mathbf{B}^*) +  \widehat{\mathbf{W}} \widehat{\mathbf{C}} -   \mathbf{W}  \mathbf{C}^* =\mathbf{X} \mathbf{\Delta}^{\widehat{B}} +  \mathbf{\Delta}^{WC} \\
	= & \mathbf{P}_{\mathbf{X}} (\mathbf{X} \mathbf{\Delta}^{\widehat{B}} + \mathbf{\Delta}^{WC}) \mathbf{P}_{RS(\mathbf{B}^*, \mathbf{C}^*)} 
	+  \mathbf{P}_{\mathbf{X}} (\mathbf{X} \widehat{\mathbf{B}} +  \widehat{\mathbf{W}} \widehat{\mathbf{C}}) \mathbf{P}^{\bot}_{RS(\mathbf{B}^*, \mathbf{C}^*)} + \mathbf{P}_{\mathbf{X}}^{\bot} \mathbf{\Delta}^{WC}\mathbf{P}_{RS(\mathbf{C}^*)} + \mathbf{P}_{\mathbf{X}}^{\perp} \widehat{\mathbf{W}} \widehat{\mathbf{C}} \mathbf{P}^{\bot}_{RS(\mathbf{C}^*)} \\
	=&: \mathbf{\Delta}_1 + \mathbf{\Delta}_2 + \mathbf{\Delta}_3 + \mathbf{\Delta}_4,
\end{align*}
where $\sum_{i = 1}^4 \|\mathbf{\Delta}_i\|_F^2 = \| \mathbf{X} \mathbf{\Delta}^{\widehat{B}} +  \mathbf{\Delta}^{WC} \|_F^2$. Thus, 
\begin{align}\label{eq: 4-Delta}
	\langle  \mathbf{E}, \,\mathbf{X} (\widehat{\mathbf{B}}-\mathbf{B}^*) +  \widehat{\mathbf{W}} \widehat{\mathbf{C}} -   \mathbf{W}  \mathbf{C}^* \rangle
	= \langle  \mathbf{E}, \,\mathbf{\Delta}_1\rangle + \langle  \mathbf{E}, \,\mathbf{\Delta}_2\rangle + \langle  \mathbf{E}, \,\mathbf{\Delta}_3\rangle + \langle  \mathbf{E}, \,\mathbf{\Delta}_4\rangle.
\end{align}
Note that $CS( \mathbf{\Delta}_1) \subset CS(\mathbf{X})$, $CS( \mathbf{\Delta}_2) \subset CS(\mathbf{X})$, $r(\mathbf{\Delta}_1) \leq r(\mathbf{C}^*) + r(\mathbf{B}^*)$, $r(\mathbf{\Delta}_2) \leq r(\widehat{ \mathbf{B}}) + r(\widehat{\mathbf{C}})$,  $r(\mathbf{\Delta}_3) \leq r(\mathbf{C}^*)$, and $r(\mathbf{\Delta}_4) \leq r(\widehat{\mathbf{C}})$, where $CS(\cdot)$ is the column space of a given matrix. These four terms on the right-hand side of \eqref{eq: 4-Delta} can be bounded using arguments similar to those in Lemmas \ref{lem1} and 
\ref{lem 2}.
For example, we can use the following result to handle $ \langle \mathbf{E}, \mathbf{\Delta}_2 \rangle$.
\begin{align*}
	\langle \mathbf{E},  \mathbf{\Delta}_2 \rangle - & a^{-1}\|\mathbf{\Delta}_2 \|_F^2 - bLP_3 \{r(\widehat{\mathbf{B}}, \widehat{\mathbf{C}}), r(\mathbf{X}), r(\widehat{\mathbf{F}})\} \\
	\leq & \| \mathbf{\Delta}_2 \|_F \langle \mathbf{E},  \mathbf{\Delta}_2/ \|\mathbf{\Delta}_2 \|_F \rangle - 2(b/a)^{1/2} \| \mathbf{\Delta}_2 \|_F \{L P_3 ( r(\widehat{\mathbf{B}}, \widehat{\mathbf{C}}), r(\mathbf{X}), r(\widehat{\mathbf{F}})) \}^{1/2} \\
	= & 2(\|\mathbf{\Delta}\|_F/\sqrt{a_2})\left\{(\sqrt{a_2}/2)\left[\langle \mathbf{E},  \mathbf{\Delta}_2/ \|\mathbf{\Delta}_2 \|_F \rangle - 2\{bL P_3 ( r(\widehat{\mathbf{B}}, \widehat{\mathbf{C}}), r(\mathbf{X}), r(\widehat{\mathbf{F}}))/a \}^{1/2}\right]\right\}\\
	\leq & a_2^{-1} \| \mathbf{\Delta}_2 \|_F^2 + (a_2/4)\left[\langle \mathbf{E},  \mathbf{\Delta}_2/\|\mathbf{\Delta}_2 \|_F \rangle  - 2(b/a)^{1/2} \{L P_3 ( r(\widehat{\mathbf{B}}, \widehat{\mathbf{C}}), r(\mathbf{X}), r(\widehat{\mathbf{F}})) \}^{1/2}\right]^2,
\end{align*}
where the first inequality follows from the arithmetic mean-geometric mean inequality, the second inequality holds from the fact that $2xy\leq x^2+y^2$ for two real numbers $x$ and $y$, and $a, b, L, a_2$ are positive constants. 
Using arguments similar to those in Lemmas \ref{lem1} and 
\ref{lem 2}, with $4b > a$, we can bound $\langle \mathbf{E},  \mathbf{\Delta}_2 \rangle$ such that for any $t>0$, the event $\langle \mathbf{E},  \mathbf{\Delta}_2 \rangle - a^{-1}\|\mathbf{\Delta}_2 \|_F^2 - bLP_3 \{r(\widehat{\mathbf{B}}, \widehat{\mathbf{C}}), r(\mathbf{X}), r(\widehat{\mathbf{F}})\} \leq \| \mathbf{\Delta}_2 \|_F^2 / a_2 + a_2 t^2 \sigma^2 /4$ occurs with probability at least $1 - \tilde{c} \exp(-ct^2)$, where $P_3 \{r(\widehat{\mathbf{B}}, \widehat{\mathbf{C}}), r(\mathbf{X}), r(\widehat{\mathbf{F}})\}\leq (p + q - \widehat{r}) ( \widehat{k} + \widehat{r})$. 
Following the lines of the proof of Theorem 2 in \cite{She2017}, we can show that for any constants $a, b, a_2$ satisfying $4b > a$, the following event
\begin{align}\label{eq: PE-pt2}
	& 2 \langle  \mathbf{E}, \,   \mathbf{X} (\widehat{\mathbf{B}}-\mathbf{B}^*) +   \widehat{\mathbf{W}}\widehat{\mathbf{C}} -  \mathbf{W} \mathbf{C}^*  \rangle \nonumber\\
	\leq & 2 (a^{-1} + a_2^{-1}) 
	\| \mathbf{X} (\widehat{\mathbf{B}}-\mathbf{B}^*) + \widehat{\mathbf{W}}\widehat{\mathbf{C}} -  \mathbf{W} \mathbf{C}^*  \|_F^2  + 8bL \sigma^2 \{ P(\mathbf{B}^*, \mathbf{C}^*) +  P(\widehat{\mathbf{B}}, \widehat{\mathbf{C}})\}
\end{align}
occurs with probability at least $1 - \tilde{c}_1 n^{-c_1}$ for some $\tilde{c}_1, c_1 > 0$ and a sufficiently large constant $L$. 

Now, we will bound the first two terms on the right-hand side of \eqref{eq: PE}. Let $\gamma$ and $\gamma'$ be constants satisfying $0 < \gamma < 1$ and $\gamma^{'} >0$. 
Denote by $\Omega_2 = \{(1 - \gamma) nq \sigma^2 \leq \|\mathbf{E}\|_F^2 \leq (1 + \gamma^{'}) nq \sigma^2\}$ an event. 
It follows from \cite{Laurent2000} that $\mathrm{Pr}(\Omega_2^c)\leq \tilde{c}_2 \exp (- c_2 nq )$, where $\tilde{c}_2$ and $c_2$ are constants depending on $\gamma$ and $\gamma'$, implying that the event $\Omega_2$ occurs with probability at least $1-\tilde{c}_2 \exp (- c_2 nq )$. On the event $\Omega_2$, we have 
\begin{align}\label{eq: PE-pt1}
	&\frac{ A \| \mathbf{E}\|_F^2}{ nq \sigma^2  - A \sigma^2 P(\mathbf{B}^*, \mathbf{C}^*)} \sigma^2 P(\mathbf{B}^*, \mathbf{C}^*) - \frac{0.5A \| \mathbf{E}\|_F^2}{ nq \sigma^2} \sigma^2 P(\widehat{\mathbf{B}}, \widehat{\mathbf{C}}) \nonumber\\
	\leq & \frac{ (1 + \gamma')A A_3}{ A_3 - A} \sigma^2 P(\mathbf{B}^*, \mathbf{C}^*) - 0.5(1 - \gamma)A \sigma^2 P(\widehat{\mathbf{B}}, \widehat{\mathbf{C}}).
\end{align}
In view of \eqref{eq: PE}, \eqref{eq: PE-pt2}, and \eqref{eq: PE-pt1}, applying the union bound gives
\begin{align*}
	& \| \mathbf{X} (\widehat{\mathbf{B}}-\mathbf{B}^*) +  \widehat{\mathbf{W}} \widehat{\mathbf{C}} - \mathbf{W} \mathbf{C}^* \|_F^2 \nonumber\\
	\leq & 2 (a^{-1} + a_2^{-1}) \| \mathbf{X} (\widehat{\mathbf{B}}-\mathbf{B}^*) + \widehat{\mathbf{W}} \widehat{\mathbf{C}} -  \mathbf{W}  \mathbf{C}^* \|_F^2 \\
	& \quad + \left[\frac{ (1 + \gamma')A A_3}{ A_3 - A} + 8bL\right] \sigma^2 P(\mathbf{B}^*, \mathbf{C}^*) + \left[8bL - 0.5(1 - \gamma)A\right] \sigma^2 P(\widehat{\mathbf{B}}, \widehat{\mathbf{C}}).
\end{align*}
With $A_3$ large enough, we can choose $a, a_2, b, A$ such that $(1/a + 1/a_2) < 1/2$, $4b > a$, and $16bL \leq (1- \gamma)A$. 
This completes the proof of Theorem~\ref{theo3}.

\hfill $\square$


\section*{Appendix B: Some lemmas and their Proofs}

\setcounter{equation}{0}
\renewcommand{\theequation}{B.\arabic{equation}} %

\begin{lemma} \label{lem1}
	For any given $1 \leq r \leq p \wedge q$, define $\Gamma_{r,\, K} = \{ (\bB, \bC) \in \mathbb{R}^{p \times q} \times \mathbb{R}^{K \times q}: r(\bB) \leq r\}$. Then under Condition \ref{con2}, 
	there exist universal constants $L_0>0$, $a>0$, and $\tilde{c}, c> 0$ such that the following event
	\begin{align}
		\sup_{(\mathbf{B}, \mathbf{C}) \in \Gamma_{r,\,K}}  & \left\{ 2 \langle \bE, \bX \bB +  \bW \bC \rangle- a^{-1} \| \bX \bB +  \bW \bC \|_F^2 
		-  a L_0 \sigma^2 \left[(p+q-r)(r + K) + qK\right] \right\} \geq 4a \sigma^2 t  \label{or1}
	\end{align}
	occurs with probability at most $\tilde{c}\exp(-c t)$.
\end{lemma}


\noindent{{\bf Proof of Lemma~\ref{lem1}}}. 
Let $\mathbf{M}^{BC}  = ( \bB^{\top}, \mathbf{C}^{\top})^{\top}\in \mathbb{R}^{(p+K) \times q}$ and $\mathbf{F}  =  (\mathbf{X},  \mathbf{W} )\in \mathbb{R}^{n \times (p + K)}$. Then $\mathbf{F} \mathbf{M}^{BC}=\bX\bB +  \mathbf{W} \mathbf{C} $ and the term $2\langle \bE, \bX \bB + \bW \bC \rangle$ can be rewritten as
\begin{align} \label{epsilon norm} 
	2\langle \bE, \bX \bB + \bW \bC \rangle & = 2\langle \bE, \mathbf{P}_{\mathbf{X}} (\bX \bB +  \mathbf{W} \mathbf{C} ) \rangle + 2 \langle \bE, \mathbf{P}_{\mathbf{X}}^{\bot} \bW \bC \rangle 
	\equiv 2\langle \bE, \bmm_1 \rangle + 2\langle \bE, \bmm_2 \rangle,
\end{align}
where $\bmm_1=\mathbf{P}_{\mathbf{X}} \mathbf{F} \mathbf{M}^{BC}$ and $\bmm_2=\mathbf{P}_{\mathbf{X}}^{\bot} \bW \bC$.  By the Pythagorean theorem, we have 
\begin{align}\label{eq: m1+m2}
	\| \bmm_1\|_F^2 + \|\bmm_2 \|_F^2 = \|\bX \bB +  \bW \bC \|_F^2.
\end{align}

Define 
$\mathscr{A}_1 = \left\{ \sup_{(\mathbf{B}, \mathbf{C}) \in \Gamma_{r,\, K}}\left\{2 \langle \bE, \bmm_1\rangle - a^{-1} \| \bmm_1\|_F^2 -2a L (p+q-r)(r + K)\right\} \geq  2 a \sigma^2 t\right\}$ and $\mathscr{A}_2 = \left\{ \sup_{(\mathbf{B}, \mathbf{C}) \in \Gamma_{r,\, K}}\left\{2 \langle \bE, \bmm_2\rangle - a^{-1} \| \bmm_2\|_F^2 -2a L qK\right\}\geq  2 a  \sigma^2 t \right\}$,
where $L>0$ and $a>0$ are two constants. The key step of our proof is to show that 
\begin{align}
	 \mathrm{Pr}(\mathscr{A}_1) = & \mathrm{Pr} \left( \sup_{(\mathbf{B}, \mathbf{C}) \in \Gamma_{r,\, K}} \left\{2 \langle \bE, \bmm_1\rangle - a^{-1} \| \bmm_1\|_F^2 -2a L (p+q-r)(r + K)\right\} \geq  2 a \sigma^2 t \right) \nonumber\\
	\leq & \tilde{c}_1 \exp(-ct), \label{m1-bound}\\
	\mathrm{Pr}(\mathscr{A}_2) =&\mathrm{Pr} \left( \sup_{(\mathbf{B}, \mathbf{C}) \in \Gamma_{r,\, K}} \left\{2 \langle \bE, \bmm_2\rangle - a^{-1} \| \bmm_2\|_F^2 -2a L qK\right\}\geq  2 a  \sigma^2 t \right) \leq \tilde{c}_1 \exp(-ct).  \label{m2-bound}
\end{align}
Once these two probability bounds are obtained, it follows from \eqref{epsilon norm}, \eqref{eq: m1+m2}, and the union bound that 
\begin{align*}
	\mathrm{Pr} & \left( \sup_{(\mathbf{B}, \mathbf{C}) \in \Gamma_{r,\, K}} \left\{2 \langle \bE, \bX \bB + \bC \rangle  - a^{-1}\| \bX \bB + \bC\|_F^2  
	-  2aL\sigma^2\{ (p+q-r)(r + K) + qK \right\} \geq 4a \sigma^2 t \right) \\
	& \leq \mathrm{Pr}(\mathscr{A}_1) + \mathrm{Pr}(\mathscr{A}_2)
	\leq  2\tilde{c}_1 \exp(-c t).
\end{align*}
Then the desired result \eqref{or1} follows immediately by taking $L_0=2L$ and $\tilde{c}=2\tilde{c}_1$.

In what follows, we will show \eqref{m1-bound} and \eqref{m2-bound} hold. Note that $r(\mathbf{B}) \leq r \leq q, r(\mathbf{C}) \leq K \land q$, and $r ( \mathbf{M}^{BC}) \leq r(\mathbf{B}) + r(\mathbf{C}) \leq r + (K \land q ) $. Let
$P_1 \{ \tilde{r}, r(\mathbf{X}),r(\mathbf{F})\} = \sigma^2 \{  r(\mathbf{X})\land r(\mathbf{F}) + q - \tilde{r})\}\tilde{r}$ with $\tilde{r}=r+(K \land q )$. Then, we have
\begin{align*}
	2 \langle \bE, \bmm_1\rangle - & a^{-1}\| \bmm_1\|_F^2 -2a L P_1  \{ \tilde{r}, r(\mathbf{X}),r(\mathbf{F})\} \\
	= & 2 \langle \bE, \bmm_1 / \|\bmm_1 \|_F \rangle \| \bmm_1 \|_F - (2a)^{-1}\| \bmm_1\|_F^2 -\left[(2a)^{-1}\| \bmm_1\|_F^2 + 2a L P_1  \{ \tilde{r}, r(\mathbf{X}),r(\mathbf{F})\}\right] \\
	\leq & 2 \langle \bE, \bmm_1 / \|\bmm_1 \|_F \rangle \| \bmm_1 \|_F - (2a)^{-1} \| \bmm_1\|_F^2 -2 \| \bmm_1\|_F  L^{1/2} P_1^{1/2}  \{ \tilde{r}, r(\mathbf{X}),r(\mathbf{F})\}  \\
	= &  2\left(\| \bmm_1\|_F/\sqrt{2a}\,\right)\left[\sqrt{2a}\left(\langle \bE, \bmm_1 / \|\bmm_1 \|_F \rangle-L^{1/2} P_1^{1/2}  \{ \tilde{r},r(\mathbf{X}),r(\mathbf{F})\}\right)\right]- (2a)^{-1} \| \bmm_1\|_F^2\\
	\leq & (2a)^{-1} \| \bmm_1\|_F^2 +2a \left[\langle \bE, \bmm_1 / \|\bmm_1 \|_F \rangle -  L^{1/2} P_1^{1/2}  \{ \tilde{r}, r(\mathbf{X}),r(\mathbf{F}) \} \right]^2 - (2a)^{-1} \| \bmm_1\|_F^2 \\
	= & 2a \left[\langle \bE, \bmm_1 / \|\bmm_1 \|_F \rangle -  L^{1/2} P_1^{1/2} \{ \tilde{r}, r(\mathbf{X}),r(\mathbf{F})\} \right]^2,
\end{align*}
where the first inequality follows from the arithmetic mean-geometric mean inequality and the second inequality holds from the fact that $2xy\leq x^2+y^2$ for two real numbers $x$ and $y$.  This implies that the inequality $\sup_{(\mathbf{B}, \mathbf{C}) \in \Gamma_{r,\, K}}\left\{\langle \bE, \bmm_1 / \|\bmm_1 \|_F \rangle - L^{1/2} P_1^{1/2} \{ \tilde{r}, r(\mathbf{X}),r(\mathbf{F})\}\right\}\geq \sigma t^{1/2}$ holds for any $t>0$ conditional on the event $\{\sup_{(\mathbf{B}, \mathbf{C}) \in \Gamma_{r,\, K}}$ $\{2 \langle \bE, \bmm_1\rangle - a^{-1}\| \bmm_1\|_F^2 -2a L P_1  \{ \tilde{r}, r(\mathbf{X}),r(\mathbf{F})\}\}\geq 2 a \sigma^2 t\}$. Therefore, we have
\begin{align}\label{eq: m1-P1}
	\mathrm{Pr} &  \left( \sup_{(\mathbf{B}, \mathbf{C}) \in \Gamma_{r,\, K}} \left\{2 \langle \bE, \bmm_1\rangle - a^{-1} \| \bmm_1\|_F^2 -2a L P_1 \{ \tilde{r}, r(\mathbf{X}), r(\mathbf{F} ) \}\right\} \geq  2 a \sigma^2 t \right) \nonumber\\
	& \leq \mathrm{Pr} \left( \sup_{(\mathbf{B}, \mathbf{C}) \in \Gamma_{r,\, K}}\left\{\langle \bE, \bmm_1 / \|\bmm_1 \|_F \rangle - L^{1/2} P_1^{1/2} \{ \tilde{r}, r(\mathbf{X}),r(\mathbf{F})\}\right\}\geq \sigma t^{1/2} \right) 
	\leq  \tilde{c}_1 \exp(-ct),
\end{align}
where the last inequality follows from Lemma~\ref{lem 2}. 
Recall that $P_1 \{ \tilde{r}, r(\mathbf{X}),r(\mathbf{F})\} = \sigma^2 \{  r(\mathbf{X})\land r(\mathbf{F}) + q - \tilde{r})\}\tilde{r}$ with $\tilde{r}=r+(K \land q )$. 
Since $r(\bX) \leq p$ and $r(\mathbf{F}) \leq n \land (p + K)$, we have 
$P_1 \{ \tilde{r}, r(\mathbf{X}),r(\mathbf{F})\} \leq \sigma^2 [  p + q - (K+r)](K+r)$ when $K<q $ and $P_1 \{ \tilde{r}, r(\mathbf{X}),r(\mathbf{F})\} \leq  \sigma^2 (  p -r)(q+r)$ when $K\geq q $. Thus, $P_1 \{ \tilde{r}, r(\mathbf{X}),r(\mathbf{F})\} $ can be bounded by $(p + q - r)( r + K)$, that is,
\begin{align}\label{eq: P1-bound}
	P_1 \{ \tilde{r}, r(\mathbf{X}),r(\mathbf{F})\} \leq (p + q - r)( r + K).
\end{align}

In view of \eqref{eq: m1-P1} and \eqref{eq: P1-bound}, we have
\begin{align*}
	\mathrm{Pr} & \left( \sup_{(\mathbf{B}, \mathbf{C}) \in \Gamma_{r,\, K}} \left\{2 \langle \bE, \bmm_1\rangle - a^{-1} \| \bmm_1\|_F^2 -2a L (p + q - r)( r + K)\right\} \geq  2 a \sigma^2 t\right) \\
	& \leq \mathrm{Pr} \left( \sup_{(\mathbf{B}, \mathbf{C}) \in \Gamma_{r,\, K}} \left\{2 \langle \bE, \bmm_1\rangle - a^{-1} \| \bmm_1\|_F^2 -2a L P_1 \{ \tilde{r}, r(\mathbf{X}), r(\mathbf{F} ) \}\right\} \geq  2 a \sigma^2 t\right) 
	\leq  \tilde{c}_1 \exp(-ct).
\end{align*}
This shows that the probability bound \eqref{m1-bound} holds.

It thus remains to prove \eqref{m2-bound}. Using similar arguments as those used to prove~\eqref{eq: m1-P1} and Lemma~\ref{lem 2}, we can show that 
\begin{align}\label{eq: m2-P2}
	&\mathrm{Pr} \left( \sup_{(\mathbf{B}, \mathbf{C}) \in \Gamma_{r,\, K}} \left\{2 \langle \bE, \bmm_2\rangle - a^{-1} \| \bmm_2\|_F^2 -2a L P_2 \{r(\mathbf{C}), r(\mathbf{X}), r( \bW )\} \right\}\geq 2a \sigma^2 t\right) \nonumber\\
	\leq & \tilde{c}_1 \exp(-ct),
\end{align}
where $P_2 \{r(\mathbf{C}), r(\mathbf{X}), r(\bW )\} = \sigma^2 [ \{n -r(\mathbf{X})\} \land r(\bW) + q - r(\mathbf{C})]r(\mathbf{C}) $.
Under the condition $p+K\leq n$, we have 
$r (\bW) \leq n \land K =K$ and thus $P_2 \{r(\mathbf{C}), r(\mathbf{X}), r(\bW )\} \leq  \sigma^2 [ K + q - r(\mathbf{C})]r(\mathbf{C})$.   
This, together with the fact that $r(\bC) \leq K \land q$, yields $P_2 \{r(\mathbf{C}), r(\mathbf{X}), r(\bW )\} \leq  \sigma^2 qK$.  Combining this with~\eqref{eq: m2-P2} leads to
\begin{align*}
	& \mathrm{Pr} \left( \sup_{(\mathbf{B}, \mathbf{C}) \in \Gamma_{r,\, K}} \left\{2 \langle \bE, \bmm_2\rangle - a^{-1} \| \bmm_2\|_F^2 -2a L \sigma^2qK \right\}\geq 2a \sigma^2 t\right) \\
	\leq & \mathrm{Pr} \left( \sup_{(\mathbf{B}, \mathbf{C}) \in \Gamma_{r,\, K}} \left\{2 \langle \bE, \bmm_2\rangle - a^{-1} \| \bmm_2\|_F^2 -2a L P_2 \{r(\mathbf{C}), r(\mathbf{X}), r( \bW )\} \right\}\geq 2a \sigma^2 t\right) \\
	\leq & \tilde{c}_1 \exp(-ct)
\end{align*}
This shows that the probability bound \eqref{m2-bound} holds and completes the proof of Lemma~\ref{lem1}.

\hfill $\square$

\begin{lemma}\label{lem 2}
	Suppose each row in $\bE$ is sub-Gaussian with mean zero and $\psi_2$-norm bounded by $\sigma$. Given $\mathbf{X} \in \mathbb{R}^{n \times p}$, $\mathbf{W}\in \mathbb{R}^{n \times K}$, and $\mathbf{F} =  (\mathbf{X}, \mathbf{W}) \in \mathbb{R}^{n \times (p+K)}$, define $\widetilde{\Gamma}_{r} = \{ \bM \in \mathbb{R}^{n \times q }: \|\bM \|_F \leq 1, r(\bM) \leq r, CS(\bM) \subset CS(\mathbf{P}_{\mathbf{X}}, \mathbf{F})\}$ with $1 \leq r \leq p \wedge q$ and $ p + K \leq n$. Let
	\begin{align*}
		P_1 \{r, r(\mathbf{X}),r(\mathbf{F})\} = \sigma^2 \{r(\mathbf{X})\land r(\mathbf{F}) + q -r\}r.
	\end{align*}
	Then for any $t \geq 0$, we have
	\begin{align*}
		\mathrm{Pr} \left(\sup_{ \mathbf{M} \in \widetilde{\Gamma}_{r}} \langle \bE, \bM \rangle \geq t\sigma + L P_1^{1/2} \{r, r(\mathbf{X}),r(\mathbf{F})\} \right) \leq \tilde{c} \exp(-ct^2),
	\end{align*}
	where $L, \tilde{c}, c > 0$ are universal constants.
\end{lemma}

\noindent{{\bf Proof of Lemma~\ref{lem 2}}}. 
Recall that all rows of the matrix $\bE$ are independent.
Since each row in $\bE$ is sub-Gaussian with mean zero and scale bounded by $\sigma$, it is easy to show that $\langle \bE, \bM \rangle$ is a mean-centered sub-Gaussian random variable with scale bounded by $\sigma \| \bM \|_F$ for any fixed matrix $\bM$.
Therefore, $\{ \langle \bE, \bM \rangle : \bM \in \widetilde{\Gamma}_r\}$ is a stochastic process with sub-Gaussian increments. The induced metric on $\widetilde{\Gamma}_r$, given by $d (\bM_1, \bM_2) = \sigma \|\bM_1 -\bM_2 \|_F$, is Euclidean. 

We can bound 
$\sup_{ \mathbf{M} \in \widetilde{\Gamma}_{r}} \langle \bE, \bM \rangle$
by using Dudley's entropy integral~\citep{Talagrand2014}.  To this end, we first 
compute the metric entropy
$\mathrm{ln} [\mathcal{N} (\bE, \widetilde{\Gamma}_r, d)]$, where $\mathcal{N} (\bE,\widetilde{\Gamma}_r, d)$ is the smallest cardinality of an $\varepsilon$-net that covers $\widetilde{\Gamma}_r$ under $d$. Motivated by \cite{Recht2010}, we characterize each matrix in $\widetilde{\Gamma}_r$ using its row/column spaces. Denote by $\mathbb{O}^{q \times r} = \{\bV \in \mathbb{R}^{q \times r}: \bV^\top \bV = \bI\}$ the set of column-orthogonal matrices of size $q \times r$. Given $\bM \in \widetilde{\Gamma}_r$, its row space must be contained in an $r$-dimensional subspace in $\mathbb{R}^q$.  Hence, we have
\begin{align}\label{USV}
	\bM = \bU\bSig \bV^{\top},
\end{align}
where $\bV \in \mathbb{O}^{q \times r}$, $\bSig \in \mathbb{ R}^{ ((p+K) \land  r(\mathbf{X}) \land  r(\mathbf{F})) \times r}=\mathbb{ R}^{ (  r(\mathbf{X}) \land  r(\mathbf{F})) \times r}$, and $\bP_{\mathbf{U}} = \bP_{(\mathbf{X},\, \mathbf{F})}$. Note that $RS (\bM)$ is a point on the Grassmann manifold (denoted by $\bG_{q,r}$) of all $r$-dimensional subspaces of $\mathbb{R}^q$. Equipped with metric $d^{''}$ which is the operator norm, i.e., $\| \bV_1 \bV_1^{\top} - \bV_2 \bV_2^{\top}\|_2$ for any $\bV_1, \bV_2 \in \mathbb{O}^{q \times r}$, $\mathcal{N}(\varepsilon, \bG_{q,r}, d^{''}) \leq (C_1/ \varepsilon)^{r(q-r)}$, where $C_1$ is a universal constant \citep{Szarek1982}. Moreover, it is easy to see 
that $\bSig = \bU^{\top} \bM \bV$ is in a unit ball of dimensionality $(r(\mathbf{X}) \land  r(\mathbf{F})) \times r$, denote by $\bB_{((r(\mathbf{X})  \land  r(\mathbf{F}))\times r}$. By a standard volume argument, $\mathcal{N} (\varepsilon, \bB_{( r(\mathbf{X})  \land  r(\mathbf{F})) \times r}, \|\cdot\|_F) \leq (C_0 / \varepsilon)^{( r(\mathbf{X}) \land  r(\mathbf{F})) \times r}$, where $C_0$ is a universal constant. Under the metric $d$, we claim that $\mathrm{ln} [\mathcal{N}(\varepsilon, \widetilde{\Gamma}_r, d)] \leq \{ (r(\mathbf{X}) \land  r(\mathbf{F}))r + r(q-r)\} \mathrm{ln} (C\sigma/\varepsilon)$. In fact, for any $\bM_1 \in \widetilde{\Gamma}_r$, we can write $\bM_1 = \bU_1 \bSig_1 \bV_1^{\top}$ according to \eqref{USV} and find $\bV_2$ and $\bSig_2$ such that $\| \bV_1 \bV_1^{\top} - \bV_2 \bV_2^{\top}\|_2 \leq \varepsilon$ and $\| \bSig_1 \bV_1^{\top} \bV_2 - \bSig_2\|_F \leq \varepsilon$. Then, 
we have 
\begin{align*}
	&\| \bM_1 - \bM_2\|_F \leq \| \bM_1 - \bM_1 \bV_2 \bV_2^{\top}\|_F + \| \bU_1 \bSig_1 \bV_1^{\top} \bV_2 \bV_2^{\top} - \bU_1 \bSig_2 \bV_2^{\top} \|_F \\
	\leq & \left[\mathrm{tr} \{\bM_1^{\top} \bM_1 (\mathbf{P}_{\mathbf{V}_1} - \mathbf{P}_{\mathbf{V}_2})^2 \}\right]^{1/2} + \| \bSig_1 \bV_1^{\top} \bV_2 - \bSig_2 \|_F  \leq \left(\| \bM_1\|_F^2 \| \mathbf{P}_{\mathbf{V}_1} - \mathbf{P}_{\mathbf{V}_2} \|_2^2\right)^{1/2} + \varepsilon \leq 2 \varepsilon
\end{align*}
for any $\bM_2 = \bU_1 \bSig_2 \bV_2^{\top}$, where $\mathrm{tr}(\cdot)$ is the trace of a given square matrix. It follows from  Dudley's integral bound that $\mathrm{Pr}\left( \sup_{\mathbf{M}  \in \widetilde{\Gamma}_r } \langle \bM, \bE \rangle  \geq t \sigma + L \int_0^{\sigma} \left[\mathrm{ln}\{\mathcal{N} (\varepsilon, \widetilde{\Gamma}_r, d)\}\right]^{1/2}  \ \textrm{d}\varepsilon \right) \leq \tilde{c} \exp(-c t^2)$.

Simple computation gives
\begin{align*}
	\int_0^{\sigma} \left[\mathrm{ln}\{\mathcal{N} (\varepsilon, \widetilde{\Gamma}_r, d)\}\right]^{1/2}  \ \textrm{d} \varepsilon & \lesssim \sigma \left\{ (r(\mathbf{X}) \land  r(\mathbf{F}))r + (q-r)r \right\}^{1/2} 
	\lesssim P_1 \{r, r(\mathbf{X}),r(\mathbf{F})\}^{1/2}.
\end{align*}
This completes the proof of Lemma \ref{lem 2}.
\hfill $\square$


\newpage
\setcounter{page}{1}
\setcounter{section}{0}
\setcounter{equation}{0}


\begin{center}{\bf \Large Supplementary Material to ``Simultaneous Heterogeneity and Reduced-Rank Learning for
		Multivariate Response Regression"}
	
\bigskip

Jie Wu$^a$, Bo Zhang$^b$, Daoji Li$^c$ and Zemin Zheng$^b$

$^a$School of Big Data and Statistics, Anhui University, Hefei, Anhui 230601, China

$^b$International Institute of Finance, School of Management, University of Science and Technology of China, Hefei, Anhui 230026, China

$^c$College of Business and Economics, California State University, Fullerton, CA, 92831, United States

\end{center}

\bigskip


This supplementary material consists of three parts. The first part  provides the proof of Proposition~\ref{pro1}, the second part contains additional technical details on the proof of Theorem~\ref{theo-pro}, and the third part shows the detailed ADMM algorithm for the heterogeneous multivariate response regression problem discussed in Section~\ref{sec.dicus}.

\section{Proof of Proposition~\ref{pro1}}\label{SuppC}
\renewcommand{\theequation}{C.\arabic{equation}}
\setcounter{equation}{0}

Before presenting the proof of Proposition \ref{pro1}, we state a basic property of the optimization problem \eqref{initial goal}. Recall that the optimization problem \eqref{initial goal} in Section \ref{subsec: method} is equivalent to the following rank-constrained minimization problem
\begin{align*}
	L(\bA, \bB, \bdelta, \bV) 
	=&\frac{1}{2}\sum_{i = 1}^{n} \|\by_i - \bB^{\top} \bx_i- \ba^{\top}_i \|_2^2
	+ \sum_{1\leq i < j \leq n} p_{\gamma} ( \|\ba_i - \ba_j \|_2, \lambda)  
	+ \sum_{1 \leq i < j \leq n} \langle \bv_{ij}, (\ba_i - \ba_j - \bdelta_{ij}) \rangle \\
	&\quad \quad + \frac{\vartheta}{2}\| \ba_i - \ba_j - \bdelta_{ij} \|_2^2 \\
	& \mbox{subject to}\,\, \ \ba_i - \ba_j - \bdelta_{ij} = \mathbf{0}\,\,\mbox{and}\,\, \mbox{rank}(\bB) \leq r.
\end{align*}
To solve this optimization problem, we update the matrices $\bA, \bB$, $\bdelta$, and $\bV$ iteratively. In review of 
\eqref{step1}, 
given any $\lambda \geq 0$ and $r \geq 0$, we have that accumulation point of $ (\bA^{(m+1)}, \bB^{(m+1)})$ satisfies that $$L(\bA^{(m+1)}, \bB^{(m)}, \bdelta^{(m)}, \bV^{(m)}) \leq L(\bA^{(m)}, \bB^{(m)}, \bdelta^{(m)}, \bV^{(m)})$$ and $L(\bA^{(m)}, \bB^{(m+1)}, \bdelta^{(m)}, \bV^{(m)}) \leq L(\bA^{(m)}, \bB^{(m)}, $ $\bdelta^{(m)}, \bV^{(m)})$ for any non-negative integer $m$.  In addition, $L(\bA^{(m+1)},$ $ \bB^{(m+1)}, \bdelta, \bV^{(m)})$ is convex with respect to $\bdelta$. By applying the block coordinate descent results from Theorem 4.1 of \cite{Tseng2001}, we conclude that the sequence $(\bA^{(m+1)}, \bB^{(m+1)},$ $ \bdelta^{(m+1)})$ has a limit point. To simplify the technical presentation, 
we slightly abuse notation by denoting the limit point as $(\bA^*, \bB^*, \bdelta^*)$.

We are now ready to present the proofs. By the definition of $\bdelta^{(m+1)}$ in \eqref{step2}, we have
\begin{eqnarray*}
	L(\bA^{(m+1)},\bB^{(m+1)},\bdelta^{(m+1)},\bV^{(m)})\le	L(\bA^{(m+1)},\bB^{(m+1)},\bdelta,\bV^{(m)})
\end{eqnarray*}
for any $\bdelta$. Define
\begin{align*}
	f^{m+1} & = \inf_{\mathbf{\Delta} \mathbf{A}^{(m+1)} - \mathbf{\bdelta} = 0} \left\{2^{-1}\|\bY - \bX\bB^{(m+1)} - \bA^{(m+1)} \|_F^2 + \sum_{1 \leq i < j \leq n} p_{\gamma} ( \| \bdelta_{ij}\|_2, \lambda )\right\} \\
	& = \inf_{\mathbf{\Delta} \mathbf{A}^{(m+1)} - \mathbf{\bdelta} = 0} L(\bA^{(m+1)},\bB^{(m+1)},\bdelta,\bV^{(m)}),
\end{align*}
where $\mathbf{\Delta} = \{(\mathbf{e}_i - \mathbf{e}_j)^{\top}, i< j \}^{\top}$ is an $[0.5 n(n-1)] \times n$ matrix. Then we have
\begin{align}\label{f(m+1)}
	L(\bA^{(m+1)},\bB^{(m+1)},\bdelta^{(m+1)},\bV^{(m)}) \le f^{m+1}.
\end{align}
Let $t$ be an integer.  It follows from ~\eqref{step3} that $\bV^{(m+1)} = \bV^{(m)} + \vartheta (\mathbf{\Delta} \bA^{(m+1)} - \bdelta^{(m+1)})$, leading to 
\begin{align*}
	\bV^{(m+t-1)} = \bV^{(m)} + \vartheta \sum_{i = 1}^{t-1}  (\mathbf{\Delta} \bA^{(m+i)} - \bdelta^{(m+i)}).
\end{align*}
Therefore, we have
\begin{align*}
	& L(\bA^{(m+t)}, \bB^{(m+t)}, \bdelta^{(m+t)}, \bV^{(m+t-1)}) \\
	= & 2^{-1}\|\bY - \bX\bB^{(m+t)} - \bA^{(m+t)} \|^2_F + \langle \bV^{(m+t-1)}, \mathbf{\Delta} \bA^{(m+t)} - \bdelta^{(m+t)}\rangle + 2^{-1}\vartheta \|\mathbf{\Delta} \bA^{(m+t)} - \bdelta^{(m+t)} \|_2^2 \\
	&\quad\quad + \sum_{1\leq i<j \leq n} p_{\gamma} (\|\bdelta_{ij}^{(m+t)} \|_2,\lambda)  \\
	= & 2^{-1}\|\bY - \bX\bB^{(m+t)} - \bA^{(m+t)} \|_F^2 + \langle \bV^{(m)}, \mathbf{\Delta} \bA^{(m+t)} -\bdelta^{(m +t)} \rangle  \\
	&\quad \quad+ \vartheta \sum_{i = 1}^{t-1} \langle \mathbf{\Delta} \bA^{(m+i)} - \bdelta^{(m+i)}, \mathbf{\Delta} \bA^{(m+t)} - \bdelta^{(m+t)} \rangle \\
	& \quad\quad + 2^{-1}\vartheta \|\mathbf{\Delta} \bA^{(m+t)} - \bdelta^{(m+t)} \|_2^2 + \sum_{1 \leq i<j \leq n} p_{\gamma} ( \| \bdelta_{ij}^{(m+t)}\|_2, \lambda) \\
	\leq & f^{m+t},
\end{align*}
where the last inequality follows from the same argument used in the proof of \eqref{f(m+1)}.  

Let $f^*= \lim\limits_{m \rightarrow \infty} f^{m+1}$. Recall that 
$(\bA^*, \bB^*, \bdelta^*)$ is the limit point of the sequence $(\bA^{(m+1)}, \bB^{(m+1)},$ $ \bdelta^{(m+1)})$. Thus, we have
\begin{align*}
	f^* & = \lim_{m \rightarrow \infty} f^{m+1}= \lim_{m \rightarrow \infty} f^{m+t} 
	= \inf_{\mathbf{\Delta} \mathbf{A}^* - \mathbf{\delta} = 0} \left\{ 2^{-1}\|\bY - \bX \bB^* - \bA^* \|_F^2 +
	\sum_{1 \leq i<j \leq n} p_{\gamma} (\|\bdelta_{ij} \|_2, \lambda)   \right\},
\end{align*}
and for all $t \geq 0$
\begin{align*}
	& \lim_{m \rightarrow \infty} L(\bA^{(m+t)}, \bB^{(m+t)}, \bdelta^{(m+t)}, \bV^{(m+t-1)}) \\
	= & 2^{-1}\|\bY - \bX\bB^* - \bA^* \|_F^2 + \sum_{1\leq i < j \leq n} p_{\gamma} (\| \bdelta_{ij}^*\|_2, \lambda) + \lim_{m \rightarrow \infty} \langle \bV^{(m)}, \mathbf{\Delta} \bA^* - \bdelta^* \rangle \\
	&\quad\quad+ (t - 2^{-1})\vartheta \|\mathbf{\Delta} \bA^* - \bdelta^* \|_2^2 \\
	\leq & f^*
\end{align*}
holds. 
Hence, by the definition of the primal residual $\mathbf{r}^{(m)} = \mathbf{\Delta} \bA^{(m)} - \bdelta^{(m)}$, we have $\lim\limits_{m \rightarrow \infty} \|\mathbf{r}^{(m)} \|_F^2 = \|\mathbf{\Delta} \bA^* - \bdelta^* \|_F^2 = 0$.

By definition, $\bA^{(m+1)}$ minimizes $L (\bA, \bB^{(m)}, \bdelta^{(m)}, \bV^{(m)})$. Thus, we have 
\begin{align*}
	\partial  L(\bA^{(m+1)}, \bB^{(m)}, \bdelta^{(m)}, \bV^{(m)})/\partial \bA = 0.
\end{align*}
Note that
\begin{align} 
	& \partial L(\bA^{(m+1)}, \bB^{(m)}, \bdelta^{(m)}, \bV^{(m)})/ \partial \bA \nonumber \\
	=&  (\bX \bB^{(m)} +  \bA^{(m+1)} - \bY)  + \mathbf{\Delta}^{\top} \bV^{(m)} + \vartheta \mathbf{\Delta}^{\top} (\mathbf{\Delta} \bA^{(m+1)} - \bdelta^{(m)}) \nonumber \\ 
	= & (\bX \bB^{(m)} + \bA^{(m+1)} - \bY)  + \mathbf{\Delta}^{\top} \left[\bV^{(m)} + \vartheta (\mathbf{\Delta} \bA^{(m+1)} - \bdelta^{(m)})\right] \label{der L}\\ 
	= & ( \bX \bB^{(m)} + \bA^{(m+1)} - \bY)  + \mathbf{\Delta}^{\top} \left[\bV^{(m+1)}  - \vartheta (\mathbf{\Delta} \bA^{(m+1)} - \bdelta^{(m+1)}) +\vartheta (\mathbf{\Delta} \bA^{(m+1)} - \bdelta^{(m)})\right] \nonumber\\ 
	= &  (\bX \bB^{(m)} +  \bA^{(m+1)} - \bY) + \mathbf{\Delta}^{\top} \bV^{(m+1)} + \vartheta \mathbf{\Delta}^{\top} (\bdelta^{(m+1)} - \bdelta^{(m)}),  \nonumber 
\end{align}
where the second last step follows from \eqref{step3}.
Therefore, by the definition of the dual residual  $\mathbf{s}^{(m+1)} = \vartheta \mathbf{\Delta}^{\top} (\bdelta^{(m+1)} - \bdelta^{(m)})$, we have
\begin{align}\label{eq:B3}
	\bs^{(m+1)} 
	= -  (\bX \bB^{(m)} + \bA^{(m+1)} - \bY ) - \mathbf{\Delta}^{\top} \bV^{(m+1)} .
\end{align}
Since $\|\mathbf{\Delta} \bA^* - \bdelta^*\|_F^2 = 0$, by \eqref{der L}, we have
\begin{align*}
	& \lim_{m \rightarrow \infty} \partial L(\bA^{(m+1)}, \bB^{(m)}, \bdelta^{(m)}, \bV^{(m)})/\partial \bA \\
	= & \lim_{m \rightarrow \infty} (\bX \bB^{(m)} + \bA^{(m+1)} - \bY) + \mathbf{\Delta}^{\top} \bV^{(m)} \\
	= & (\bX \bB^* + \bA^* - \bY) + \mathbf{\Delta}^{\top} \bV^*  = \bzero.
\end{align*}
This, together with \eqref{eq:B3}, entails
$\lim\limits_{m \rightarrow \infty} \bs ^{(m+1)} =-[(\bX \bB^* + \bA^* - \bY) + \mathbf{\Delta}^{\top} \bV^*]= \bzero$.


\section{More Details on the Proof of Results in Steps 1 and 2 in Theorem~\ref{theo-pro}}\label{SuppD}
\renewcommand{\theequation}{D.\arabic{equation}}
\setcounter{equation}{0}

Recall that in the proof of Theorem~\ref{theo-pro}, we showed that $(\widehat{\bA}^{or}, \widehat{\bB}^{or})$ is a strictly local minimizer of the objective function \eqref{initial goal} with probability approaching one through the following two steps.
\begin{description}
	\item[Step 1.] On the event $E_1$, for any $((\bA^0)^{\top}, \bB^{\top})^{\top} \in \Theta$ and $ ((\bA^0)^{\top}, (\bB)^{\top})^{\top} \neq  ((\widehat{\bA}^{or})^{\top},$ $  (\widehat{\bB}^{or})^{\top})^{\top}$, we have $Q_n (\bA^0, \bB) > Q_n (\widehat{\bA}^{or}, \widehat{\bB}^{or})$.
	\item[Step 2.] There exists an event $E_2$ such that $\mathrm{Pr}(E_2^C) \leq 2 qn^{-2}$, where $E_2^C$ is the complement of event $E_2$. On $E_1 \cap E_2$, there is a neighborhood of $((\widehat{\bA}^{or})^{\top}, (\widehat{\bB}^{or})^{\top})^{\top}$, denoted by $\Theta_n$, such that $Q_n(\bA, \bB ) \geq Q_n (\bA^0, \bB)$ for any $(\bA^{\top}, \bB^{\top})^{\top} \in \Theta \cap \Theta_n$ for sufficiently large $n$.
\end{description}
In this section, we will provide detailed proofs for the results in these two steps.  To prove the result in Step 1, we first show that $P_n^{\psi} (T^{*}(\bA)) = C_n $ for any $\bA \in \Theta$, where $C_n$ is a constant and does not depend on $\bA$. Let $T^{*} (\bA) = \bC = (\bc_1,\dots,\bc_K)^{\top}$. It suffices to show that $\|\bc_k - \bc_{k^{'}} \|_2 > a_1 \lambda$ for all $k \neq k^{'}$. Then by Condition \ref{con1}, $\rho (\|\bc_k - \bc_{k^{'}} \|_2)$ is a constant yielding that $P_n^{\psi} (T^{*}( \bA))$ is a constant. To proceed, note that
\begin{align}\label{eq: C1}
	\| \bc_k - \bc_{k^{'}}\|_2 & = \| \bc_k - \bc_k^* + \bc_k^* - \bc^*_{k^{'}} + \bc^*_{k^{'}} - \bc_{k^{'}} \|_2 \geq \| \bc^* _k - \bc^* _{k^{'}}\|_2 - 2\sup_k \| \bc_k - \bc^* _{k}\|_2
\end{align}
and
\begin{align}\nonumber
	\sup_k \| \bc_k - \bc^* _{k}\|_2^2 & = \sup_k \| |\psi_k|^{-1} \sum_{i \in \psi_k} \ba_i - \bc^* _{k}\|_2^2 = \sup_k \| |\psi_k|^{-1} \sum_{i \in \psi_k} (\ba_i - \ba^* _{i})\|_2^2 \\ \nonumber
	& =  \sup_k |\psi_k|^{-2} \| \sum_{i \in \psi_k} (\ba_i - \ba^* _{i})\|_2^2 \leq \sup_k |\psi_k|^{-1} \sum_{i \in \psi_k} \| (\ba_i - \ba^* _{i})\|_2^2 \\ \label{star error}
	& \leq \sup_i \| (\ba_i - \ba^* _{i})\|_2^2 \leq \phi_{n}^2,
\end{align}
where the last inequality follows from the definition of the set $\Theta$. 
Combining \eqref{eq: C1} with \eqref{star error} yields
\begin{align*}
	\| \bc_k - \bc_{k^{'}}\|_2 
	\geq b_n - 2 \phi_{n} > a_1 \lambda
\end{align*}
for all $k$ and $k^{'}$, where the last inequality is due to the assumption that $b_n > a_1 \lambda \gg \phi_n$.  Therefore, we have $P_n^{\psi}( T^*(\bA)) = C_n$, and as a result $Q_n^{\psi}( T^*(\bA), \bB) = L_n^{\psi} (T^*(\bA), \bB) + C_n$ with $r(\bB) \leq r $ for all $(\bA^{\top}, \bB^{\top})^{\top} \in \Theta$. Given $T^*(\bA)$, we have $L_n^{\psi} (T^*(\bA), \bB) \geq L_n^{\psi} (T^*(\bA), \widehat{\bB}^{or})$ since $\widehat{\bB}^{or}$ is a global minimizer of $L_n^{\psi} (T^*(\bA), \bB)$ subject to $r(\bB) \leq r$. Moreover, $L_n^{\psi} (T^*(\bA), \widehat{\bB}^{or}) > L_n^{\psi} (\widehat{\bC}^{or}, \widehat{\bB}^{or})$ since $\widehat{\bC}^{or}$ is the unique minimizer of $L_n^{\psi} (T^*(\bA), \bB)$ with fixed $\bB$. Hence $Q_n^{\psi} (T^*(\bA), \bB) > Q_n^{\psi} (\widehat{\bC}^{or}, \widehat{\bB}^{or})$ for all $T^*(\bA) \neq \widehat{\bC}^{or}$. By (\ref{Q relationship}), we have $ Q_n^{\psi} (\widehat{\bC}^{or},\widehat{\bB}^{or}) =  Q_n (\widehat{\bA}^{or},\widehat{\bB}^{or})$ and $Q_n^{\psi} (T^*(\bA), $ $\bB)= Q_n (T^{-1} (T^*(\bA)), \bB) = Q_n (\bA^0, \bB)$. Therefore, $Q_n ( \bA^0, \bB) > Q_n (\widehat{\bA}^{or}, \widehat{\bB}^{or})$ for all $\bA^0 \neq \widehat{\bA}^{or}$, and the result in Step 1 is proved.

Next we prove the result in Step 2. For a positive sequence $t_n$, let $\Theta_n = \{\ba_i : \sup_i \|\ba_i - \widehat{\ba}^{or}_i \|_2 \leq t_n \}$. For $ (\bA^{\top}, \bB^{\top})^{\top} \in \Theta \cap \Theta_n$, by Taylor's expansion, we have
\begin{align*}
	Q_n(\bA, \bB) - Q_n (\bA^0, \bB) = \Gamma_1 + \Gamma_2\,\,\mbox{and}\,\,\ \mbox{rank}(\bB) \leq r,
\end{align*}
where
\begin{align}\label{Gamma1}
	& \Gamma_1 = -  (\bY - (\bI, \bX) ((\bA^m)^{\top},(\bB)^{\top})^{\top}) \cdot (\bA - \bA^0), 
	\\ \nonumber
	& \Gamma_2 = \sum_{i = 1}^{n} \frac{\partial P_n (\bA^m)}{\partial \ba^{\top}_i} \cdot (\ba_i - \ba_i^0).
\end{align}
where $`` \cdot "$ denotes a dot product and $\bA^m = \alpha \bA + (1 - \alpha)\bA^0$ for some constant $\alpha \in (0,1)$. Now we first handle $\Gamma_2$.
\begin{align} \nonumber
	\Gamma_2 & = \lambda \sum_{j > i } \rho^{'} (\|\ba^m_i - \ba^m_j \|_2)\|\ba^m_i - \ba^m_j \|_2^{-1}(\ba^m_i - \ba^m_j ) \cdot (\ba_i - \ba_i^0) \\ \nonumber
	& \quad\quad + \lambda \sum_{j < i } \rho^{'} (\|\ba^m_i - \ba^m_j \|_2)\|\ba^m_i - \ba^m_j \|_2^{-1}(\ba^m_i - \ba^m_j ) \cdot (\ba_i - \ba_i^0) \\
	& = \lambda \sum_{j > i} \rho^{'} (\|\ba^m_i - \ba^m_j \|_2)\|\ba^m_i - \ba^m_j \|_2^{-1}(\ba^m_i - \ba^m_j ) \cdot (\ba_i - \ba_i^0) \\\nonumber
	& \quad\quad+ \lambda \sum_{i < j } \rho^{'} (\|\ba^m_j - \ba^m_i \|_2)\|\ba^m_j - \ba^m_i \|_2^{-1}(\ba^m_j - \ba^m_i ) \cdot (\ba_j - \ba_j^0) \\\nonumber
	& = \lambda \sum_{i < j } \rho^{'} (\|\ba^m_i - \ba^m_j \|_2)\|\ba^m_i - \ba^m_j \|_2^{-1}(\ba^m_i - \ba^m_j ) \cdot \{(\ba_i - \ba_i^0) - (\ba_j - \ba_j^0)\}.
\end{align}
When $i,j \in \psi_k$, we have $\ba^0_i = \ba^0_j$ and $\ba_i^m - \ba^m_j = \alpha (\ba_i - \ba_j)$. Thus,
\begin{align*}
	\Gamma_2  
	& =\lambda \sum_{k=1}^K \sum_{i,j \in \psi_k, i< j} \rho^{'} (\|\ba_i^m - \ba_j^m \|_2) \| \ba_i^m - \ba_j^m\|_2^{-1} (\ba^m_i - \ba^m_j) \cdot (\ba_i - \ba_j) \\
	& +  \lambda \sum_{k<k^{'}} \sum_{i \in \psi_k, j \in \psi_{k^{'}}} \rho^{'} (\|\ba_i^m - \ba_j^m \|_2) \| \ba_i^m - \ba_j^m\|_2^{-1} (\ba^m_i - \ba^m_j) \cdot \{ (\ba_i - \ba_i^0) - ( \ba_j - \ba^0_j)\}.
\end{align*}
Moreover,
\begin{align}
	\sup_i \|\ba^*_i - \ba^0_i \|_2^2 = \sup_k \| \bc_k^* - \bc_k\|_2^2 \leq \phi_n^2,
\end{align}
where the last inequality follows from (\ref{star error}). Since $\ba_i^m=\alpha\ba_i + (1-\alpha)\ba_i^0$, we have
\begin{align} \label{m-star}
	\sup_i \|\ba_i^m - \ba_i^* \|_2 \leq \alpha \sup_i \|\ba_i - \ba_i^* \|_2 + (1 - \alpha) \sup_i \| \ba_i^0 - \ba_i^*\|_2 \leq  \phi_n
\end{align}
Hence for $k \neq k^{'}$, $i \in \psi_k$, $j^{'} \in \psi_{k^{'}}$,
\begin{align*}
	\| \ba_i^m - \ba_j^m\|_F \geq \min_{i \in \psi_k, j^{'}\in \psi_{k^{'}}} \| \ba_i^* - \ba_j^*\|_2 - 2 \max_i \| \ba_i^m - \ba_i^*\|_2 \geq b_n - 2 \phi_n > a_1 \lambda,
\end{align*}
and thus $\rho^{'} (\|\ba_i^m - \ba_j^m \|_2) = 0$. Therefore,
\begin{align*}
	\Gamma_2 & = \lambda \sum_{k=1}^K \sum_{i,j \in \psi_k, i< j} \rho^{'} (\|\ba_i^m - \ba_j^m \|_2) \| \ba_i^m - \ba_j^m\|_2^{-1} (\ba^m_i - \ba^m_j) \cdot (\ba_i - \ba_j) \\
	& = \lambda \sum_{k = 1}^K \sum_{i,j \in \psi_k, i<j} \rho^{'}  (\|\ba_i^m - \ba_j^m \|_2) \| \ba_i - \ba_j\|_2.
\end{align*}
where the last step holds because $\ba_i^m - \ba^m_j = \alpha (\ba_i - \ba_j)$ for $i,j \in \psi_k$. Furthermore, following the same argument as in \eqref{star error},
we have
\begin{align*}
	\sup_i \|\ba_i^0 - \widehat{\ba}_i^{or} \|_2 = \sup_{k} \|\bc_k - \widehat{\bc}_k^{or} \|_2 \leq \sup_i \|\ba_i - \widehat{\ba}_i^{or} \|_2.
\end{align*}
Then for $i, j \in \psi_k$, we have
\begin{align*}
	& \sup_i \|\ba_i^m - \ba_j^m \|_2 \leq 2 \sup_i \| \ba_i^m - \ba_i^0\|_2 \leq 2 \sup_i \|\ba_i - \ba_i^0 \|_2 \\
	 \leq & 2 (\sup_i \|\ba_i - \widehat{\ba}_i^{or} \|_2 + \| \widehat{\ba}_i^{or} - \ba_i^0\|_2) \leq 4\sup_i \|\ba_i - \widehat{\ba}_i^{or} \|_2 \leq 4 t_n,
\end{align*}
where the last inequality follows from the definition of $\Theta_n$. Hence, $\rho^{'} (\| \ba_i^m - \ba_j^m \|_2) \geq \rho^{'} (4 t_n)$ by concavity of $\rho(\cdot)$. As a result,
\begin{align}
	\Gamma_2 \geq \sum_{k = 1}^K \sum_{i,j \in \psi_k, i < j } \lambda \rho^{'} (4 t_n) \| \ba_i - \ba_j\|_2.
\end{align}

Next we handle $\Gamma_1$. 
Let
\begin{align*}
	\bQ = (\bQ_1^{\top}, \dots, \bQ_n^{\top}) = (\bY - \bX \bB -  \bA^m)^{\top} .
\end{align*}
Then we have
\begin{align}\nonumber
	\Gamma_1 & = -\bQ^{\top} \cdot (\bA - \bA^0) = \sum_{k =1}^{K} \sum_{\{i,j \in \psi_k\}} \frac{ \bQ_i  \cdot (\ba_i - \ba_j) }{|\psi_k |} 
	\\
	& = - \sum_{k =1}^{K} \sum_{\{i,j \in \psi_k\}} \frac{ \bQ_i  \cdot (\ba_i - \ba_j) }{2|\psi_k |} - \sum_{k =1}^{K} \sum_{\{i,j \in \psi_k\}} \frac{ \bQ_i  \cdot (\ba_i - \ba_j)   }{2|\psi_k |} \\ \nonumber
	& = - \sum_{k =1}^{K} \sum_{\{i,j \in \psi_k\}} \frac{ (\bQ_i - \bQ_j)  \cdot (\ba_i - \ba_j) }{2|\psi_k |} \\ \nonumber
	& = - \sum_{k =1}^{K} \sum_{\{i,j \in \psi_k, i <j\}} \frac{ (\bQ_i - \bQ_j) \cdot (\ba_i - \ba_j) }{|\psi_k |}.
\end{align}
Since
\begin{align*}
	\bQ_i = \by_i - \bx_i^{\top} \bB - \ba_i^m  = \bepsilon_i + \bx_i^{\top} (\bB^*  - \bB) +  (\ba^* _i - \ba^m_i),
\end{align*}
we have
\begin{align*}
	\sup_i \| \bQ_i\|_2 \leq \sup_i \| \bepsilon_i + \bx_i^{\top} (\bB^*  - \bB) +  (\ba^*_i - \ba^m_i) \|_2.
\end{align*}
By Condition~\ref{con4}, the result in \eqref{m-star}, and $\| \bB^*  - \bB\|_F \leq \phi_n$ given in the definition of $\Theta_n$, we have
\begin{align*}
	\sup_i \| \bQ_i\|_2 \leq  \| \bepsilon_i\|_2 + C_2 \sqrt{p} \phi_n +  \phi_n ,
\end{align*}
It follows from Condition \ref{con2} that
\begin{align*}
	\mathrm{Pr}  \left\{\| \bepsilon_i\|^2_2 > 2c_1^{-1} q\mathrm{ln}(n)\right\} 
	\leq\sum_{i=1}^q \mathrm{Pr}\left\{\bepsilon_{ij}^2 \geq 2c_1^{-1} \mathrm{ln}(n) \right\} = \sum_{i=1}^q \mathrm{Pr}\left\{|\bepsilon_{ij}| \geq \sqrt{2c_1^{-1} \mathrm{ln}(n)} \,\right\} \leq 2 q n^{-2}.
\end{align*}
Thus there is an event $E_2$ such that $\mathrm{Pr}(E_2^C) \leq 2 qn^{-2}$, and on the event $E_2$ we have
\begin{align*}
	\sup_i \| \bQ_i\|_2 \leq C_3 \left[\sqrt{2c_1^{-1} } \sqrt{q\mathrm{ln}(n)} + C_2 \sqrt{p} \phi_n + C_3\sqrt{d}\phi_n\right].
\end{align*}
Then
\begin{align*}
	& |\frac{(\bQ_i - \bQ_j) \cdot (\ba_i - \ba_j)}{| \psi_k|}| \\
	& \leq | \psi_{\min}|^{-1} 2\sup_i \bQ_i \cdot (\ba_i - \ba_j)  \leq | \psi_{\min}|^{-1} 2 \sup_i \|\bQ_i \|_2 \| \ba_i - \ba_j\|_2 \\
	& \leq 2 |\psi_{\min}|^{-1}\left[\sqrt{2c_1^{-1} } \sqrt{q\mathrm{ln}(n)} + C_2 \sqrt{p} \phi_n + \phi_n\right]  \| \ba_i - \ba_j\|_F.
\end{align*}
Therefore, we have
\begin{align*}
	& Q_n(\bC, \bB) - Q_n (\bC^0, \bB) \\
	\geq & \sum_{k =1}^{K} \sum_{\{i,j \in \psi_k, i <j\}} \left[\lambda \rho^{'} (4 t_n)  - 2 |\psi_{\min}|^{-1}  ( \sqrt{2c_1^{-1} } \sqrt{q\mathrm{ln}(n)} + C_2 \sqrt{p} \phi_n +  \phi_n) \right]  \| \ba_i - \ba_j\|_2.
\end{align*}
Let $t_n = o(1)$, then $\rho^{'}(4 t_n) \rightarrow 1$. Since $\lambda \gg \phi_n$, $|\psi_{\min}| \gg \sigma^2 [ (p+q-r^*)(r^* + K) + qK + \mathrm{ln}(n)]$, $|\psi_{\min}|^{-1}  p  = o(1)$, and $|\psi_{\min}|^{-1} = o(1)$, we have $\lambda \gg |\psi_{\min}|^{-1}\sqrt{q \mathrm{ln}(n)}$, $\lambda \gg |\psi_{\min} |^{-1} \sqrt{p} \phi_n$ and $\lambda \gg |\psi_{\min}|^{-1} \phi_n$. Therefore, $Q_n(\bC, \bB) - Q_n (\bC^0, \bB) \geq 0$ for sufficiently large $n$, and the result in Step 2 is proved.

\medskip


\section{Algorithm for model~\eqref{HTE-mod3} }\label{supp.algo}\label{SuppE}
\renewcommand{\theequation}{E.\arabic{equation}}
\setcounter{equation}{0}

In this section, we present the estimation 
procedure for heterogeneous multivariate response regression model~\eqref{HTE-mod3} and provide the detailed ADMM algorithm to obtain the estimators $\widehat{\mathbf{A}}_i$ and $ \widehat{\mathbf{B}}$.
Recall that the heterogeneous multivariate response model~\eqref{HTE-mod3} is given by
\begin{align*} 
	\by_i = \bA_i^{\top} \bz_i + \bB^{\top} \bx_i + \bepsilon_i,
\end{align*}
where $\bz_i$ is a $d$-dimensional feature vector, $\bA_i \in \mathbb{R}^{d \times q}$ denotes the heterogeneous coefficient matrices of the feature vectors and $\bB \in \mathbb{R}^{p \times q}$ characterizes the association between responses and other predict variables in $\bx_i$. In particular, 
$\bB \in \mathbb{R}^{p \times q}$ is assumed to be of low-rank, i.e., $\mbox{rank}(\bB)=r^{*}$ with $r^{*}\leq \min\{p, q\}$. In this model setup, our goal is to accurately estimate the regression coefficient matrices $\bB$, and heterogeneous coefficient matrices $\bA_i$, and correctly identify the subgroup outcomes including the number of subgroups $K$ and the group indicators $w_{ik}$. To achieve this goal, we can develop a rank-constrained pairwise penalized procedure for simultaneous subgroup identification and coefficient matrix estimation by solving the following optimization problem 
\begin{align*} 
	(\widehat{\bA}, \widehat{\bB}) =&  \arg \min_{\mathbf{A}, \mathbf{B}} \frac{1}{2} \sum_{i = 1}^{n} \|\by_i - \bB^{\top} \bx_i - \bA^{\top}_i \bz_i \|_2^2 + \sum_{1 \leq i < j \leq n} p_{\gamma} ( \|\bA_i - \bA_j \|_F, \lambda)   \\
	& \mbox{subject to}\,\, \mbox{rank}(\bB) \leq r,
\end{align*}
where $\lambda>0$ is a regularization parameter and $r$ is a given positive integer.

By introducing a new parameter $\bdelta_{ij} = \bA_i - \bA_j$, augmented Lagrangian method can be adopted and the objective function is reformulated as
\begin{align*}
	L(\bA, \bB, \bdelta, \bV)
	& = \frac{1}{2} \sum_{i = 1}^{n} \|\by_i - \bx_i^{\top} \bB - \bz_i^{\top} \bA_i  \|_2^2 + \sum_{1 \leq i < j \leq n} p_{\gamma} ( \|\bdelta_{ij} \|_F, \lambda) + \sum_{1 \leq i < j \leq n} \langle  \bv_{ij}, (\bA_i - \bA_j - \bdelta_{ij}) \rangle \\
	& + \frac{\vartheta}{2}\sum_{1 \leq i < j \leq n} \| \bA_i - \bA_j - \bdelta_{ij} \|_F^2 \,\,  \mbox{subject to}\,\,  \mbox{rank}(\bB) \leq r,
\end{align*}
where $\vartheta$ is a penalty parameter, $\langle \bA, \bB \rangle = \rm{tr} (\bA^{\top} \bB)$ denotes the matrix inner product and $\rm{tr}(\cdot)$ denotes the trace, and $\bV = \{\bv_{ij}^{\top}, i < j \}^{\top}$ is a $[0.5 n(n-1)d] \times q$ Lagrange multiplier matrix or dual variable matrix with $\bv_{ij}$ a $d \times q$ matrix. This augmented Lagrangian function can be solved by an alternating direction method of multipliers (ADMM) algorithm.

Specifically, given $\bdelta^{(m)}$, $\bV^{(m)}$ at the $m$th iteration, we update the algorithm by the following three steps: 
\begin{align}\label{sup.step1}
	& (\mathbf{A}^{(m+1)},\mathbf{B}^{(m+1)} )= \arg \min_{\mathbf{A}, \mathbf{B}} L( \mathbf{A}, \mathbf{B}, \bdelta^{(m)}, \mathbf{V}^{(m)}) \,\,  \mbox{subject to}\,\, \mbox{rank}(\bB) \leq r \\ \label{sup.step2}
	& \bdelta^{(m + 1)} = \arg \min_{\bdelta} L (\bA^{(m +1)}, \bB^{(m +1)}, \bdelta, \bV^{(m)}), \\ \label{sup.step3}
	&\bv_{ij}^{(m + 1)} = \bv_{ij}^{(m)} + \vartheta (\bA_i^{(m + 1)} - \bA_j^{(m +1)} - \bdelta_{ij}^{(m + 1)}).
\end{align}
In the first step, the objective function in the problem \eqref{sup.step1} can be simplified as
\begin{align*}
	f(\bA, \bB) 
	=&  \frac{1}{2} \sum_{i = 1}^{n} \|\by_i - \bx_i^{\top} \bB -  \bz_i^{\top}\bA_i \|_2^2  
	+ \frac{\vartheta}{2} \sum_{1 \leq i < j \leq n} \|\bA_i - \bA_j - \bdelta_{ij}^{(m)} + \vartheta^{-1} \mathbf{v}_{ij}^{(m)} \|_F^2 + C_n \\
	& \mbox{subject to}\,\,  \mbox{rank}(\bB) \leq r,
\end{align*}
where $C_n$ is a constant independent of $(\bA,\bB)$. Through some algebra, we rewrite $f(\bA,\bB)$ in a matrix notation as: 
\begin{align} \label{sup.Delta}
	f(\bA, \bB) 
	=&  \frac{1}{2} \|\bY - \bX \bB - \bZ\bA \|_F^2 
	+ \frac{\vartheta}{2} \|\mathbf{\Delta} \bA - \bdelta^{(m)} - \vartheta^{-1} \mathbf{V}^{(m)} \|_F^2 + C_n \nonumber\\
	& \mbox{subject to}\,\,  \mbox{rank}(\bB) \leq r,
\end{align}
where $\mathbf{\Delta} = \{(\mathbf{e}_i - \mathbf{e}_j)^{\top} \otimes \mathbf{I}_d, i< j \}$ is an $[0.5 n(n-1)d] \times nd$ matrix and 
$\mathbf{e}_i$ is an $n \times 1$ vector with 1 for the $i$th entry and 0 for all other entries.

Due to the rank constraint on $\bB$, 
we apply a block coordinate descent method proposed in \citep{Tseng2001}.  
Then the two matrices $\bA^{(m + 1)}$ and $\bB^{(m+1)}$ can be updated alternately while holding the other fixed, as follows:
\begin{align}\label{sup.C update}
	&\mathbf{A}^{(m + 1)} = ( \mathbf{Z}^{\top} \mathbf{Z} + \vartheta \mathbf{\Delta}^{\top}\mathbf{\Delta})^{-1} [ \mathbf{Z}^{\top}(\mathbf{Y} - \mathbf{X} \mathbf{B}^{(m)}) + \vartheta \mathbf{\Delta}^{\top} (\bdelta^{(m)} - \vartheta^{-1} \bV^{(m)})],    \\ \label{sup.B update}
	&\mathbf{B}^{(m + 1)} = (\mathbf{X}^{\top} \mathbf{X})^{-1} \mathbf{X}^{\top}(\mathbf{Y} - \mathbf{Z}\mathbf{A}^{(m+1)}) \bQ_{V(\mathbf{X}, \mathbf{Y} - \mathbf{Z} \mathbf{A}^{(m+1)}, r)}.
\end{align}
Here $\bQ_{\mathbf{X}}= \mathbf{X}(\mathbf{X}^{\top} \mathbf{X})^{-} \mathbf{X}^{\top}$ denotes the orthogonal projection matrix onto the range of $\bX$ and $V(\mathbf{X}, \mathbf{Y} - \mathbf{Z} \mathbf{A}^{(m+1)}, r)$ is formed by the leading $r$ eigenvectors of $(\bY - \bZ \bA^{(m+1)})^{\top} \bQ_{\mathbf{X}} (\bY - \bZ \bA^{(m+1)})$. Obviously, the procedure updating $\bA$ in \eqref{sup.C update} performs simple matrix operations and the procedure updating $\bB$ in \eqref{sup.B update} performs reduced-rank regression on the adjusted response matrix $\bY - \bZ \bA^{(m+1)}$.

In the second step, after discarding the terms independent of $\bdelta$ in the problem (\ref{sup.step2}), the objective function is
\begin{align*}
	\sum_{1 \leq i < j \leq n} \langle \bv^{(m)}_{ij}, (\bA^{(m)}_i - \bA^{(m)}_j - \bdelta_{ij}) \rangle + \frac{\vartheta}{2} \|\bA^{(m)}_i - \bA^{(m)}_j - \bdelta_{ij} \|_F^2 + p_{\gamma} (\| \bdelta_{ij}\|_F, \lambda).
\end{align*}
Denote by $\bzeta_{ij}^{(m)} = (\bA_i^{(m+1)} - \bA_j^{(m+1)}) + \vartheta^{-1} \bv_{ij}^{(m)}$. For the MCP with $\gamma > 1/\vartheta$, the solution is
\begin{equation*}
	\bdelta^{(m+1)}_{ij} =
	\left\{
	\begin{array}{ll}
		\frac{S (\bzeta^{(m)}_{ij}, \lambda /\vartheta)}{1 - 1/(\gamma \vartheta)},  &  \ \|\bzeta_{ij}^{(m)}\|_F \leq \gamma \lambda, \\
		\bzeta_{ij}^{(m)}, &  \ \|\bzeta_{ij}^{(m)}\|_F > \gamma \lambda,
	\end{array}
	\right.
\end{equation*}
where $S(\bz, t) = (1 - t/\| \bz\|_F)_+ \bz$ is a groupwise soft thresholding operator. For the SCAD penalty with $\gamma > 1/\vartheta + 1$, the solution is
\begin{equation*}
	\bdelta^{(m+1)}_{ij} =
	\left\{
	\begin{array}{ll}
		S (\bzeta^{(m)}_{ij}, \lambda /\vartheta),
		& \ \|\bzeta_{ij}^m\|_F \leq \lambda + \lambda/\vartheta, \\
		\frac{S (\bzeta^{(m)}_{ij}, \gamma \lambda / ((\gamma - 1)\vartheta))}{1 - 1/((\gamma -1 )\vartheta))}, &  \ \lambda + \lambda/\vartheta \leq \|\bzeta_{ij}^{(m)}\|_F \leq \gamma \lambda, \\
		\bzeta_{ij}^{(m)}, &  \ \|\bzeta_{ij}^{(m)}\|_F > \gamma \lambda .
	\end{array}
	\right.
\end{equation*}
In the third step, the update of the dual variable $\bv_{ij}$ is given in (\ref{sup.step3}). Note that $\mathbf{r}^{(m)} = \mathbf{\Delta} \bA^{(m)} - \bdelta^{(m)}$ and $\mathbf{s}^{(m)} = \vartheta \mathbf{\Delta}^{\top} (\bdelta^{(m)} - \bdelta^{(m+1)})$ are the primal residual and the dual residual in the ADMM algorithm, respectively. The iterations continue until the stopping criterion is reached. To be specific, we stop the algorithm when the primal residual $\mathbf{r}^{(m+1)} = \mathbf{\Delta} \bA^{(m+1)} - \bdelta^{(m+1)}$ is close to zero such that $\| \mathbf{r}^{(m+1)} \|_F < \varepsilon$ for some small value $\varepsilon$.

\textbf{Initial values}. To facilitate the update of the ADMM algorithm, proper initial values need to be specified. For this purpose, we consider the following fusion criterion
\begin{align*}
	L_R (\bA, \bB) = \frac{1}{2} \| \bY- \bZ \bA  - \bX \bB \|_F^2 + \frac{\lambda^*}{2} \sum_{1 \leq i \leq j \leq n} \| \bA_i - \bA_j\|_F^2,
\end{align*}
where $\lambda^*$ is a small tuning parameter. Similar to \citep{Ma2020, Hu2021}, we use $\lambda^* = 0.001$ in our simulation studies. Recall that $\mathbf{\Delta}$ is defined in \eqref{Delta}. Then $ L_R (\bA, \bB)$ can be written using a matrix notation as
\begin{align*}
	L_R (\bA, \bB) = \frac{1}{2} \| \bY- \bZ \bA  - \bX \bB \|_F^2 + \frac{\lambda^*}{2}\| \mathbf{\Delta}  \bA \|_F^2.
\end{align*}
Correspondingly, the solutions are
\begin{align*}
	&\bA_R (\lambda^*) = [ \bZ^{\top} (\bI_n - \bQ_{\mathbf{X}}) \bZ + \lambda^* \mathbf{\Delta}^{\top} \mathbf{\Delta}]^{-1} \bZ^{\top}  (\bI_n - \bQ_{\mathbf{X}}) \bY, \\ 
	&\bB_R (\lambda^*) = (\bX^{\top} \bX)^{-1} \bX^{\top} (\bY - \bZ \bA_R (\lambda^*)),
\end{align*}
respectively. Hence, we take $\bA^{(0)} = \bA_{R}(\lambda^*)$, $\bB^{(0)} = \bB_R (\lambda^*)$, $\bdelta^{(0)} = \mathbf{\Delta}\bA^{(0)}$, and $\bV^{(0)}= \mathbf{0}$. Finally, we can partition the sample into subgroups based on the estimator $\widehat{\mathbf{A}}= (\widehat{\mathbf{A}}_1^{\top}, \dots$, $\widehat{\mathbf{A}}_n^{\top})^{\top}$. Let $\{ \widehat{\bC}_1, \dots, \widehat{\bC}_{\widehat{K}}\}$ be the distinct values of $\widehat{\bA}_i$ and $\widehat{\mathcal{A}}_k=\{i:\widehat{\bA}_i=\widehat{\bC}_k,1\le i\le n\}$, $1\le k\le \widehat{K}$. Then $\widehat{\mathcal{A}}_1 \cup \dots \cup \widehat{\mathcal{A}}_{\widehat{K}}$ constitutes an estimated subgroup partition of $\{1,\dots,n\}$.


\begin{thebibliography}{48}
	\expandafter\ifx\csname natexlab\endcsname\relax\def\natexlab#1{#1}\fi
	\providecommand{\bibinfo}[2]{#2}
	\ifx\xfnm\relax \def\xfnm[#1]{\unskip,\space#1}\fi
	\bibitem[{Anderson(1951)}]{Anderson1951}
	\bibinfo{author}{T.~W. Anderson}, \bibinfo{title}{Estimating linear
		restrictions on regression coefficients for multivariate normal
		distributions}, \bibinfo{journal}{The Ann. Math. Statist.}
	\bibinfo{volume}{22} (\bibinfo{year}{1951}) \bibinfo{pages}{327--351}.
	\bibitem[{Bickel et~al.(2009)Bickel, Ritov and Tsybakov}]{Bickel2009}
	\bibinfo{author}{P.~J. Bickel}, \bibinfo{author}{Y.~Ritov},
	\bibinfo{author}{A.~Tsybakov}, \bibinfo{title}{Simultaneous analysis of lasso
		and dantzig selector}, \bibinfo{journal}{Ann. Stat.} \bibinfo{volume}{37}
	(\bibinfo{year}{2009}) \bibinfo{pages}{1705--1732}.
	\bibitem[{Bunea et~al.(2011)Bunea, She and Wegkamp}]{Bunea2011}
	\bibinfo{author}{F.~Bunea}, \bibinfo{author}{Y.~She}, \bibinfo{author}{M.~H.
		Wegkamp}, \bibinfo{title}{Optimal selection of reduced rank estimation of
		high-dimensional matrices}, \bibinfo{journal}{Ann. Stat.}
	\bibinfo{volume}{39} (\bibinfo{year}{2011}) \bibinfo{pages}{1282--1309}.
	\bibitem[{Bunea et~al.(2012)Bunea, She and Wegkamp}]{Bunea2012}
	\bibinfo{author}{F.~Bunea}, \bibinfo{author}{Y.~She}, \bibinfo{author}{M.~H.
		Wegkamp}, \bibinfo{title}{Joint variable and rank selection for parsimonious
		estimation of high-dimensional matrices}, \bibinfo{journal}{Ann. Stat.}
	\bibinfo{volume}{40} (\bibinfo{year}{2012}) \bibinfo{pages}{2359--2388}.
	\bibitem[{Cai et~al.(2024)Cai, Li, Zhou, Yin and Zhang}]{cai2024subgroup}
	\bibinfo{author}{T.~Cai}, \bibinfo{author}{J.~Li}, \bibinfo{author}{Q.~Zhou},
	\bibinfo{author}{S.~Yin}, \bibinfo{author}{R.~Zhang},
	\bibinfo{title}{Subgroup detection based on partially linear additive
		individualized model with missing data in response},
	\bibinfo{journal}{Comput. Stat. Data Anal.} \bibinfo{volume}{192}
	(\bibinfo{year}{2024}) \bibinfo{pages}{107910}.
	\bibitem[{Chen et~al.(2012)Chen, Chan and Stenseth}]{Chen2012a}
	\bibinfo{author}{K.~Chen}, \bibinfo{author}{K.-S. Chan}, \bibinfo{author}{N.~C.
		Stenseth}, \bibinfo{title}{Reduced rank stochastic regression with a sparse
		singular value decomposition}, \bibinfo{journal}{J. Roy. Stat. Soc. B.}
	\bibinfo{volume}{74} (\bibinfo{year}{2012}) \bibinfo{pages}{203--221}.
	\bibitem[{Chen et~al.(2022{\natexlab{a}})Chen, Dong, Xu and
		Zheng}]{chen2022fast}
	\bibinfo{author}{K.~Chen}, \bibinfo{author}{R.~Dong}, \bibinfo{author}{W.~Xu},
	\bibinfo{author}{Z.~Zheng}, \bibinfo{title}{Fast stagewise sparse factor
		regression}, \bibinfo{journal}{J. Mach. Learn. Res} \bibinfo{volume}{23}
	(\bibinfo{year}{2022}{\natexlab{a}}) \bibinfo{pages}{1--45}.
	\bibitem[{Chen et~al.(2019)Chen, Huang, Chan and Yau}]{Chen2019}
	\bibinfo{author}{K.~Chen}, \bibinfo{author}{R.~Huang}, \bibinfo{author}{N.~H.
		Chan}, \bibinfo{author}{C.~Y. Yau}, \bibinfo{title}{Subgroup analysis of
		zero-inflated poisson regression model with applications to insurance data},
	\bibinfo{journal}{Insur. Math. Econ.} \bibinfo{volume}{86}
	(\bibinfo{year}{2019}) \bibinfo{pages}{8--18}.
	\bibitem[{Chen et~al.(2022{\natexlab{b}})Chen, Wang and Yan}]{rrpack}
	\bibinfo{author}{K.~Chen}, \bibinfo{author}{W.~Wang}, \bibinfo{author}{J.~Yan},
	\bibinfo{title}{\textit{rrpack: Reduced-Rank Regression}},
	\bibinfo{journal}{R package version 0.1-13}
	(\bibinfo{year}{2022}{\natexlab{b}}).
	\bibitem[{Chen and Huang(2012)}]{Chen2012sparse}
	\bibinfo{author}{L.~Chen}, \bibinfo{author}{J.~Z. Huang},
	\bibinfo{title}{Sparse reduced-rank regression for simultaneous dimension
		reduction and variable selection}, \bibinfo{journal}{J. Amer. Statist.
		Assoc.} \bibinfo{volume}{107} (\bibinfo{year}{2012})
	\bibinfo{pages}{1533--1545}.
	\bibitem[{Crump et~al.(2008)Crump, Hotz, Imbens and Mitnik}]{Crump2008}
	\bibinfo{author}{R.~K. Crump}, \bibinfo{author}{V.~J. Hotz},
	\bibinfo{author}{G.~W. Imbens}, \bibinfo{author}{O.~A. Mitnik},
	\bibinfo{title}{Nonparametric tests for treatment effect heterogeneity},
	\bibinfo{journal}{Rev. Econ. Stat.} \bibinfo{volume}{90}
	(\bibinfo{year}{2008}) \bibinfo{pages}{389--405}.
	\bibitem[{Dong et~al.(2021)Dong, Li and Zheng}]{dong2021parallel}
	\bibinfo{author}{R.~Dong}, \bibinfo{author}{D.~Li}, \bibinfo{author}{Z.~Zheng},
	\bibinfo{title}{Parallel integrative learning for large-scale multi-response
		regression with incomplete outcomes}, \bibinfo{journal}{Comput. Stat. Data
		Anal.} \bibinfo{volume}{160} (\bibinfo{year}{2021}) \bibinfo{pages}{107243}.
	\bibitem[{Fan and Li(2001)}]{Fan2001}
	\bibinfo{author}{J.~Fan}, \bibinfo{author}{R.~Li}, \bibinfo{title}{Variable
		selection via nonconcave penalized likelihood and its oracle properties},
	\bibinfo{journal}{J. Amer. Stat. Assoc.} \bibinfo{volume}{96}
	(\bibinfo{year}{2001}) \bibinfo{pages}{1348--1360}.
	\bibitem[{Guo and He(2021)}]{guo2021inference}
	\bibinfo{author}{X.~Guo}, \bibinfo{author}{X.~He}, \bibinfo{title}{Inference on
		selected subgroups in clinical trials}, \bibinfo{journal}{J. Amer. Stat.
		Assoc.} \bibinfo{volume}{116} (\bibinfo{year}{2021})
	\bibinfo{pages}{1498--1506}.
	\bibitem[{Hastie and Tibshirani(1996)}]{Hastie1996}
	\bibinfo{author}{T.~Hastie}, \bibinfo{author}{R.~Tibshirani},
	\bibinfo{title}{Discriminant analysis by gaussian mixtures},
	\bibinfo{journal}{J. R. Stat. Soc. Ser. B} \bibinfo{volume}{58}
	(\bibinfo{year}{1996}) \bibinfo{pages}{155--176}.
	\bibitem[{Hu et~al.(2021)Hu, Huang, Liu, Sun and Zhao}]{Hu2021}
	\bibinfo{author}{X.~Hu}, \bibinfo{author}{J.~Huang}, \bibinfo{author}{L.~Liu},
	\bibinfo{author}{D.~Sun}, \bibinfo{author}{X.~Zhao}, \bibinfo{title}{Subgroup
		analysis in the heterogeneous cox model}, \bibinfo{journal}{Stat. Med.}
	\bibinfo{volume}{40} (\bibinfo{year}{2021}) \bibinfo{pages}{739--757}.
	\bibitem[{Izenman(1975)}]{Izenman1975}
	\bibinfo{author}{A.~J. Izenman}, \bibinfo{title}{Reduced-rank regression for
		the multivariate linear model}, \bibinfo{journal}{J. Multivar. Anal.}
	\bibinfo{volume}{5} (\bibinfo{year}{1975}) \bibinfo{pages}{248--264}.
	\bibitem[{Kasahara and Shimotsu(2015)}]{Kasahara2015}
	\bibinfo{author}{H.~Kasahara}, \bibinfo{author}{K.~Shimotsu},
	\bibinfo{title}{Testing the number of components in normal mixture regression
		models}, \bibinfo{journal}{J. Amer. Stat. Assoc.} \bibinfo{volume}{110}
	(\bibinfo{year}{2015}) \bibinfo{pages}{1632--1645}.
	\bibitem[{Koltchinskii et~al.(2011)Koltchinskii, Lounici and
		Tsybakov}]{Koltchinskii2011}
	\bibinfo{author}{V.~Koltchinskii}, \bibinfo{author}{K.~Lounici},
	\bibinfo{author}{A.~B. Tsybakov}, \bibinfo{title}{Nuclear-norm penalization
		and optimal rates for noisy low-rank completion}, \bibinfo{journal}{Ann.
		Stat.} \bibinfo{volume}{39} (\bibinfo{year}{2011})
	\bibinfo{pages}{2302--2329}.
	\bibitem[{Laurent and Massart(2000)}]{Laurent2000}
	\bibinfo{author}{B.~Laurent}, \bibinfo{author}{P.~Massart},
	\bibinfo{title}{Adaptive estimation of a quadratic functional by model
		selection}, \bibinfo{journal}{Ann. Stat.} \bibinfo{volume}{28}
	(\bibinfo{year}{2000}) \bibinfo{pages}{1302--1338}.
	\bibitem[{Li and Chen(2010)}]{Li2010}
	\bibinfo{author}{P.~Li}, \bibinfo{author}{J.~Chen}, \bibinfo{title}{Testing the
		order of a finite mixture}, \bibinfo{journal}{J. Amer. Stat. Assoc.}
	\bibinfo{volume}{105} (\bibinfo{year}{2010}) \bibinfo{pages}{1084--1092}.
	\bibitem[{Li et~al.(2021)Li, Song, Zhang, Zhu and Zhu}]{li2021clusterwise}
	\bibinfo{author}{T.~Li}, \bibinfo{author}{X.~Song}, \bibinfo{author}{Y.~Zhang},
	\bibinfo{author}{H.~Zhu}, \bibinfo{author}{Z.~Zhu},
	\bibinfo{title}{Clusterwise functional linear regression models},
	\bibinfo{journal}{Comput. Stat. Data Anal.} \bibinfo{volume}{158}
	(\bibinfo{year}{2021}) \bibinfo{pages}{107192}.
	\bibitem[{Liao et~al.(2022)Liao, Chung and Hsieh}]{liao2022}
	\bibinfo{author}{H.~Liao}, \bibinfo{author}{Y.~Chung},
	\bibinfo{author}{M.~Hsieh}, \bibinfo{title}{Glutamate: A multifunctional
		amino acid in plants}, \bibinfo{journal}{Plant Sci.} \bibinfo{volume}{318}
	(\bibinfo{year}{2022}) \bibinfo{pages}{111238}.
	\bibitem[{Liu and Lin(2019)}]{Liu2019}
	\bibinfo{author}{L.~Liu}, \bibinfo{author}{L.~Lin}, \bibinfo{title}{Subgroup
		analysis for heterogeneous additive partially linear models and its
		application to car sales data}, \bibinfo{journal}{Comput. Stat. Data Anal.}
	\bibinfo{volume}{138} (\bibinfo{year}{2019}) \bibinfo{pages}{239--259}.
	\bibitem[{Liu et~al.(2023)Liu, Yang, Liu, Jia, Chen, Sun and
		Ma}]{liu2023fusion}
	\bibinfo{author}{M.~Liu}, \bibinfo{author}{J.~Yang}, \bibinfo{author}{Y.~Liu},
	\bibinfo{author}{B.~Jia}, \bibinfo{author}{Y.-F. Chen},
	\bibinfo{author}{L.~Sun}, \bibinfo{author}{S.~Ma}, \bibinfo{title}{A fusion
		learning method to subgroup analysis of alzheimer's disease},
	\bibinfo{journal}{J. Appl. Stat.} \bibinfo{volume}{50} (\bibinfo{year}{2023})
	\bibinfo{pages}{1686--1708}.
	\bibitem[{Liu et~al.(2022)Liu, Ma and Chen}]{liu2022multivariate}
	\bibinfo{author}{X.~Liu}, \bibinfo{author}{S.~Ma}, \bibinfo{author}{K.~Chen},
	\bibinfo{title}{Multivariate functional regression via nested reduced-rank
		regularization}, \bibinfo{journal}{J. Comput. Graph. Stat.}
	\bibinfo{volume}{31} (\bibinfo{year}{2022}) \bibinfo{pages}{231--240}.
	\bibitem[{Lu et~al.(2021)Lu, Qin, Zhu and Tu}]{Lu2021}
	\bibinfo{author}{W.~Lu}, \bibinfo{author}{G.~Qin}, \bibinfo{author}{Z.~Zhu},
	\bibinfo{author}{D.~Tu}, \bibinfo{title}{Multiply robust subgroup
		identification for longitudinal data with dropouts via median regression},
	\bibinfo{journal}{J. Multivar. Anal.} \bibinfo{volume}{181}
	(\bibinfo{year}{2021}) \bibinfo{pages}{104691}.
	\bibitem[{Ma et~al.(2023)Ma, Liu, Xu and Yang}]{ma2023subgroup}
	\bibinfo{author}{H.~Ma}, \bibinfo{author}{C.~Liu}, \bibinfo{author}{S.~Xu},
	\bibinfo{author}{J.~Yang}, \bibinfo{title}{Subgroup analysis for functional
		partial linear regression model}, \bibinfo{journal}{Can. J. Stat.}
	\bibinfo{volume}{51} (\bibinfo{year}{2023}) \bibinfo{pages}{559--579}.
	\bibitem[{Ma and Huang(2017)}]{Shu2017}
	\bibinfo{author}{S.~Ma}, \bibinfo{author}{J.~Huang}, \bibinfo{title}{A concave
		pairwise fusion approach to subgroup analysis}, \bibinfo{journal}{J. Amer.
		Stat. Assoc.} \bibinfo{volume}{112} (\bibinfo{year}{2017})
	\bibinfo{pages}{410--423}.
	\bibitem[{Ma et~al.(2020)Ma, Huang, Zhang and Liu}]{Ma2020}
	\bibinfo{author}{S.~Ma}, \bibinfo{author}{J.~Huang},
	\bibinfo{author}{Z.~Zhang}, \bibinfo{author}{M.~Liu},
	\bibinfo{title}{Exploration of heterogeneous treatment effects via concave
		fusion}, \bibinfo{journal}{Int J Biostat.} \bibinfo{volume}{16}
	(\bibinfo{year}{2020}) \bibinfo{pages}{0026}.
	\bibitem[{Recht et~al.(2010)Recht, Fazel and Parrilo}]{Recht2010}
	\bibinfo{author}{B.~Recht}, \bibinfo{author}{M.~Fazel}, \bibinfo{author}{P.~A.
		Parrilo}, \bibinfo{title}{Guaranteed minimum-rank solutions of linear matrix
		equations via nuclear norm minimization}, \bibinfo{journal}{SIAM Review}
	\bibinfo{volume}{52} (\bibinfo{year}{2010}) \bibinfo{pages}{471--501}.
	\bibitem[{Reinsel et~al.(2022)Reinsel, Velu and Chen}]{reinsel2022multivariate}
	\bibinfo{author}{G.~C. Reinsel}, \bibinfo{author}{R.~P. Velu},
	\bibinfo{author}{K.~Chen}, \bibinfo{title}{\textit{Multivariate reduced-rank
			regression: theory, methods and applications}}, \bibinfo{publisher}{Second
		Editon, New York, Springer}, \bibinfo{year}{2022}.
	\bibitem[{She(2017)}]{She17}
	\bibinfo{author}{Y.~She}, \bibinfo{title}{Selective factor extraction in high
		dimensions}, \bibinfo{journal}{Biometrika} \bibinfo{volume}{104}
	(\bibinfo{year}{2017}) \bibinfo{pages}{97--110}.
	\bibitem[{She and Chen(2017)}]{She2017}
	\bibinfo{author}{Y.~She}, \bibinfo{author}{K.~Chen}, \bibinfo{title}{Robust
		reduced-rank regression}, \bibinfo{journal}{Biometrika} \bibinfo{volume}{104}
	(\bibinfo{year}{2017}) \bibinfo{pages}{633--647}.
	\bibitem[{Szarek(1982)}]{Szarek1982}
	\bibinfo{author}{S.~J. Szarek}, \bibinfo{title}{Nets of grassmann manifold and
		orthogonal groups}, \bibinfo{journal}{In Proceedings of Banach Spaces
		Workshop} \bibinfo{volume}{169} (\bibinfo{year}{1982})
	\bibinfo{pages}{169--185}.
	\bibitem[{Talagrand(2014)}]{Talagrand2014}
	\bibinfo{author}{M.~Talagrand}, \bibinfo{title}{Upper and lower bounds for
		stochastic processes: modern methods and classical problems},
	\bibinfo{publisher}{Springer Science $\&$ Business Media},
	\bibinfo{year}{2014}.
	\bibitem[{Tibshirani(1996)}]{tibshirani1996regression}
	\bibinfo{author}{R.~Tibshirani}, \bibinfo{title}{Regression shrinkage and
		selection via the lasso}, \bibinfo{journal}{J. R. Stat. Soc. Ser. B}
	\bibinfo{volume}{58} (\bibinfo{year}{1996}) \bibinfo{pages}{267--288}.
	\bibitem[{Tseng(2001)}]{Tseng2001}
	\bibinfo{author}{P.~Tseng}, \bibinfo{title}{Convergence of a block coordinate
		descent method for nondifferentiable minimization}, \bibinfo{journal}{J.
		Optim. Theory Appl.} \bibinfo{volume}{109} (\bibinfo{year}{2001})
	\bibinfo{pages}{475--494}.
	\bibitem[{Uematsu et~al.(2019)Uematsu, Fan, Chen, Lv and Lin}]{Uematsu2019}
	\bibinfo{author}{Y.~Uematsu}, \bibinfo{author}{Y.~Fan},
	\bibinfo{author}{K.~Chen}, \bibinfo{author}{J.~Lv}, \bibinfo{author}{W.~Lin},
	\bibinfo{title}{{SOFAR}: large-scale association network learning},
	\bibinfo{journal}{IEEE Trans. Inf. Theory} \bibinfo{volume}{65}
	(\bibinfo{year}{2019}) \bibinfo{pages}{4924--4939}.
	\bibitem[{Wang et~al.(2016)Wang, Gosik, Li, Lindasay and Wu}]{Wang2016}
	\bibinfo{author}{N.~Wang}, \bibinfo{author}{K.~Gosik}, \bibinfo{author}{R.~Li},
	\bibinfo{author}{B.~Lindasay}, \bibinfo{author}{R.~Wu}, \bibinfo{title}{A
		block mixture model to map eqtls for gene clustering and networking},
	\bibinfo{journal}{Sci Rep} \bibinfo{volume}{6} (\bibinfo{year}{2016})
	\bibinfo{pages}{21193}.
	\bibitem[{Wille et~al.(2004)Wille, Zimmermann, Vranov$\acute{a}$,
		F$\ddot{u}$rholz, Laule, Bleuler, Hennig, Preli$\acute{c}$, von Rohr, Thiele,
		Zitzler, Gruissem and B$\ddot{u}$hlmann}]{Wille2004}
	\bibinfo{author}{A.~Wille}, \bibinfo{author}{P.~Zimmermann},
	\bibinfo{author}{E.~Vranov$\acute{a}$},
	\bibinfo{author}{A.~F$\ddot{u}$rholz}, \bibinfo{author}{O.~Laule},
	\bibinfo{author}{S.~Bleuler}, \bibinfo{author}{L.~Hennig},
	\bibinfo{author}{A.~Preli$\acute{c}$}, \bibinfo{author}{P.~von Rohr},
	\bibinfo{author}{L.~Thiele}, \bibinfo{author}{E.~Zitzler},
	\bibinfo{author}{W.~Gruissem}, \bibinfo{author}{P.~B$\ddot{u}$hlmann},
	\bibinfo{title}{Sparse graphical gaussian modeling of the isoprenoid gene
		network in arabidopsis thaliana}, \bibinfo{journal}{Genome Biol.}
	\bibinfo{volume}{5} (\bibinfo{year}{2004}) \bibinfo{pages}{R92}.
	\bibitem[{Yan et~al.(2021)Yan, Yin and Zhao}]{Yan2021}
	\bibinfo{author}{X.~Yan}, \bibinfo{author}{G.~Yin}, \bibinfo{author}{X.~Zhao},
	\bibinfo{title}{Subgroup analysis in censored linear regression},
	\bibinfo{journal}{Stat. Sin.} \bibinfo{volume}{31} (\bibinfo{year}{2021})
	\bibinfo{pages}{1027--1054}.
	\bibitem[{You et~al.(2015)You, Si, Zhang, Zeng, Leung and Li}]{You2015}
	\bibinfo{author}{Z.~You}, \bibinfo{author}{Y.~W. Si},
	\bibinfo{author}{D.~Zhang}, \bibinfo{author}{X.~X. Zeng},
	\bibinfo{author}{S.~C.~H. Leung}, \bibinfo{author}{T.~Li}, \bibinfo{title}{A
		decision-making framework for precision marketing}, \bibinfo{journal}{Expert
		Syst. Appl.} \bibinfo{volume}{42} (\bibinfo{year}{2015})
	\bibinfo{pages}{3357--3367}.
	\bibitem[{Yuan et~al.(2007)Yuan, Ekici, Lu and Monteiro}]{Yuan2007}
	\bibinfo{author}{M.~Yuan}, \bibinfo{author}{A.~Ekici}, \bibinfo{author}{Z.~Lu},
	\bibinfo{author}{R.~Monteiro}, \bibinfo{title}{Dimension reduction and
		coefficient estimation in multivariate linear regression},
	\bibinfo{journal}{J. R. Stat. Soc. Ser. B} \bibinfo{volume}{69}
	(\bibinfo{year}{2007}) \bibinfo{pages}{329--346}.
	\bibitem[{Zhang(2010)}]{Zhang2010}
	\bibinfo{author}{C.~Zhang}, \bibinfo{title}{Nearly unbiased variable selection
		under minimax concave penalty}, \bibinfo{journal}{Ann. Stat.}
	\bibinfo{volume}{38} (\bibinfo{year}{2010}) \bibinfo{pages}{894--942}.
	\bibitem[{Zhang et~al.(2022)Zhang, Zhang, Ma and Fang}]{Zhang2022}
	\bibinfo{author}{X.~Zhang}, \bibinfo{author}{Q.~Zhang},
	\bibinfo{author}{S.~Ma}, \bibinfo{author}{K.~Fang}, \bibinfo{title}{Subgroup
		analysis for high-dimensional functional regression}, \bibinfo{journal}{J.
		Multivar. Anal.} \bibinfo{volume}{192} (\bibinfo{year}{2022})
	\bibinfo{pages}{105100}.
	\bibitem[{Zhang et~al.(2019)Zhang, Wang and Zhu}]{Zhangsinica2019}
	\bibinfo{author}{Y.~Zhang}, \bibinfo{author}{H.~J. Wang},
	\bibinfo{author}{Z.~Zhu}, \bibinfo{title}{Robust subgroup identification},
	\bibinfo{journal}{Stat. Sin.} \bibinfo{volume}{29} (\bibinfo{year}{2019})
	\bibinfo{pages}{1837--1889}.
	\bibitem[{Zhu et~al.(2025)Zhu, Xu and Fan}]{zhu2023simultaneous}
	\bibinfo{author}{X.~Zhu}, \bibinfo{author}{G.~Xu}, \bibinfo{author}{J.~Fan},
	\bibinfo{title}{Simultaneous estimation and group identification for network vector autoregressive model with heterogeneous nodes}, \bibinfo{journal}{J.
		Econom.}  \bibinfo{volume}{249}(\bibinfo{year}{2025}) \bibinfo{pages}{105564}.
		

	
\end{thebibliography}
\end{document}